\documentclass[aip,preprint,pof,amsmath,amssymb]{revtex4-1}
\usepackage[abs]{overpic}
\usepackage{graphicx}
\usepackage{dcolumn}
\usepackage{bm}
\usepackage{placeins}
\usepackage{subfig}
\usepackage{color}

\begin{document}


\title{Forward and backward in time dispersion of fluid and inertial particles in isotropic turbulence}

\author{Andrew D. Bragg}
\email{adbragg265@gmail.com}
\altaffiliation{Present address: Applied Mathematics \& Plasma Physics Group, Los Alamos National Laboratory, Los Alamos, NM 87545, USA.}
\author{Peter J. Ireland}
\author{Lance R. Collins}%
\affiliation{%
Sibley School of Mechanical \& Aerospace Engineering, Cornell University, Ithaca, NY 14853}%
\affiliation{International Collaboration for Turbulence Research}

\date{\today}

\begin{abstract}

In this paper we investigate both theoretically and numerically the forward in time (FIT) and backward in time (BIT) dispersion of fluid and inertial particle pairs in isotropic turbulence.  Fluid particles are known to separate faster BIT than FIT in three-dimensional turbulence, and we find that inertial particles do the same.  However, we find that the irreversibility in the inertial particle dispersion is in general much stronger than that for fluid particles.  For example, the ratio of the BIT to FIT mean-square separation can be up to an order of magnitude larger for inertial particles than for the fluid particles.  We also find that for both the inertial and fluid particles the irreversibility becomes stronger as the scale of their separation decreases.  Regarding the physical mechanism for the irreversibility, we argue that whereas the irreversibility of fluid particle-pair dispersion can be understood in terms of a directional bias arising from the energy transfer process in turbulence, inertial particles experience an additional source of irreversibility arising from the non-local contribution to their velocity dynamics, a contribution which vanishes  in the limit ${St\to0}$, where $St$ is the particle Stokes number.  For each given initial (final, in the backward in time case) separation $\bm{r}^0$ there is an optimum value of $St$ for which the dispersion irreversibility is strongest, as such particles are optimally affected by both sources of irreversibility.  We derive analytical expressions for the BIT, mean-square separation of inertial particles and compare the predictions with numerical data obtained from a $Re_{\lambda}\approx 580$ DNS of particle-laden isotropic turbulent flow.  The small-time theory, which in the dissipation range is valid for times $\leq\max[St\tau_\eta,\tau_\eta]$ (where $\tau_\eta$ is the Kolmogorov timescale), we find excellent agreement between the theoretical predictions and the DNS.  The theory for long-times is in good agreement with the DNS provided that $St$ is small enough so that the inertial particle motion at long-times may be considered as a perturbation about the fluid particle motion, a condition which would in fact be satisfied for arbitrary $St$ at sufficiently long-times in the limit ${Re_{\lambda}\to\infty}$. 

\end{abstract}

\maketitle


%
%

\section{Introduction}

The relative dispersion of fluid particles in turbulent flows has been a subject of intense investigation since the pioneering studies of Taylor \cite{taylor22} and Richardson \cite{richardson26}.  The subject has attracted great interest both because of the theoretical challenges it poses and also because of its importance in environmental problems such as the way pollutants in the atmosphere and in oceans spread out \cite{csanady,berloff02} (see \cite{bourgoin06} for further examples).

The traditional scenario involves forward in time (FIT) dispersion, that is, the variation in time of pairs of fluid particles which have a given initial separation.  Much of the work has focused on the mean-square separation $\langle\vert\bm{r}^{f}(t)\vert^{2}\rangle_{\bm{r}^\prime}$, where $\bm{r}^{f}(t)$ is the fluid particle relative separation vector and  $\langle\cdot\rangle_{\bm{r}^\prime}$ denotes an ensemble average conditioned on $\bm{r}^{f}(t^\prime)=\bm{r}^\prime$ with $t^\prime\leq t$.  Several theoretical predictions for $\langle\vert\bm{r}^{f}(t)\vert^{2}\rangle_{\bm{r}^\prime}$ for varying $\bm{r}^\prime$ and $t$ have been developed, which we shall discuss in \S\ref{FPDT}.  For extensive reviews of this topic see \cite{sawford01,salazar09}.

In \cite{sawford05} the backward in time (BIT) dispersion of fluid particles was investigated and compared with the FIT dispersion.  BIT dispersion concerns the behavior of particle pairs which \emph{arrive} at a given location at a given time and which were dispersed at times in the past (i.e. a given end condition, in contrast to FIT dispersion where it is a given initial condition).  The BIT mean-square separation may be denoted as $\langle\vert\bm{r}^{f}(t^\prime)\vert^{2}\rangle_{\bm{r}}$, where $\bm{r}^{f}(t^\prime)$ is the fluid particle relative separation vector and  $\langle\cdot\rangle_{\bm{r}}$ denotes an ensemble average conditioned on $\bm{r}^{f}(t)=\bm{r}$ with $t^\prime\leq t$.  The simulations in \cite{sawford05} of fluid particle relative dispersion in 3D Navier-Stokes turbulence show that BIT is faster than FIT dispersion, which has also been found in experiments \cite{berg06a}.

A point worth emphasizing to avoid confusion is that in BIT dispersion, the underlying dynamical system is not actually evolving backward in time (a scenario which would be physically uninteresting since time runs forward in physical systems).  The dispersion is BIT only in the sense that one is considering the positions of particles at earlier times $t^\prime$, given their position at a later time $t$, but the dynamical evolution according to which the particle state evolved from time $t^\prime$ to time $t$ is the standard forward in time evolution.  Sawford \emph{et al.} \cite{sawford05} note that it is BIT dispersion, not FIT dispersion that is connected to turbulent mixing processes, which serves to emphasize the physical relevance of studying BIT dispersion.

Compared to the relative dispersion of fluid particles, that of inertial particles has only recently begun to be investigated.  The most comprehensive study to date is that of Bec \emph{et al.} \cite{bec10b} where they used DNS data to investigate the FIT dispersion of inertial particles and also developed mean-field theoretical descriptions of the dispersion process.  They found that for small-times the inertial particles undergo a ballistic separation, driven by their initial velocities.  In the long-time limit, they found that the dispersion tends to the fluid particle Richardson $t^3$ law, but with an inertial correction that decays like $t^{-1}$.  An experimental study \cite{gibert10} also observed a ballistic separation for the inertial particles at small-times, but they were unable to measure the dispersion at long-times.  Theoretical work on the FIT dispersion of inertial particles has also been done for limiting cases such as $St\gg1$ \cite{fouxon08} and time-uncorrelated flows \cite{bec08}.  The present study is inspired by \cite{bec10b} and takes the study to a next step by considering the BIT dispersion of inertial particles, comparing this to the FIT dispersion and seeking to provide both theoretical predictions and physical explanations for irreversability of the inertial particle-pair dispersion.  

As already noted, in \cite{sawford05} it was emphasized that turbulent mixing problems are physically related to BIT, not FIT dispersion.  A motivation for the present work is therefore that it will lead to advances in our understanding concerning the way particle inertia affects mixing processes in turbulent velocity fields.

Another motivation for this work is that it provides insight into how the inertial particle relative velocity theory in \cite{pan10} might be improved.  Since inertial particles posses a memory timescale, their relative velocities are influenced by the fluid velocity field that they have encountered along their path-history, and this depends upon the location of the particle pairs at times in the past, i.e. their BIT dispersion.  However, the authors in \cite{pan10} note that an investigation into the BIT dispersion of inertial particle-pairs has not yet been undertaken, and therefore in their theory they approximate the BIT dispersion by the known FIT dispersion which was examined in \cite{bec10b}.  Under the assumption that FIT and BIT dispersion are not equivalent for inertial particles, in \cite{bragg14c} it was suggested that the approximation of their equivalence in \cite{pan10} could be a source of error in the relative velocity predictions from the theory.  This highlights the need to understand and predict the BIT dispersion of inertial particles in turbulence.

The outline of the paper is as follows.  In \S\ref{GEGS} we construct the general expressions which describe the mean-square separation of inertial particles in a turbulent flow field that will be used in subsequent sections to derive closed, analytic predictions for the particle dispersion.  In \S\ref{FPD} we consider the relative dispersion of fluid particles both theoretically and using DNS simulations.  In \S\ref{IPD} we present the additional irreversibility mechanism that inertia introduces to the dispersion, derive theoretical predictions for the mean-square dispersion at small and long-times, and then consider DNS data for these quantities, against which we test the theoretical predictions.

\section{Governing equations and general solutions}
\label{GEGS}

In this section we construct the exact, but unclosed, expressions for the FIT and BIT mean square separation of the inertial particles that will be used in subsequent sections as the basis from which closed, analytical expressions for these quantities are derived.

We consider the relative dispersion of monodisperse inertial particles that are small, $d/\eta\ll1$ (where $d$ is the particle diameter, and $\eta$ is the Kolmogorov lengthscale), dense $\rho_p/\rho_f\gg1$ (where $\rho_p$ and $\rho_f$ are the particle and fluid material densities, respectively) and subject to linear drag forces only.  The latter approximation is generally considered to be appropriate for describing the dynamics of water droplets in air (e.g. \cite{shaw03}).  It is possible that for some of the larger particles considered in this present study, the linear drag approximation may not be valid.  However, the linear drag approximation will serve as a first approximation for understanding and predicting the BIT dispersion of inertial particles, and we hope that in future work more realistic equations of motion could be considered.  Furthermore, excluding the range of very large particle Reynolds numbers, we expect that non-linear drag effects will only change the dispersion quantitatively and that the essential physical aspects of BIT dispersion will not be qualitatively affected by the liner drag approximation.

The equation governing the relative motion of particles satisfying the aforementioned requirements is then obtained from the simplified form of the Maxey-Riley equation~\cite{maxey83}
\begin{align}
\ddot{\bm{r}}^{p}(t)=\dot{\bm{w}}^{p}(t)=\frac{1}{\tau_{p}}\Big(\Delta\bm{u}^{p}(t)-\bm{w}^{p}(t)\Big),
\label{reom}
\end{align}
where $\bm{r}^{p}(t),\bm{w}^{p}(t)$ are the inertial particle-pair relative position and relative velocity vectors, $\tau_{p}$ is the momentum response time of the particles (we will also use the Stokes number later in the paper $St\equiv\tau_p/\tau_\eta$, where $\tau_\eta$ is the fluid Kolmogorov timescale), $\Delta\bm{u}^{p}(t)\equiv\Delta\bm{u}(\bm{x}^{p}(t),\bm{r}^{p}(t),t)$ is the difference between the fluid velocity field evaluated at the positions of the two particles and $\bm{x}^{p}(t)$ is the position of the reference particle.  The formal solution to (\ref{reom}) may be written as (for isotropic $\Delta\bm{u}$)
\begin{align}
\bm{r}^{p}(t)&=\bm{r}^{p}(t^\prime)+G(t-t^\prime)\bm{w}^{p}(t^\prime)+\tau_{p}^{-1}\int\limits_{t^\prime}^{t}G(t-s)\Delta\bm{u}^{p}(s)ds,\quad t^\prime\leq t,\label{rpsol}\\
\bm{w}^{p}(t)&=\dot{G}(t-t^\prime)\bm{w}^{p}(t^\prime)+\tau_{p}^{-1}\int\limits_{t^\prime}^{t}\dot{G}(t-s)\Delta\bm{u}^{p}(s)ds,\quad t^\prime\leq t,
\label{wpsol}
\end{align}
where $G$ is the Green function for the equation of motion for $\bm{r}^{p}(t)$ 
\begin{align}
G(t-t^\prime)&\equiv\tau_{p}\Big(1-\exp[-\tau_{p}^{-1}(t-t^\prime)]\Big),
\label{G}\\
\dot{G}(t-t^\prime)&\equiv\exp[-\tau_{p}^{-1}(t-t^\prime)].
\label{Gdot}
\end{align}
The FIT dispersion PDF is defined as
\begin{align}
\varrho^{F}(\bm{r},t\vert\bm{r}^\prime,t^\prime)\equiv\Big\langle\delta(\bm{r}^{p}(t)-\bm{r})\Big\rangle_{\bm{r}^\prime},	
\label{fitPDF}
\end{align}
and for BIT dispersion
\begin{align}
\varrho^{B}(\bm{r}^\prime,t^\prime\vert\bm{r},t)\equiv\Big\langle\delta(\bm{r}^{p}(t^\prime)-\bm{r}^\prime)\Big\rangle_{\bm{r}},
\label{bitPDF}
\end{align}
and in each case $t^\prime\leq t$.  The notation $\langle{\cdot}\rangle_{\bm{r}^\prime}$ and $\langle{\cdot}\rangle_{\bm{r}}$ in (\ref{fitPDF}) and (\ref{bitPDF}) denote conditional ensemble averaging; conditioned on $\bm{r}^{p}(t^\prime)=\bm{r}^\prime$ in the FIT case (`initial-time conditioning') and $\bm{r}^{p}(t)=\bm{r}$ in the BIT case (`end-time conditioning').  In this paper we are interested in the mean-square separation behavior rather than the full dispersion PDF.  From (\ref{fitPDF}) and (\ref{bitPDF}) we may define the FIT and BIT mean-square separation 
\begin{align}
\Big\langle\vert\bm{r}^{p}(t)\vert^{2}\Big\rangle_{\bm{r}^\prime}\equiv\int\limits_{\bm{r}}\bm{r}\bm{\cdot}\bm{r}\varrho^{F}(\bm{r},t\vert\bm{r}^\prime,t^\prime)d\bm{r},	
\label{MSfit}
\end{align}
\begin{align}
\Big\langle\vert\bm{r}^{p}(t^\prime)\vert^{2}\Big\rangle_{\bm{r}}\equiv\int\limits_{\bm{r}^\prime}\bm{r}^{\prime}\bm{\cdot}\bm{r}^{\prime}\varrho^{B}(\bm{r}^\prime,t^\prime\vert\bm{r},t)d\bm{r}^{\prime}.	
\label{MSbit}
\end{align}
We may construct an exact expression for (\ref{MSfit}) using (\ref{rpsol}) 
\begin{align}
\begin{split}
\Big\langle\vert\bm{r}^{p}(t)\vert^{2}\Big\rangle_{\bm{r}^\prime}=\,&\bm{r}^\prime\bm{\cdot}\bm{r}^\prime+2G(t-t^\prime)\bm{r}^\prime\bm{\cdot}\Big\langle \bm{w}^{p}(t^\prime)\Big\rangle_{\bm{r}^\prime}+2\tau_{p}^{-1}\bm{r}^\prime\bm{\cdot}\int\limits_{t^\prime}^{t}G(t-s)\Big\langle\Delta\bm{u}^{p}(s)\Big\rangle_{\bm{r}^\prime}ds\\
&+G^{2}(t-t^\prime)\Big\langle\bm{w}^{p}(t^\prime)\bm{\cdot}\bm{w}^{p}(t^\prime)\Big\rangle_{\bm{r}^\prime}+2\tau_{p}^{-1}G(t-t^\prime)\int\limits_{t^\prime}^{t}G(t-s)\Big\langle\bm{w}^{p}(t^\prime)\bm{\cdot}\Delta \bm{u}^{p}(s)\Big\rangle_{\bm{r}^\prime}ds\\
&+\tau_{p}^{-2}\int\limits_{t^\prime}^{t}\int\limits_{t^\prime}^{t}G(t-s)G(t-s^\prime)\Big\langle\Delta\bm{u}^{p}(s)\bm{\cdot}\Delta\bm{u}^{p}(s^\prime)\Big\rangle_{\bm{r}^\prime}ds^\prime\,ds,
\label{FITmsr}
\end{split}
\end{align}

and rearranging (\ref{rpsol}) for $\bm{r}^{p}(t^\prime)$ we may construct the expression for (\ref{MSbit}) 
\begin{align}
\begin{split}
\Big\langle\vert\bm{r}^{p}(t^\prime)\vert^{2}\Big\rangle_{\bm{r}}=\,&\bm{r}\bm{\cdot}\bm{r}-2G(t-t^\prime)\bm{r}\bm{\cdot}\Big\langle \bm{w}^{p}(t^\prime)\Big\rangle_{\bm{r}}-2\tau_{p}^{-1}\bm{r}\bm{\cdot}\int\limits_{t^\prime}^{t}G(t-s)\Big\langle\Delta\bm{u}^{p}(s)\Big\rangle_{\bm{r}}ds\\
&+G^{2}(t-t^\prime)\Big\langle\bm{w}^{p}(t^\prime)\bm{\cdot}\bm{w}^{p}(t^\prime)\Big\rangle_{\bm{r}}+2\tau_{p}^{-1}G(t-t^\prime)\int\limits_{t^\prime}^{t}G(t-s)\Big\langle \bm{w}^{p}(t^\prime)\bm{\cdot}\Delta \bm{u}^{p}(s)\Big\rangle_{\bm{r}}ds\\
&+\tau_{p}^{-2}\int\limits_{t^\prime}^{t}\int\limits_{t^\prime}^{t}G(t-s)G(t-s^\prime)\Big\langle\Delta\bm{u}^{p}(s)\bm{\cdot}\Delta\bm{u}^{p}(s^\prime)\Big\rangle_{\bm{r}}ds^\prime\,ds.
\label{BITmsr}
\end{split}
\end{align}
Since we are interested in statistically stationary systems, we may set the `conditioning time' to zero and consider the dispersion behavior as a function of time separation.  In the FIT case this amounts to setting $t^\prime=0$ and in the BIT case setting $t=0$.  Further, since $t^\prime\leq t$ we may re-write (\ref{FITmsr}) and (\ref{BITmsr}) as
\begin{align}
\begin{split}
\Big\langle\vert\bm{r}^{p}(\mathcal{T})\vert^{2}\Big\rangle_{\bm{r}^0}=\,&\bm{r}^{0}\bm{\cdot}\bm{r}^{0}+2G(\mathcal{T})\bm{r}^{0}\bm{\cdot}\Big\langle \bm{w}^{p}(0)\Big\rangle_{\bm{r}^0}+2\tau_{p}^{-1}\bm{r}^{0}\bm{\cdot}\int\limits_{0}^{\mathcal{T}}G(\mathcal{T}-s)\Big\langle\Delta\bm{u}^{p}(s)\Big\rangle_{\bm{r}^0}ds\\
&+G^{2}(\mathcal{T})\Big\langle\bm{w}^{p}(0)\bm{\cdot}\bm{w}^{p}(0)\Big\rangle_{\bm{r}^0}+2\tau_{p}^{-1}G(\mathcal{T})\int\limits_{0}^{\mathcal{T}}G(\mathcal{T}-s)\Big\langle\bm{w}^{p}(0)\bm{\cdot}\Delta\bm{u}^{p}(s)\Big\rangle_{\bm{r}^0}ds\\
&+\tau_{p}^{-2}\int\limits_{0}^{\mathcal{T}}\int\limits_{0}^{\mathcal{T}}G(\mathcal{T}-s)G(\mathcal{T}-s^\prime)\Big\langle\Delta\bm{u}^{p}(s)\bm{\cdot}\Delta\bm{u}^{p}(s^\prime)\Big\rangle_{\bm{r}^0}ds^\prime\,ds,
\label{FITmsr2}
\end{split}
\end{align}
where $\mathcal{T}=t-t^\prime$ with $t^\prime=0$, $t^\prime\leq t$ so that $\mathcal{T}\geq0$ and 
\begin{align}
\begin{split}
\Big\langle\vert\bm{r}^{p}(-\mathcal{T})\vert^{2}\Big\rangle_{\bm{r}^0}=\,&\bm{r}^{0}\bm{\cdot}\bm{r}^{0}-2G(\mathcal{T})\bm{r}^{0}\bm{\cdot}\Big\langle \bm{w}^{p}(-\mathcal{T})\Big\rangle_{\bm{r}^0}-2\tau_{p}^{-1}\bm{r}^{0}\bm{\cdot}\int\limits^{0}_{-\mathcal{T}}G(-s)\Big\langle\Delta\bm{u}^{p}(s)\Big\rangle_{\bm{r}^0}ds\\
&+G^{2}(\mathcal{T})\Big\langle\bm{w}^{p}(-\mathcal{T})\bm{\cdot}\bm{w}^{p}(-\mathcal{T})\Big\rangle_{\bm{r}^0}+2\tau_{p}^{-1}G(\mathcal{T})\int\limits^{0}_{-\mathcal{T}}G(-s)\Big\langle\bm{w}^{p}(-\mathcal{T})\bm{\cdot}\Delta \bm{u}^{p}(s)\Big\rangle_{\bm{r}^0}ds\\
&+\tau_{p}^{-2}\int\limits^{0}_{-\mathcal{T}}\int\limits^{0}_{-\mathcal{T}}G(-s)G(-s^\prime)\Big\langle\Delta\bm{u}^{p}(s)\bm{\cdot}\Delta\bm{u}^{p}(s^\prime)\Big\rangle_{\bm{r}^0}ds^\prime\,ds,
\label{BITmsr2}
\end{split}
\end{align}
where again $\mathcal{T}=t-t^\prime$, but now with $t=0$, $t^\prime\leq t$ so that $\mathcal{T}\geq0$ and $t^\prime=-\mathcal{T}$, and $\langle{\cdot}\rangle_{\bm{r}^0}$ denotes an ensemble average conditioned on $\bm{r}^{p}(\mathcal{T}=0)=\bm{r}^0$.  In going from (\ref{BITmsr}) to (\ref{BITmsr2}), we note that\[G(t-t^\prime)=\tau_{p}\Big(1-\exp[-\tau_{p}^{-1}(t-t^\prime)]\Big)=\tau_{p}\Big(1-\exp[-\tau_{p}^{-1}\mathcal{T}]\Big)=G(\mathcal{T}),\]
and\[G(t-s)=\tau_{p}\Big(1-\exp[-\tau_{p}^{-1}(t-s)]\Big)=\tau_{p}\Big(1-\exp[\tau_{p}^{-1}s]\Big)=G(-s).\]We may develop theoretical descriptions for the FIT and BIT dispersion of inertial particles by applying closure approximations to (\ref{FITmsr2}) and (\ref{BITmsr2}) to construct closed analytical solutions.  

Note that we have not constructed (\ref{BITmsr2}) using the solutions to the time-reversed form of the equations of motion, which is commonly done when analyzing BIT problems.  We have chosen not to construct the BIT results via this method because we believe it hinders the physical understanding of the problem since, as explained earlier, in BIT dispersion the dynamical system is not actually evolving backward in time.  Rather, we have therefore constructed (\ref{BITmsr2}) in a manner consistent with how the same statistics would be obtained in an experiment or DNS where the BIT statistics are constructed by recording the trajectories of the particles (which are being evolved using the standard forward-in-time equations of motion) and then subsequently evaluate the BIT statistics based on the particle trajectory histories.  

In addition we will also consider the dispersion of fluid particles whose equation of relative motion is simply
\begin{align}
\dot{\bm{r}}^f(t)=\Delta\bm{u}^f(t),	
\end{align}
with solution
\begin{align}
\bm{r}^f(t)=\bm{r}^f(t')+\int\limits_{t'}^t\Delta\bm{u}^f(s)\,ds,\label{rfsol}	
\end{align}
where $\bm{r}^f(t)$ is the relative separation between two fluid particles and $\Delta\bm{u}^{f}(t)\equiv\Delta\bm{u}(\bm{x}^{f}(t),\bm{r}^{f}(t),t)$ is the difference in the fluid velocity evaluated at the positions of the two particles, $\bm{x}^f(t)$ being the position of the reference fluid particle.  Here and throughout the superscript `$p$' denotes that the variable is defined along inertial particle trajectories, and superscript`$f$' is used to denote that the variable is defined along fluid particle trajectories.  Following the same steps as was used to derive (\ref{FITmsr2}) and (\ref{BITmsr2}) we obtain for the fluid particles
\begin{align}
\begin{split}
\Big\langle\vert\bm{r}^{f}(\mathcal{T})\vert^{2}\Big\rangle_{\bm{r}^0}=&\bm{r}^{0}\bm{\cdot}\bm{r}^{0}+2\bm{r}^{0}\bm{\cdot}\int\limits_{0}^{\mathcal{T}}\Big\langle\Delta\bm{u}^{f}(s)\Big\rangle_{\bm{r}^0}ds+\int\limits_{0}^{\mathcal{T}}\int\limits_{0}^{\mathcal{T}}\Big\langle\Delta\bm{u}^{f}(s)\bm{\cdot}\Delta\bm{u}^{f}(s^\prime)\Big\rangle_{\bm{r}^0}ds^\prime\,ds,
\label{FITmsr2FP}
\end{split}
\end{align}
\begin{align}
\begin{split}
\Big\langle\vert\bm{r}^{f}(-\mathcal{T})\vert^{2}\Big\rangle_{\bm{r}^0}=&\bm{r}^{0}\bm{\cdot}\bm{r}^{0}-2\bm{r}^{0}\bm{\cdot}\int\limits^{0}_{-\mathcal{T}}\Big\langle\Delta\bm{u}^{f}(s)\Big\rangle_{\bm{r}^0}ds+\int\limits^{0}_{-\mathcal{T}}\int\limits^{0}_{-\mathcal{T}}\Big\langle\Delta\bm{u}^{f}(s)\bm{\cdot}\Delta\bm{u}^{f}(s^\prime)\Big\rangle_{\bm{r}^0}ds^\prime\,ds.
\label{BITmsr2FP}
\end{split}
\end{align}

\section{Fluid particle dispersion}
\label{FPD}

We will first consider the case of fluid particle dispersion before considering inertial particle dispersion, which is the main contribution of this paper.  Our purpose in this section is not to derive new results but to consider various results and explanations that have been previously proposed.  This will be especially helpful when we consider in \S\ref{IPD} inertial particle dispersion, which introduces additional complexities compared to the fluid particle dispersion.

\subsection{Irreversibility mechanisms}
\label{FPIM}

The FIT and BIT dispersion of fluid particles in turbulence has been considered in several studies, using theoretical, computational and experimental techniques (e.g. \cite{sawford05,berg06a,jucha14}).  These studies have revealed that in 3D turbulence, BIT dispersion is faster than FIT dispersion.  Different explanations have been given for this observed irreversibility.  In \cite{sawford05} the behavior of the odd-moments of the fluid velocity increments $\Delta\bm{u}$ in turbulence under time-reversal was used to provide an explanation for the difference between FIT and BIT dispersion.  In \cite{berg06a} the authors appealed to the behavior of the eigenvalues of the strain-rate tensor $\bm{\mathcal{S}}(\bm{x},t)\equiv (1/2)[\bm{\nabla_x}\bm{u}+(\bm{\nabla_x}\bm{u})^\top]$ under time-reversal to explain the difference.  The largest eigenvalue of $\bm{\mathcal{S}}(\bm{x},-t)$ is greater than the largest eigenvalue of $\bm{\mathcal{S}}(\bm{x},t)$, thus explaining, not only why FIT and BIT dispersion are different, but also why BIT dispersion is faster than FIT dispersion.  They also argued that since the course-grained version of $\bm{\mathcal{S}}$ exhibits similar dynamics, the same argument also applies for dispersion in the inertial range of the turbulence.

We may further clarify the irreversibility mechanism by considering the equation governing the relative separation of the fluid particle-pair which is simply $\dot{\bm{r}}^f(t)=\Delta\bm{u}(\bm{r}^f(t),t)$.  FIT dispersion corresponds to particles separating (on average) as time increases (i.e. $\dot{\bm{r}}^f(t)>\bm{0}$), whereas BIT can be thought of as particles approaching each other as time increases (i.e. $\dot{\bm{r}}^f(t)<\bm{0}$).  Since the PDF of $\Delta\bm{u}$ is negatively skewed in 3D turbulence because of the energy transfer to the small scales, the particle-pairs move together faster than they move apart, and hence BIT is faster than FIT dispersion.  This type of energy flux argument can also be quantified in the inertial range by considering a small-time expansion of the dispersion process, and such an analysis indeed predicts that BIT dispersion should be faster than FIT dispersion in 3D turbulence \cite{jucha14}.  Consistent with these arguments, fluid particle dispersion in kinematically simulated flow fields, where $\Delta\bm{u}$ has a Gaussian distribution, exhibits FIT and BIT symmetry \cite{flohr05}.
  
These arguments for FIT and BIT asymmetry based on the energy flux and associated asymmetry in the PDF for $\Delta\bm{u}$ also suggest that in 2D turbulence where there is a flux of energy towards the large scales (yielding a positively skewed PDF for $\Delta\bm{u}$), FIT dispersion should be faster than BIT dispersion, something that has been shown numerically in \cite{faber09}.  

What each of these explanations share in common is that the time-irreversibility of fluid particle dispersion in turbulence arises, fundamentally, because of the intrinsic time directionality in turbulence dynamics, a consequence of its dissipative nature.  However, although this is the physical origin of the irreversibility in Navier-Stokes turbulence, any model flow field which generates asymmetric probability density functions for $\Delta\bm{u}(\bm{r},t)$ would give rise to irreversible fluid particle-pair dispersion.  

\subsection{Theoretical results}
\label{FPDT}

Having considered how FIT and BIT differ in turbulence, we now turn to consider various theoretical predictions that have been made to describe the fluid particle-pair dispersion.  Note that throughout this paper we are considering the statistically stationary state of incompressible flow where the fluid particles are assumed to be fully mixed in the system.

The FIT and BIT mean square dispersion of fluid particles is given by (\ref{FITmsr2FP}) and (\ref{BITmsr2FP}).  We now introduce the turnover timescale of $\Delta\bm{u}$ at separation $r^0\equiv\vert\bm{r}^0\vert$, known in the context of dispersion studies as the Batchelor timescale $\tau_{r^{0}}$.  For initial separations in the dissipation regime, we take $\tau_{r^{0}}=\tau_{\eta}$, and in the inertial range $\tau_{r^{0}}=(\vert\bm{r}^{0}\vert^{2}/\langle\epsilon\rangle)^{1/3}$, where $\langle\epsilon\rangle$ is the turbulent kinetic energy dissipation rate.  In the regime ${\mathcal{T}\ll\tau_{r^0}}$ we may make the approximation 
\begin{align}
\Delta\bm{u}^{f}(\mathcal{T})\approx\Delta\bm{u}^{f}(0)+\mathcal{O}(\mathcal{T}/\tau_{r^0}),\label{uTexp}
\end{align}
(and similarly for the terms in (\ref{FITmsr2}) and (\ref{BITmsr2}) involving $s$ and $s^\prime$ in the time arguments) and introducing this approximation into (\ref{FITmsr2FP}) and (\ref{BITmsr2FP}), we obtain 
\begin{align}
\Big\langle\vert\bm{r}^{f}(\mathcal{T})\vert^{2}\Big\rangle_{\bm{r}^0}\approx\Big\langle\vert\bm{r}^{f}(-\mathcal{T})\vert^{2}\Big\rangle_{\bm{r}^0}\approx\vert\bm{r}^0\vert^{2}+\mathcal{T}^{2}\Big\langle\vert\Delta\bm{u}^{f}(0)\vert^{2}\Big\rangle_{\bm{r}^0}\label{Bal}.	
\end{align}
Notice that in the ballistic regime the fluid-particle dispersion is time-reversible: this is a consequence of (\ref{uTexp}), which ignores the dynamical evolution of $\Delta\bm{u}$ along the particle trajectories, and it is the nature of the dynamical evolution of the turbulence that gives rise to irreversibility in the dispersion process for fluid particles.

In \cite{ouellette06c} the authors consider the importance of correctly including the terms involving $\bm{r}^{0}$ in the description of the mean square separation (this is also discussed in \cite{salazar09}).  Specifically they consider the difference between (\ref{Bal}) and
\begin{align}
\Big\langle\vert\bm{r}^{f}(\mathcal{T})-\bm{r}^{0}\vert^{2}\Big\rangle_{\bm{r}^0}\approx\mathcal{T}^{2}\Big\langle\vert\Delta\bm{u}^{f}(0)\vert^{2}\Big\rangle_{\bm{r}^0}\label{Bal2}.	
\end{align}
We may write
\begin{align}
\Big\langle\vert\bm{r}^{f}(\mathcal{T})-\bm{r}^{0}\vert^{2}\Big\rangle_{\bm{r}^0}=\Big\langle\vert\bm{r}^{f}(\mathcal{T})\vert^{2}\Big\rangle_{\bm{r}^0}-2\bm{r}^{0}\bm{\cdot}\Big\langle\bm{r}^{f}(\mathcal{T})\Big\rangle_{\bm{r}^0}+\vert\bm{r}^0\vert^{2},
\end{align}
showing that a $\mathcal{T}$ dependent difference between (\ref{Bal}) and (\ref{Bal2}) arises when $\langle\bm{r}^{f}(\mathcal{T})\rangle_{\bm{r}^0}\neq\bm{r}^{0}$.  From (\ref{rfsol}) we may derive the result for ${\mathcal{T}\ll\tau_{r^0}}$
\begin{align}
\Big\langle\bm{r}^{f}(\mathcal{T})\Big\rangle_{\bm{r}^0}=\bm{r}^{0}+\mathcal{T}\Big\langle\Delta\bm{u}^{f}(0)\Big\rangle_{\bm{r}^0}+\mathcal{O}(\mathcal{T}/\tau_{r^0}),
\end{align}
giving 
\begin{align}
\Big\langle\vert\bm{r}^{f}(\mathcal{T})-\bm{r}^{0}\vert^{2}\Big\rangle_{\bm{r}^0}=\Big\langle\vert\bm{r}^{f}(\mathcal{T})\vert^{2}\Big\rangle_{\bm{r}^0}-\vert\bm{r}^{0}\vert^{2}-2\mathcal{T}\bm{r}^{0}\bm{\cdot}\Big\langle\Delta\bm{u}^{f}(0)\Big\rangle_{\bm{r}^0}+\mathcal{O}(\mathcal{T}/\tau_{r^0}).\label{I1}
\end{align}
In \cite{ouellette06c} the authors find that if they plot their experimental data for\[\Big\langle\vert\bm{r}^{f}(\mathcal{T})-\bm{r}^{0}\vert^{2}\Big\rangle_{\bm{r}^0}\Big/\mathcal{T}^{2}\Big\langle\vert\Delta\bm{u}^{f}(0)\vert^{2}\Big\rangle_{\bm{r}^0},\]they find a good collapse for ${\mathcal{T}<\tau_{r^0}}$, verifying the validity of the ballistic prediction.  However when they plot \[\Big(\Big\langle\vert\bm{r}^{f}(\mathcal{T})\vert^{2}\Big\rangle_{\bm{r}^0}-\vert\bm{r}^{0}\vert^{2}\Big)\Big/\mathcal{T}^{2}\Big\langle\vert\Delta\bm{u}^{f}(0)\vert^{2}\Big\rangle_{\bm{r}^0},\]they do not find a good collapse of the data indicating that the term of difference between the two expressions is important.  From (\ref{I1}) we see that the difference between these two expressions depends upon $\langle\Delta\bm{u}^{f}(0)\rangle_{\bm{r}^0}$ and for fully mixed fluid particles in incompressible, isotropic turbulence 
\begin{align}
\Big\langle\Delta\bm{u}^{f}(0)\Big\rangle_{\bm{r}^0}\equiv\frac{1}{\varphi(\bm{r}^0,t)}\Big\langle\Delta\bm{u}^{f}(0)\delta\Big(\bm{r}^f(0)-\bm{r}^0\Big)\Big\rangle=\Big\langle\Delta\bm{u}(\bm{r}^0,0)\Big\rangle=\bm{0},
\end{align}
where $\varphi(\bm{r}^0,t)\equiv\langle\delta(\bm{r}^f(0)-\bm{r}^0)\rangle$.  Consequently, under the ballistic approximation
\begin{align}
\Big\langle\vert\bm{r}^{f}(\mathcal{T})-\bm{r}^{0}\vert^{2}\Big\rangle_{\bm{r}^0}=\Big\langle\vert\bm{r}^{f}(\mathcal{T})\vert^{2}\Big\rangle_{\bm{r}^0}-\vert\bm{r}^{0}\vert^{2}.
\end{align}
The results in \cite{ouellette06c} for the ballistic case are therefore surprising, since in this regime there should be no difference between (\ref{Bal}) and (\ref{Bal2}).  However, the data presented in Fig.~6 of \cite{ouellette06c} implies \[\Big(\Big\langle\vert\bm{r}^{f}(\mathcal{T})\vert^{2}\Big\rangle_{\bm{r}^0}-\vert \bm{r}^{0}\vert^{2}\Big)\Big/\tau_{\eta}^{2}\Big\langle\vert\Delta\bm{u}^{f}(0)\vert\Big\rangle_{\bm{r}^{0}}\to\text{finite value as $\mathcal{T}\to0$},\] which cannot be correct (by definition $\langle\vert\bm{r}^{f}(\mathcal{T})\vert^{2}\rangle_{\bm{r}^0}-\vert \bm{r}^{0}\vert^{2}\to0$ as $\mathcal{T}\to0$).  One explanation for this may be errors introduced by the relatively large size of the bins used to construct the statistics from their experimental data, having widths $\approx 43\eta$.

In the case where ${\mathcal{T}\geq\mathcal{O}(\tau_{r^0})}$, $\langle\bm{r}^{f}(\mathcal{T})\rangle_{\bm{r}^0}$ depends upon $\int^\mathcal{T}_0 \langle\Delta\bm{u}^{f}(s)\rangle_{\bm{r}^0}\,ds$, and $\langle\Delta\bm{u}^{f}(s)\rangle_{\bm{r}^0}\neq\bm{0}$ even for isotropic turbulence.  The reason is the conditional nature of the average; particle-pairs which were at $\bm{r}^{0}$ at $\mathcal{T}=0$ and are on average separating will be experiencing positive velocity differences on average, i.e. $\langle\Delta\bm{u}^{f}(s)\rangle_{\bm{r}^0}\geq\bm{0}$.  Nevertheless, we expect that the effect of this on the prediction of $\langle\vert\bm{r}^{f}(\mathcal{T})\vert^{2}\rangle_{\bm{r}^0}$ will be small compared to the higher-order moment terms in its evolution equation.  Similar arguments also describe the BIT case, only in that case, since fluid particles are on average approaching each other towards $\bm{r}^{0}$, then $\langle\Delta\bm{u}^{f}(s)\rangle_{\bm{r}^0}\leq\bm{0}$.

Having considered the small-time behavior that gives rise to the $\mathcal{T}^2$ ballistic relationship, we now consider the finite $\mathcal{T}$ behavior, at which point the irreversibility of the disperson process becomes manifest.  The simplest way to proceed is to consider the contribution from higher order terms in the $\mathcal{T}/\tau_{r^{0}}$ expansion in (\ref{uTexp}).  Accounting for the second term in the expansion ($\propto\mathcal{T}$) gives rise to the first term describing the break in time-reversibility of the dispersion, and involves the correlation between the fluid relative velocity and relative acceleration measured at ${\mathcal{T}=0}$ \cite{jucha14}.  However, since this expansion becomes formally divergent for ${\mathcal{T}\geq\mathcal{O}(\tau_{r^0})}$, we seek alternative approximations to describe the finite $\mathcal{T}/\tau_{r^{0}}$ behavior of the dispersion.  We begin by considering the case for $\bm{r}^0$ in the dissipation range, and then for $\bm{r}^0$ in the inertial range.

In the dissipation regime $\Delta\bm{u}^f(\mathcal{T})\approx\bm{\Gamma}^f(\mathcal{T})\bm{\cdot}\bm{r}^f(\mathcal{T})$, where $\bm{\Gamma}^f(\mathcal{T})\equiv\bm{\nabla_x}\bm{u}(\bm{x}^f(\mathcal{T}),\mathcal{T})$.  In this case we have $\dot{\bm{r}}^f(\mathcal{T})=\bm{\Gamma}^f(\mathcal{T})\bm{\cdot}\bm{r}^f(\mathcal{T})$ whose solution may be expressed using the time-ordered exponential $\exp_{\mathrm{T}}[\cdot]$ as \cite{falkovich01}
\begin{align}
\bm{r}^f(\mathcal{T})=\bm{r}^f(0)\exp_{\mathrm{T}}\Bigg(\int_{0}^{\mathcal{T}}\bm{\Gamma}^f(s)\,ds\Bigg).
\end{align}
Based upon this observation that the pair separation grows exponentially in time in the dissipation range, Batchelor \cite{batchelor52a} gave an order-of-magnitude estimate for $\vert\bm{r}^{f}(\mathcal{T})\vert$ for large $\mathcal{T}/\tau_\eta$ which gives rise to the prediction
\begin{align}
\begin{split}
\Big\langle\vert\bm{r}^{f}(\mathcal{T})\vert^{2}\Big\rangle_{\bm{r}^0}=\vert\bm{r}^{0}\vert^{2}\exp[2B\tau_{\eta}^{-1}\mathcal{T}],
\label{BatchelorSmallr}
\end{split}
\end{align}
where various values for $B$ have been given (see \cite{salazar12a}).  It is important to emphasize that (\ref{BatchelorSmallr}) is only supposed to be valid for large $\mathcal{T}/\tau_\eta$; expanding the exponential gives   
\begin{align}
\begin{split}
\exp[2B\tau_{\eta}^{-1}\mathcal{T}]=1+2B\tau_{\eta}^{-1}\mathcal{T}+2B^2\tau_{\eta}^{-2}\mathcal{T}^2+\mathcal{O}(\mathcal{T}^3),
\label{BatchExpan}
\end{split}
\end{align}
which shows that (\ref{BatchelorSmallr}) is not consistent with the ballistic behavior that should be obtained in the regime $\mathcal{T}/\tau_\eta\ll 1$.

In \cite{ni13} experimental evidence for (\ref{BatchelorSmallr}) is given, and it is claimed that the result is validated for $\mathcal{T}<\tau_\eta$ and $\vert\bm{r}^0\vert$ in in the dissipation range.  However, their results appear to be problematic for several reasons.  The main issue is that the experimental results in Fig 5(b) of \cite{ni13} show that the fluid particles undergo an exponential type growth (described by (\ref{BatchelorSmallr})) for $\mathcal{T}<\tau_\eta$ and $\vert\bm{r}^0\vert$ in the dissipation range, and then \emph{subsequently} undergo a ballistic type growth for $\mathcal{T}>\tau_\eta$.  This behavior cannot be correct since the ballistic law is exact in the limit ${\mathcal{T}/\tau_{r^0}\to 0}$, provided that the fluid particles are fully mixed (so that $\langle\Delta\bm{u}^{f}(0)\rangle_{\bm{r}^0}=\bm{0}$).  The exponential growth can only occur subsequent to this in the dissipation range when finite $\mathcal{T}$ contributions to $\Delta\bm{u}^f(\mathcal{T})=\bm{\Gamma}^f(\mathcal{T})\bm{\cdot}\bm{r}^f(\mathcal{T})$ become significant.  We will consider these issues further in the next section when we consider DNS data for the fluid particle dispersion.

We now consider the case where $\bm{r}^{0}$ lies in the inertial range of scales, i.e. $\eta\ll\vert\bm{r}^{0}\vert\ll L$, where $L$ is the integral lengthscale.  In this case the Lagrangian behavior of $\Delta\bm{u}$ is more complex and is no longer linearly proportional to $\bm{r}$.  The standard approach for describing dispersion in the inertial range is to use Kolmogorov's K41 theory to describe the growth of the fluid velocity differences in (\ref{FITmsr2FP}) and (\ref{BITmsr2FP}) and obtain the result for $\mathcal{T}\gg\tau_{r^0}$ in the inertial range 
\begin{align}
\Big\langle\vert\bm{r}^{f}(\mathcal{T})\vert^{2}\Big\rangle_{\bm{r}^0}&=\mathfrak{g}^{F}\langle\epsilon\rangle\mathcal{T}^{3},\label{ROf}\\	
\Big\langle\vert\bm{r}^{f}(-\mathcal{T})\vert^{2}\Big\rangle_{\bm{r}^0}&=\mathfrak{g}^{B}\langle\epsilon\rangle\mathcal{T}^{3},\label{ROb}
\end{align}
where $\mathfrak{g}^{F}$ and $\mathfrak{g}^{B}$ are the FIT and BIT Richardson's constants, estimated from experimental data to be $\mathfrak{g}^{F}\approx 0.55$ and $\mathfrak{g}^{B}\approx 1.15$ \cite{berg06a}.  The result in (\ref{ROf}) is from Batchelor's work \cite{batchelor50}.  The result in (\ref{ROb}), the BIT equivalent of (\ref{ROf}), was proposed by Sawford \emph{et al.} \cite{sawford05}.  It is also conventional to refer to the $\mathcal{T}^{3}$ scaling law as the Richardson-Obukhov (RO) law.  Note that the use of K41 does not lead to a prediction of the relative rate of FIT and BIT dispersion; that $\mathfrak{g}^{B}>\mathfrak{g}^{F}$ in 3D turbulence is simply an empirical finding.  However, based upon the arguments given in \S\ref{FPIM}, we would in fact expect that $\mathfrak{g}^{B}>\mathfrak{g}^{F}$ in 3D turbulence. 

If $\vert\bm{r}^{0}\vert>L$ then the dispersion (following the initial ballistic separation) is diffusive and time-reversible.  At these separations the two particles experience no correlation between their motion, and so the dispersion becomes directly related to the one-particle dispersion which in stationary, homogeneous turbulence is time-reversible.  This is simply a result of the system symmetries; if $\bm{x}^f$ is the single particle position then $\langle(\bm{x}^f(t)-\bm{x}^f(0))^2\rangle$ becomes under time reversal ${\langle(\bm{x}^f(-t)-\bm{x}^f(0))^2\rangle}$.  However, because of homogeneity and stationarity, applying a time shift gives $\langle(\bm{x}^f(-t)-\bm{x}^f(0))^2\rangle=\langle(\bm{x}^f(0)-\bm{x}^f(t))^2\rangle=\langle(\bm{x}^f(t)-\bm{x}^f(0))^2\rangle$.
  
\subsection{DNS results}
\label{FPS}

We now consider results from a DNS of statistically stationary, homogeneous, isotropic turbulence against which we will test the theoretical results discussed in \S\ref{FPDT}.  We use a pseudospectral method to solve the incompressible Navier-Stokes equations
for statistically stationary isotropic turbulence in a three-dimensional periodic cube of length $2 \pi$,
\begin{equation}
\partial_t\bm{u} +\bm{\omega}\times\bm{u} 
+\bm{\nabla_x}\left(\frac{p}{\rho} + \frac{\bm{u}\cdot\bm{u}}{2}\right) 
= \nu\bm{\nabla_x}^2\bm{u} + \bm{f},
\label{eq:N-S}
\end{equation}
where $\bm{u}(\bm{x},t)$ is the fluid velocity, $\bm{\omega}(\bm{x},t)$ is the vorticity, 
$p(\bm{x},t)$ is the pressure, $\rho$ is the fluid density, $\nu$ is the kinematic viscosity, and
$\bm{f}(\bm{x},t)$ is a large-scale forcing function that is added to achieve stationary turbulence.
For this simulation, forcing was added to the first two wavenumbers in Fourier space.
Time integration is performed through a second-order, explicit Runge-Kutta scheme 
with aliasing errors removed by means of a combination of spherical truncation and phase-shifting.
The time step was chosen to achieve a CFL number of about $0.5$.

The fluid field was solved on a grid with $2048^3$ grid points on $16,384$ processors on the
Yellowstone cluster at the U.S. National Center for Atmospheric Research \cite{yellowstone}.
The three-dimensional fast Fourier transforms required for the pseudospectral solution
of (\ref{eq:N-S}) are performed in parallel with MPI using the P3DFFT library~\cite{p3dfft}.
The Taylor microscale Reynolds number $R_\lambda$ 
for our flow is about $580$ and the ratio $L/\eta \approx 800$.  The viscosity was chosen
to achieve a small-resolution $k_\mathrm{max} \eta \approx 1.7$
(where $k_\mathrm{max} = 2048 \sqrt{2}/3$ is the maximum resolved wavenumber magnitude).
This initial flow field was then evolved for about $5$ large eddy turnover times until the
flow statistics were statistically stationary.

For comparison with the theory to be presented in \S\ref{IPD}, the inertial particle equation of motion is the simplified form of the Maxey-Riley equation~\cite{maxey83}
\begin{align}
\ddot{\bm{x}}^{p}=\dot{\bm{v}}^{p}= \frac{1}{\tau_{p}}\Big(\bm{u}(\bm{x}^{p}(t),t)-\bm{v}^{p}(t)\Big),
\label{eq:velocity}
\end{align}
where $\bm{x}^{p}(t)$ and $\bm{v}^{p}(t)$ are the particle position and velocity vectors and
$\bm{u}(\bm{x}^{p}(t),t)$ is the fluid velocity at the particle position which is calculated using an eight-point B-spline interpolation~\cite{vanhinsberg12}.  Fluid particles
were also tracked by solving $\dot{\bm{x}}^{f}(t)=\bm{u}(\bm{x}^{f}(t),t)$.

A total of $18$ different particle classes were simulated, with Stokes numbers ranging from $0$ to $30$.
About $17$ million particles were tracked for each value of $St$, for a total of $300$ million particles.
At the initial time, particles were injected in the flow with a uniform distribution.
The particles were allowed to equilibrate with the statistically stationary flow field for about $5$ 
large-eddy turnover times before we began gathering statistics.
Measurement of the particle radial distributions and velocities confirmed that the 
particle field had reached a statistically stationary state after this development time.

The mean-square separation calculations are carried out over a total time of 
$100 \tau_\eta$, or about $1.6$ large eddy turnover times, and particle positions
and velocities were stored approximately every $0.1 \tau_\eta$.
\begin{figure}[ht]
\centering
\vspace{5mm}
{\begin{overpic}
[trim = 30mm 75mm 20mm 75mm,scale=0.5,clip,tics=20]{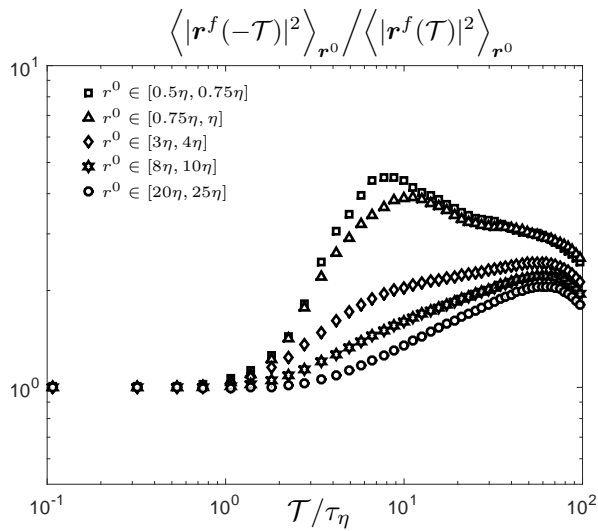}
\put(105,3){$\mathcal{T}/\tau_{\eta}$}
\put(35,163){\tiny\text{$r^{0}\in[0.5\eta,0.75\eta]$}}
\put(35,153){\tiny\text{$r^{0}\in[0.75\eta,\eta]$}}
\put(35,144){\tiny\text{$r^{0}\in[3\eta,4\eta]$}}
\put(35,135){\tiny\text{$r^{0}\in[8\eta,10\eta]$}}
\put(35,125){\tiny\text{$r^{0}\in[20\eta,25\eta]$}}
\put(60,185){\footnotesize\text{$\Big\langle\vert\bm{r}^{f}(-\mathcal{T})\vert^{2}\Big\rangle_{\bm{r}^0}\Big/\Big\langle\vert\bm{r}^{f}(\mathcal{T})\vert^{2}\Big\rangle_{\bm{r}^0}$}}
\end{overpic}}
\caption{DNS data for the ratio of the BIT to FIT mean square separation of fluid particles as a function of $\mathcal{T}$ for varying $r^0$.}
\label{FP_ratio}
\end{figure}
\FloatBarrier
Since we are considering an isotropic system, the statistics depend only upon on the separation magnitude $r^0\equiv\vert\bm{r}^0\vert$.  The results in Fig.~\ref{FP_ratio} show the ratio of the BIT to FIT mean square separation, and the results clearly show that the fluid particle dispersion is irreversible, with BIT faster than FIT dispersion.  The results also show that the peak in the ratio increases with decreasing $r^0$, indicating that the irreversibility in the dispersion becomes stronger as one goes to smaller scales.  This is consistent with the explanation given for the irreversibility in \S\ref{FPIM} since, as shown in Figure~\ref{Skew_plot}, the skewness of the field $\Delta\bm{u}(\bm{r},t)$ becomes stronger with decreasing $\bm{r}$.  It is worth pointing out that in our DNS, the dissipation spectra peaks at wavenumber $k\approx 0.12/\eta$, which roughly translates to $r\approx 37\eta$, yet the results in figure~\ref{FP_ratio} show that the strength of the irreversibility of the dispersion continues to increase as $r^0$ is decreased below $\eta$.  This serves to emphasize that it is not the amount of local dissipation, \emph{per-se}, that controls the irreversibility of the dispersion, but rather the local asymmetry of the distribution of $\Delta\bm{u}$.  The results in Fig.~\ref{FP_ratio} also show that the time it takes for the ratio to begin to increase from unity increases with increasing $r^0$.  This is because the initial ballistic motion, in which the separation is time reversible, persists for longer times as $r^0$ is increased, because $\tau_{r^0}$ increases with increasing $r^0$.  For $r^0>L$, where $L$ is the integral length scale of the flow ($L/\eta\approx 800$ for this DNS), the ratio would be unity as explained earlier.
\begin{figure}[ht]
\centering
\vspace{5mm}
{\begin{overpic}
[trim = 30mm 70mm 20mm 75mm,scale=0.5,clip,tics=20]{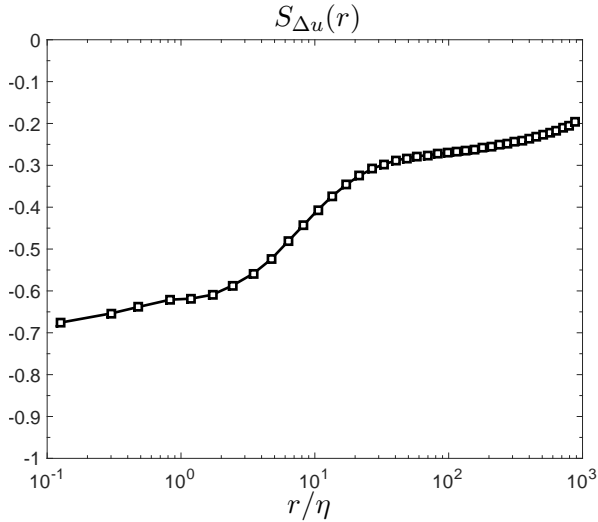}
\put(105,3){$r/\eta$}
\put(100,187){$S_{\Delta u}(r)$}
\end{overpic}}
\caption{DNS data for the skewness ${S_{\Delta u}(r)\equiv\langle[\Delta u_\parallel(r,t)]^3\rangle/\langle[\Delta u_\parallel(r,t)]^2\rangle^{3/2}}$, where $\Delta{u}_{\parallel}(r,t)\equiv r^{-1}\bm{r\cdot}\Delta\bm{u}(\bm{r},t)$.}
\label{Skew_plot}
\end{figure}
\FloatBarrier
In Fig.~\ref{Bal_scaled} we plot the DNS data for $(\langle\vert\bm{r}^{f}(\mathcal{T})\vert^{2}\rangle_{\bm{r}^0}-\vert \bm{r}^{0}\vert^{2})/\tau_{\eta}^{2}\langle\vert\Delta\bm{u}^{f}(0)\vert\rangle_{\bm{r}^{0}}$ and~$(\langle\vert\bm{r}^{f}(-\mathcal{T})\vert^{2}\rangle_{\bm{r}^0}-\vert \bm{r}^{0}\vert^{2})/\tau_{\eta}^{2}\langle\vert\Delta\bm{u}^{f}(0)\vert\rangle_{\bm{r}^{0}}$ (here and throughout, we use the DNS data for $\langle\vert\Delta\bm{u}^{f}(0)\vert\rangle_{\bm{r}^{0}}$).   The data shows a good collapse for small $\mathcal{T}$ for both the FIT and BIT cases, demonstrating the accuracy of (\ref{Bal}) for small $\mathcal{T}$, and also that the duration of the ballistic regime increases with increasing $r^{0}$.  
\begin{figure}[ht]
\centering
\vspace{5mm}
\subfloat[]
{\begin{overpic}
[trim = 30mm 75mm 20mm 75mm,scale=0.5,clip,tics=20]{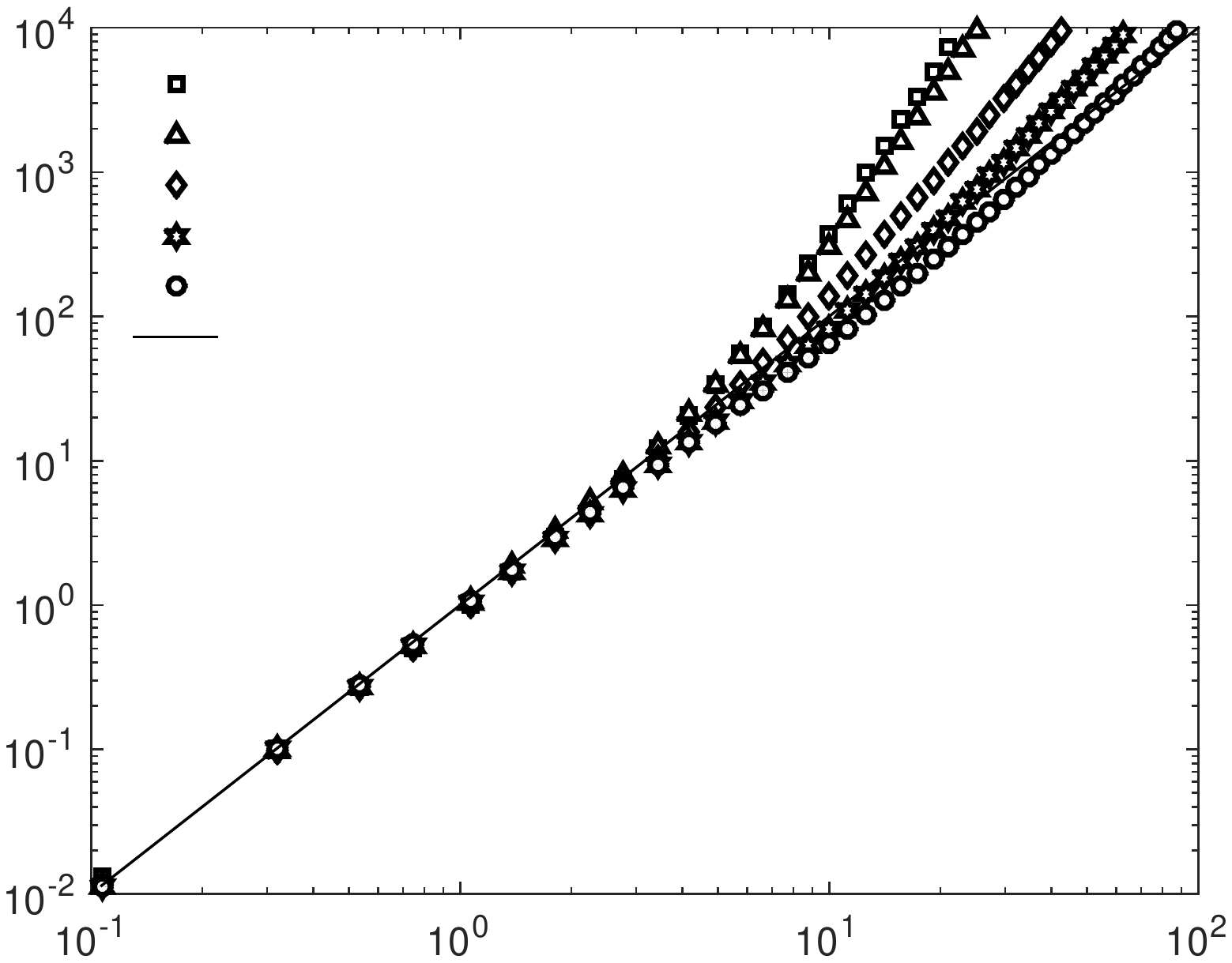}
\put(105,3){$\mathcal{T}/\tau_{\eta}$}
\put(40,165){\tiny\text{$r^{0}\in[0.5\eta,0.75\eta]$}}
\put(40,155){\tiny\text{$r^{0}\in[0.75\eta,\eta]$}}
\put(40,145){\tiny\text{$r^{0}\in[3\eta,4\eta]$}}
\put(40,135){\tiny\text{$r^{0}\in[8\eta,10\eta]$}}
\put(40,125){\tiny\text{$r^{0}\in[20\eta,25\eta]$}}
\put(41,117){\tiny\text{$(\mathcal{T}/\tau_{\eta})^{2}$}}
\put(30,190){\footnotesize\text{$\Bigg(\Big\langle\vert\bm{r}^{f}(\mathcal{T})\vert^{2}\Big\rangle_{\bm{r}^0}-\vert \bm{r}^{0}\vert^{2}\Bigg)\Bigg/\tau_{\eta}^{2}\Big\langle\vert\Delta\bm{u}^{f}(0)\vert\Big\rangle_{\bm{r}^{0}}$}}
\end{overpic}}
\subfloat[]
{\begin{overpic}
[trim = 30mm 75mm 20mm 75mm,scale=0.5,clip,tics=20]{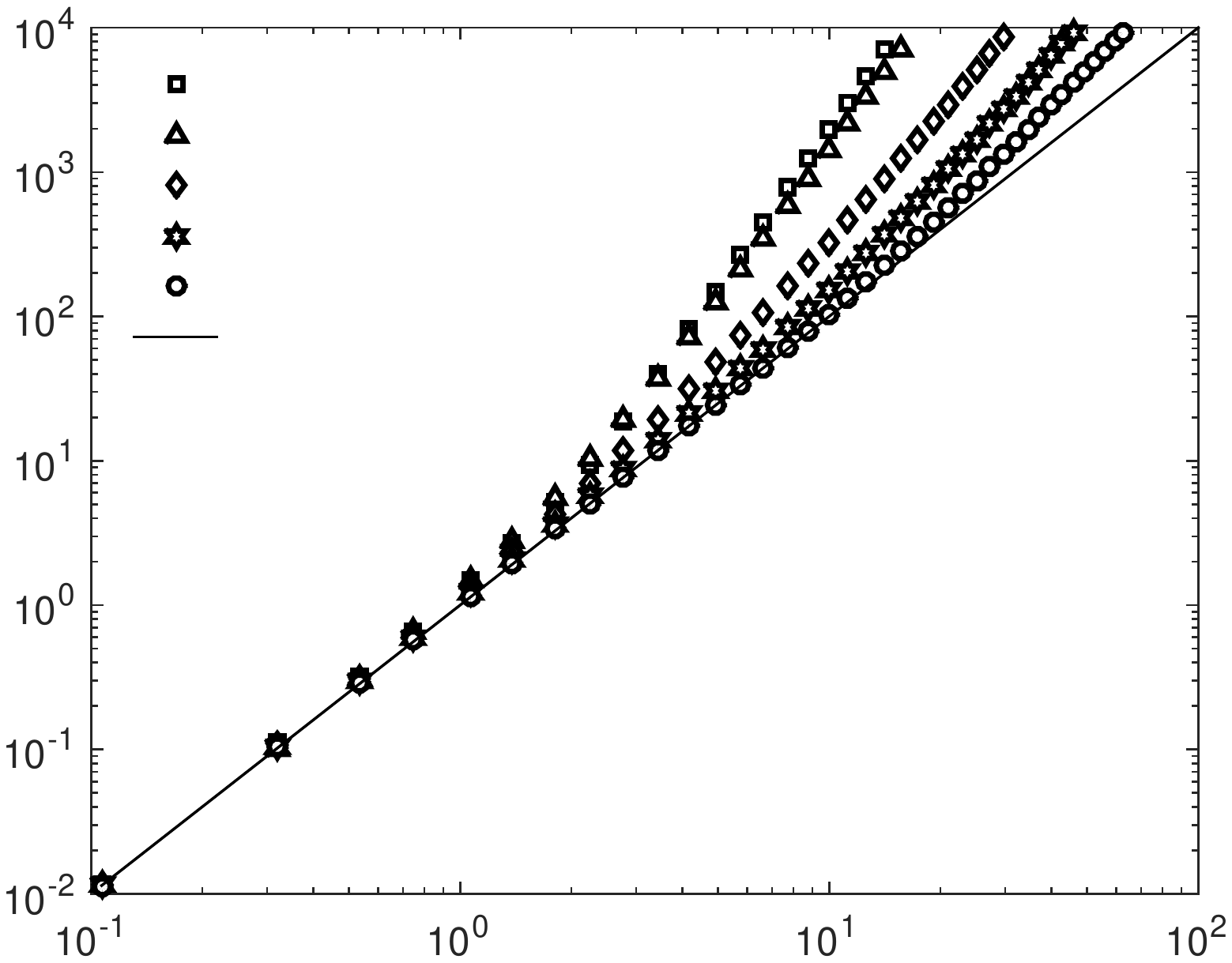}
\put(105,3){$\mathcal{T}/\tau_{\eta}$}
\put(40,165){\tiny\text{$r^{0}\in[0.5\eta,0.75\eta]$}}
\put(40,155){\tiny\text{$r^{0}\in[0.75\eta,\eta]$}}
\put(40,145){\tiny\text{$r^{0}\in[3\eta,4\eta]$}}
\put(40,135){\tiny\text{$r^{0}\in[8\eta,10\eta]$}}
\put(40,125){\tiny\text{$r^{0}\in[20\eta,25\eta]$}}
\put(41,117){\tiny\text{$(\mathcal{T}/\tau_{\eta})^{2}$}}
\put(30,190){\footnotesize\text{$\Bigg(\Big\langle\vert\bm{r}^{f}(-\mathcal{T})\vert^{2}\Big\rangle_{\bm{r}^0}-\vert \bm{r}^{0}\vert^{2}\Bigg)\Bigg/\tau_{\eta}^{2}\Big\langle\vert\Delta\bm{u}^{f}(0)\vert\Big\rangle_{\bm{r}^{0}}$}}
\end{overpic}}
\caption{DNS data for (a) FIT and (b) BIT fluid particle mean-square separation (with the initial separation subtracted), scaled by $\tau_{\eta}^{2}\langle\vert\Delta\bm{u}^{f}(0)\vert\rangle_{\bm{r}^{0}}$.}
\label{Bal_scaled}
\end{figure}
\FloatBarrier
Note, however, that in agreement with Fig.~4 in \cite{ouellette06c} we find in Fig.~\ref{Bal_scaled} (a) that for the larger initial separations, the data shows that the growth slows down for a time after an initially ballistic separation.  In Fig.~\ref{Bal_scaled_tB} we plot the fluid particle mean-square separation (with the initial separation subtracted) scaled by the ballistic prediction and plotted against $\mathcal{T}/\tau_{r^{0}}$, i.e. with time scaled by the Batchelor timescale.
%
\begin{figure}[ht]
\centering
\subfloat[]
{\begin{overpic}
[trim = 30mm 75mm 20mm 75mm,scale=0.5,clip,tics=20]{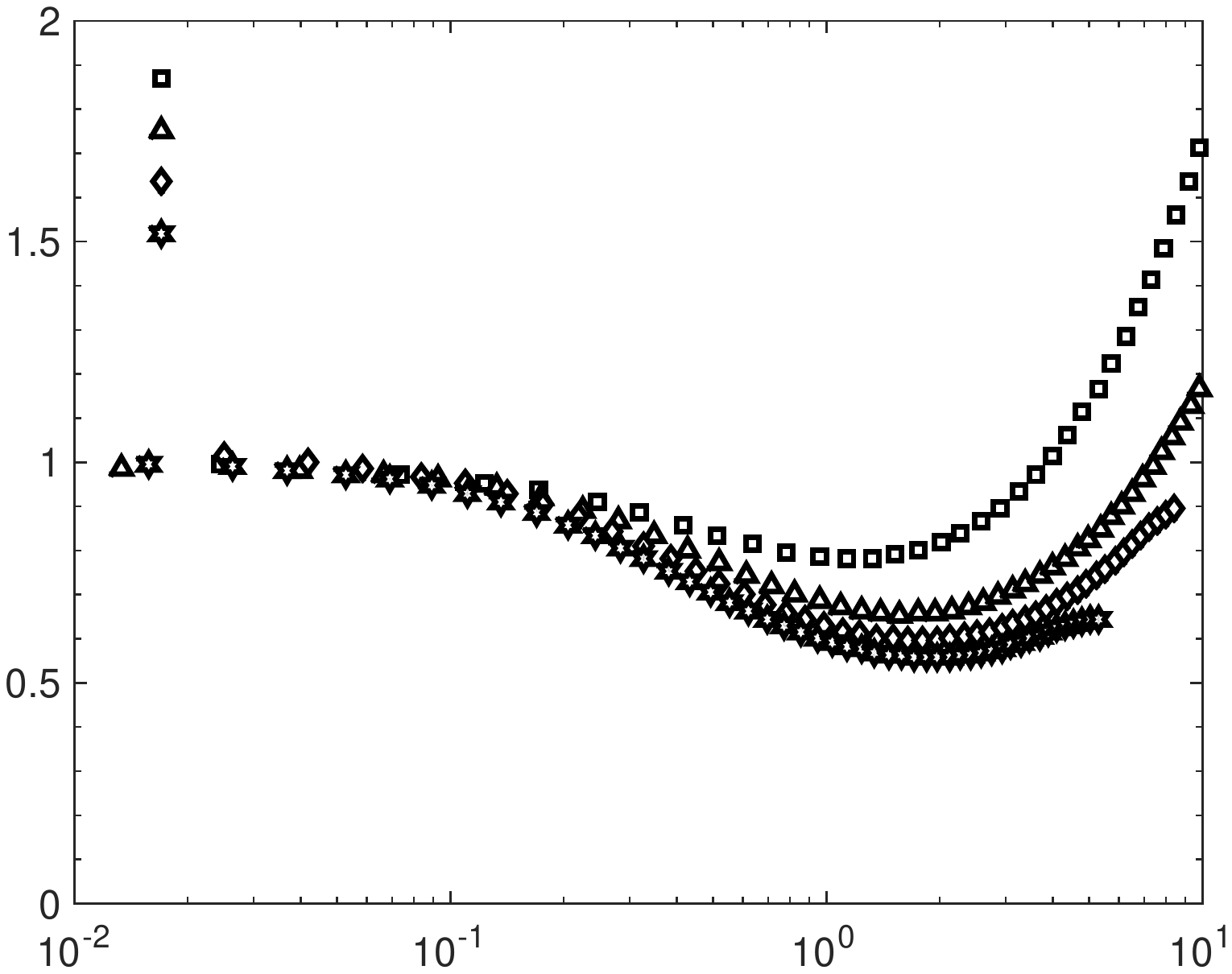}
\put(105,3){$\mathcal{T}/\tau_{r^{0}}$}
\put(38,165){\tiny\text{$r^{0}\in[8\eta,10\eta]$}}
\put(38,155){\tiny\text{$r^{0}\in[20\eta,25\eta]$}}
\put(38,145){\tiny\text{$r^{0}\in[40\eta,50\eta]$}}
\put(38,135){\tiny\text{$r^{0}\in[80\eta,100\eta]$}}
\put(30,190){\footnotesize\text{$\Bigg(\Big\langle\vert\bm{r}^{f}(\mathcal{T})\vert^{2}\Big\rangle_{\bm{r}^0}-\vert \bm{r}^{0}\vert^{2}\Bigg)\Bigg/\mathcal{T}^{2}\Big\langle\vert\Delta\bm{u}^{f}(0)\vert\Big\rangle_{\bm{r}^{0}}$}}
\end{overpic}}
\subfloat[]
{\begin{overpic}
[trim = 30mm 75mm 20mm 75mm,scale=0.5,clip,tics=20]{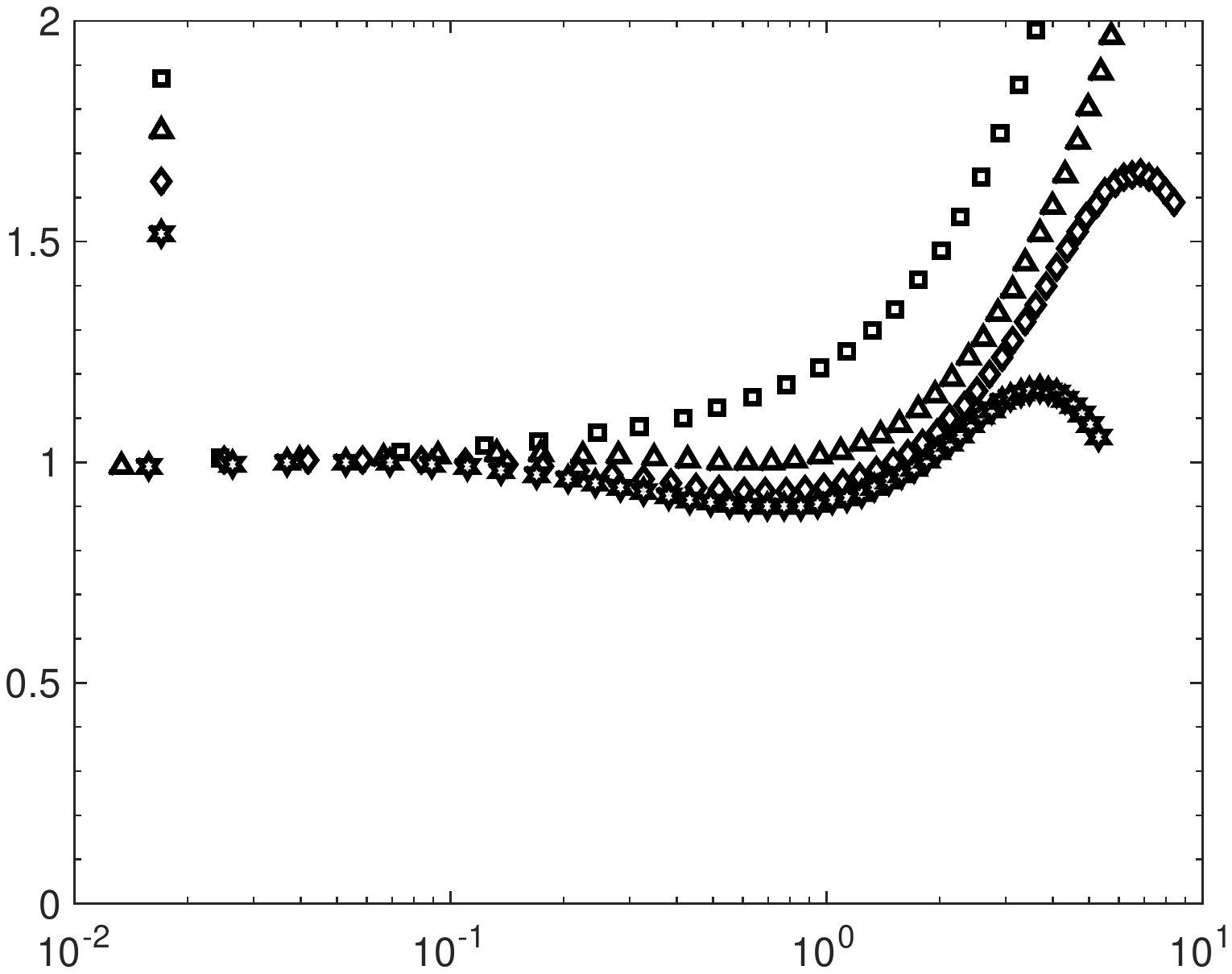}
\put(105,3){$\mathcal{T}/\tau_{r^{0}}$}
\put(38,165){\tiny\text{$r^{0}\in[8\eta,10\eta]$}}
\put(38,155){\tiny\text{$r^{0}\in[20\eta,25\eta]$}}
\put(38,145){\tiny\text{$r^{0}\in[40\eta,50\eta]$}}
\put(38,135){\tiny\text{$r^{0}\in[80\eta,100\eta]$}}
\put(30,190){\footnotesize\text{$\Bigg(\Big\langle\vert\bm{r}^{f}(-\mathcal{T})\vert^{2}\Big\rangle_{\bm{r}^0}-\vert \bm{r}^{0}\vert^{2}\Bigg)\Bigg/\mathcal{T}^{2}\Big\langle\vert\Delta\bm{u}^{f}(0)\vert\Big\rangle_{\bm{r}^{0}}$}}
\end{overpic}}
\caption{DNS data for (a) FIT and (b) BIT fluid particle mean-square separation (with the initial separation subtracted), scaled by the ballistic prediction and plotted against $\mathcal{T}/\tau_{r^{0}}$.}
\label{Bal_scaled_tB}
\end{figure}
In agreement with the experimental data in \cite{ouellette06c}, we find in Fig.~\ref{Bal_scaled_tB} (a) that the FIT dispersion of the particles is sub-ballistic for some time after $\mathcal{T}=\mathcal{O}(\tau_{r^{0}}/10)$.  The authors in \cite{ouellette06c} argue that this slowing down is not due to higher-order correction terms in the small-time series expansion, but argue that it is more likely explained in terms of the effect of the large scales on the separation.  In our case the influence of the large scales cannot be the explanation: At these times the particle separations are much smaller than the integral length scale and our data shows a temporary slowing down but then a speeding up towards a faster than ballistic separation (clearly observable in Fig.~\ref{Bal_scaled_tB} (a) for $r^{0}\in[8\eta,10\eta]$ and $r^{0}\in[20\eta,25\eta]$) that cannot be caused by the large scales.  A possible explanation for the temporary slowing down of the separation is the effect of the decorrelation of $\Delta\bm{u}^f$ along the pair trajectory (which the ballistic regime neglects through its use of ${\Delta\bm{u}^{f}(\mathcal{T})\approx\Delta\bm{u}^{f}(0)}$), which is subsequently overcome by the growth of the autocovariances of $\Delta\bm{u}^{f}$ in the inertial regime.  

The BIT results in Fig.~\ref{Bal_scaled_tB} (b) also show the slowing down but to a lesser degree.  The BIT data for $r^{0}\in[40\eta,50\eta]$ and $r^{0}\in[80\eta,100\eta]$ shows that after the initial ballistic separation their separation slows down, speeds up again to a faster than ballistic separation behavior and then finally slows down again.  This final stage of separation is likely due to the large scales since at these times the particle separations exceed the integral length scale, at which point the growth follows a diffusive law (corresponding to a line $\propto\mathcal{T}^{-1}$ in Fig.~\ref{Bal_scaled_tB}).  It is however also possible that at the largest values of $\mathcal{T}$, the results are affected by the finite box size and the periodic boundary conditions used in the DNS. 
\FloatBarrier
We now consider in more depth the case where $\bm{r}^{f}(\mathcal{T})$ lies in the dissipation range.  In \S\ref{FPD} we argued that for $\bm{r}^{f}(\mathcal{T})$ in the dissipation range, the pair separation begins with a ballistic separation growth and then may transition to some form of exponential growth.  This is in contrast to the experimental results in \cite{ni13} where they argue based on their data that the growth is first exponential and then ballistic.

\begin{figure}[ht]
\centering
\subfloat[]
{\begin{overpic}
[trim = 30mm 75mm 20mm 75mm,scale=0.5,clip,tics=20]{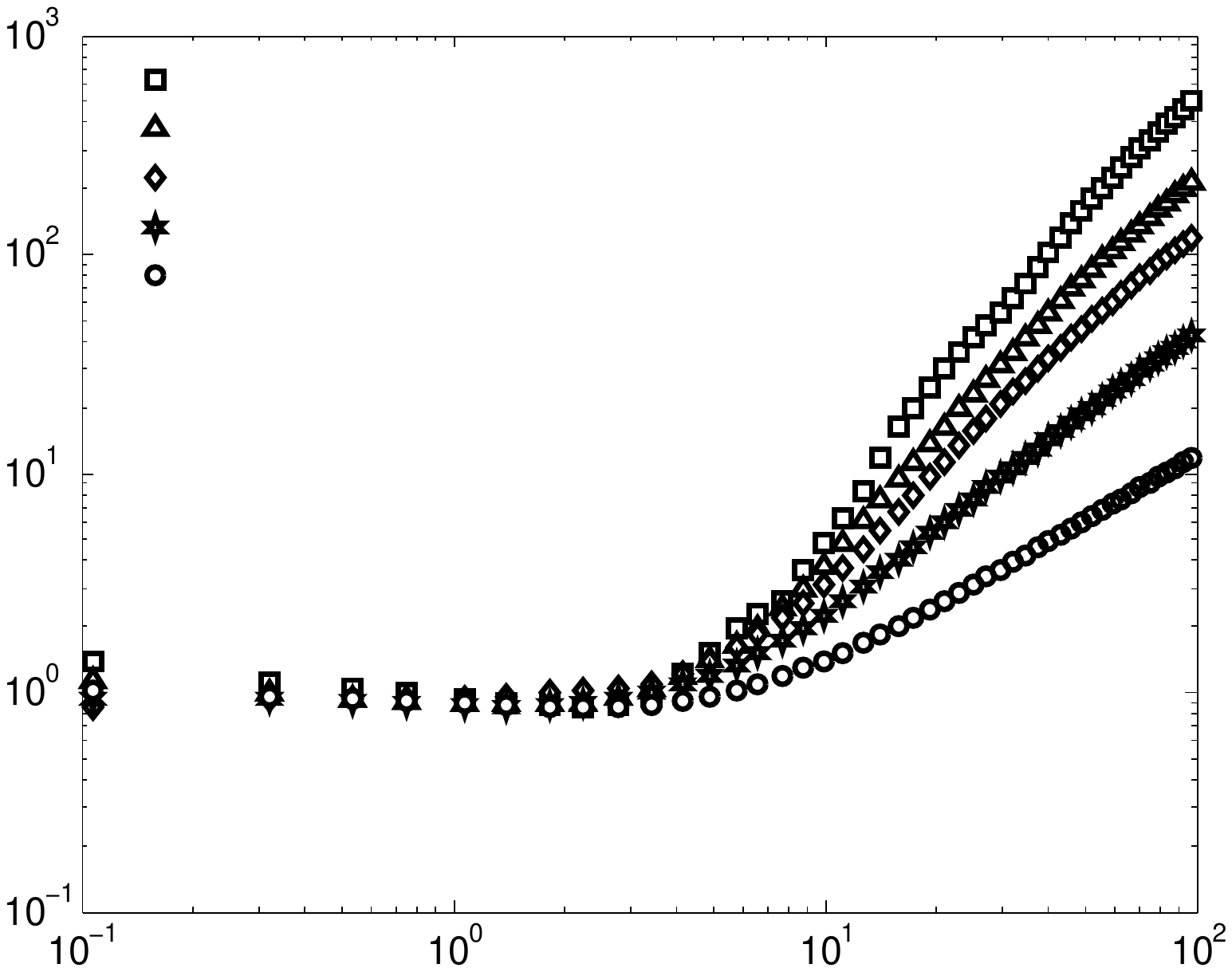}
\put(20,190){\footnotesize\text{$\Bigg(\Big\langle\vert\bm{r}^{f}(\mathcal{T})\vert^{2}\Big\rangle_{\bm{r}^0}-\vert \bm{r}^{0}\vert^{2}\Bigg)\Bigg/\mathcal{T}^{2}\Big\langle\vert\Delta\bm{u}^{f}(0)\vert\Big\rangle_{\bm{r}^{0}}$}}
\put(105,3){$\mathcal{T}/\tau_{\eta}$}
\put(32,165){\tiny\text{$r^{0}\in[0.25\eta,0.5\eta]$}}
\put(32,156){\tiny\text{$r^{0}\in[0.5,0.75\eta]$}}
\put(32,147){\tiny\text{$r^{0}\in[0.75\eta,\eta]$}}
\put(32,138){\tiny\text{$r^{0}\in[\eta,2\eta]$}}
\put(32,129){\tiny\text{$r^{0}\in[3\eta,4\eta]$}}
\end{overpic}}
\subfloat[]
{\begin{overpic}
[trim = 30mm 75mm 20mm 75mm,scale=0.5,clip,tics=20]{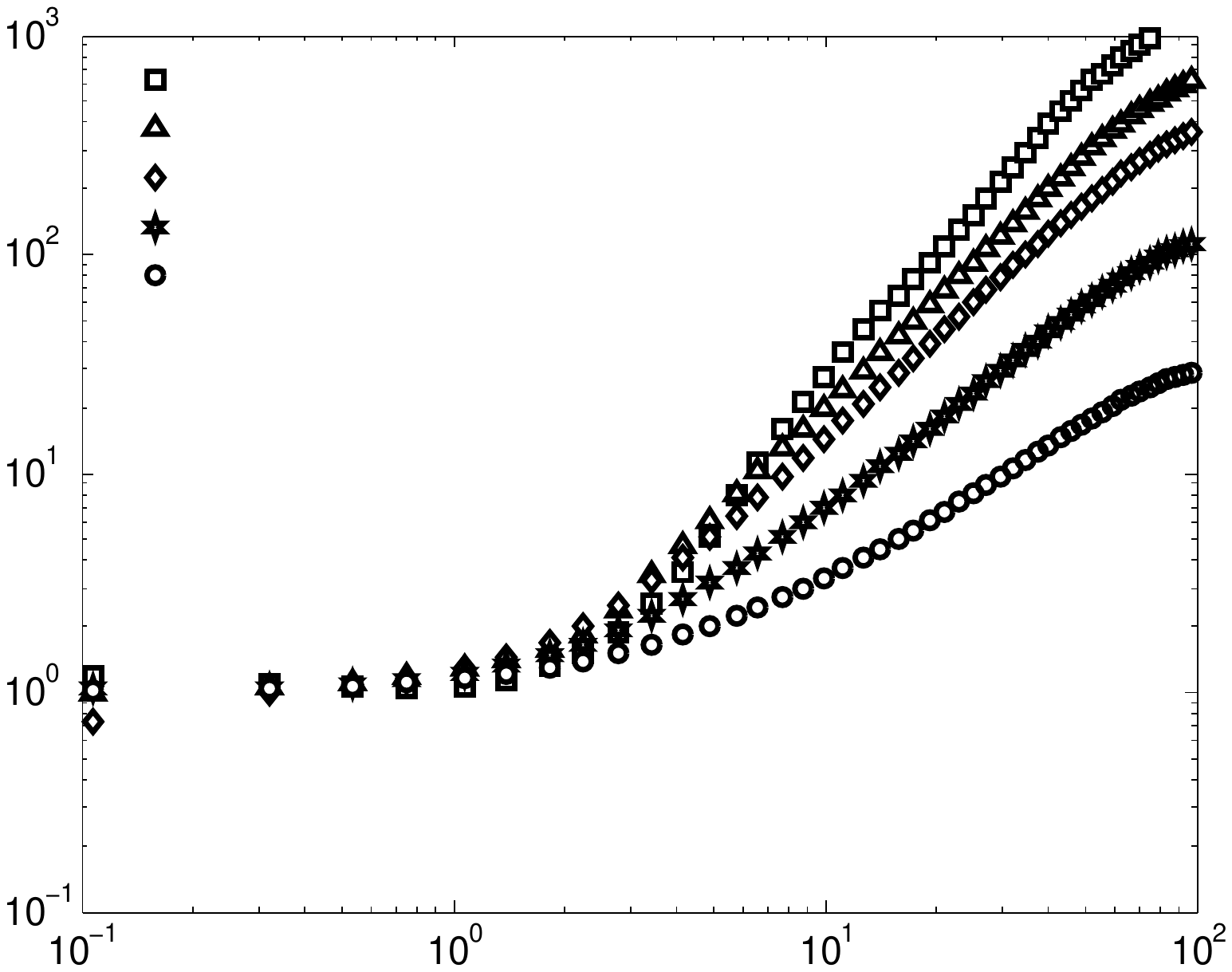}
\put(20,190){\footnotesize\text{$\Bigg(\Big\langle\vert\bm{r}^{f}(-\mathcal{T})\vert^{2}\Big\rangle_{\bm{r}^0}-\vert \bm{r}^{0}\vert^{2}\Bigg)\Bigg/\mathcal{T}^{2}\Big\langle\vert\Delta\bm{u}^{f}(0)\vert\Big\rangle_{\bm{r}^{0}}$}}
\put(105,3){$\mathcal{T}/\tau_{\eta}$}
\put(32,165){\tiny\text{$r^{0}\in[0.25\eta,0.5\eta]$}}
\put(32,156){\tiny\text{$r^{0}\in[0.5,0.75\eta]$}}
\put(32,147){\tiny\text{$r^{0}\in[0.75\eta,\eta]$}}
\put(32,138){\tiny\text{$r^{0}\in[\eta,2\eta]$}}
\put(32,129){\tiny\text{$r^{0}\in[3\eta,4\eta]$}}
\end{overpic}}
\caption{DNS data for (a) FIT and (b) BIT fluid particle mean-square separation (with the initial separation subtracted), scaled by the ballistic prediction and plotted against $\mathcal{T}/\tau_\eta$, for $r^0$ in the dissipation range.}
\label{EXP_test}
\end{figure}

\begin{figure}[ht]
\centering
\subfloat[]
{\begin{overpic}
[trim = 26mm 75mm 20mm 75mm,scale=0.5,clip,tics=20]{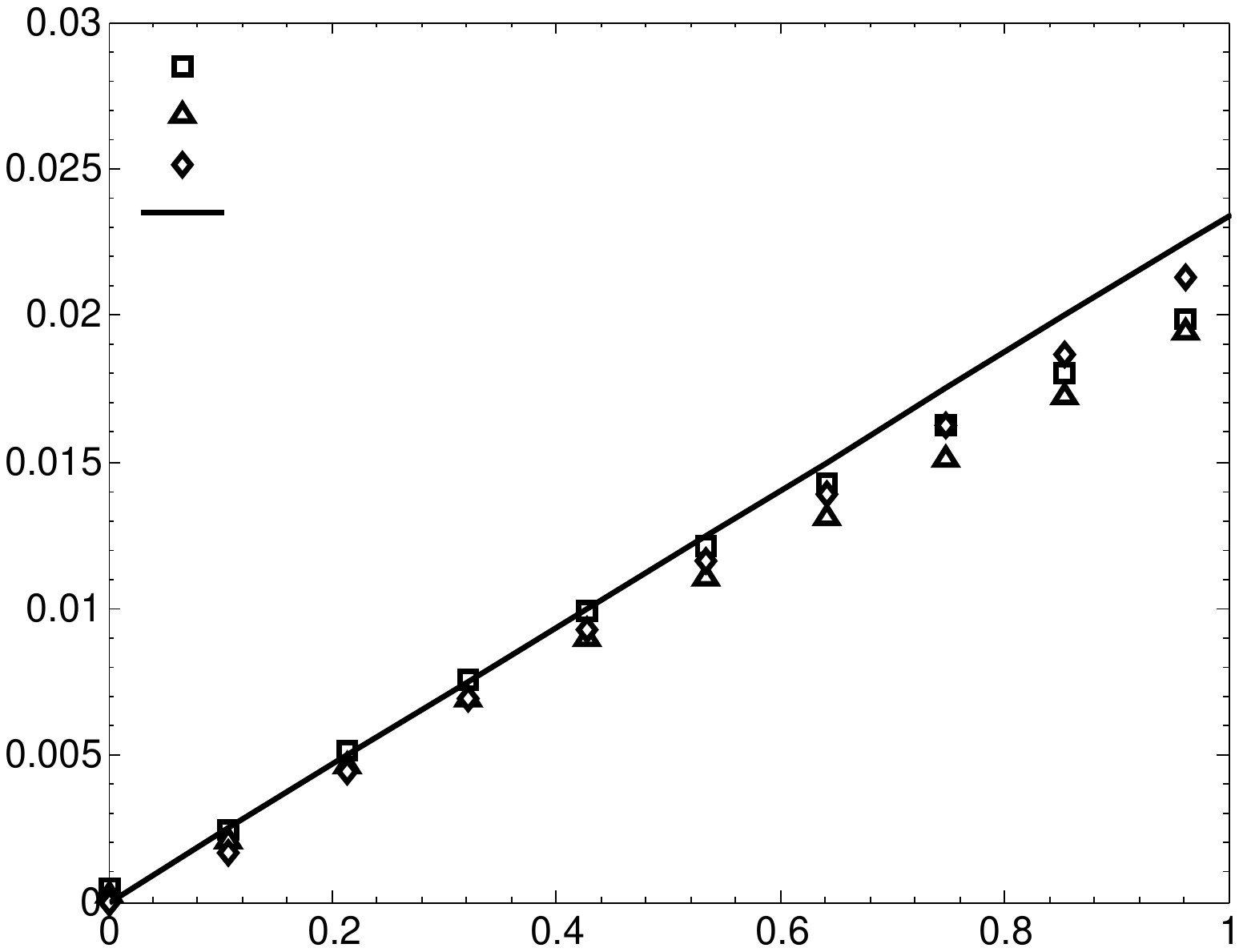}
\put(50,190){\footnotesize\text{$\dfrac{d}{d\mathcal{T}}\Big\langle\vert\bm{r}^{f}(\mathcal{T})\vert^{2}\Big\rangle_{\bm{r}^0}\Big/\Big\langle\vert\Delta\bm{u}^{f}(0)\vert\Big\rangle_{\bm{r}^{0}}$}}
\put(108,3){$\mathcal{T}/\tau_{\eta}$}
\put(42,165){\tiny\text{$r^{0}\in[0.25\eta,0.5\eta]$}}
\put(42,156){\tiny\text{$r^{0}\in[0.5,0.75\eta]$}}
\put(42,147){\tiny\text{$r^{0}\in[0.75\eta,\eta]$}}
\put(42,138){\tiny\text{$\mathcal{T}$}}
\end{overpic}}
\subfloat[]
{\begin{overpic}
[trim = 26mm 75mm 20mm 75mm,scale=0.5,clip,tics=20]{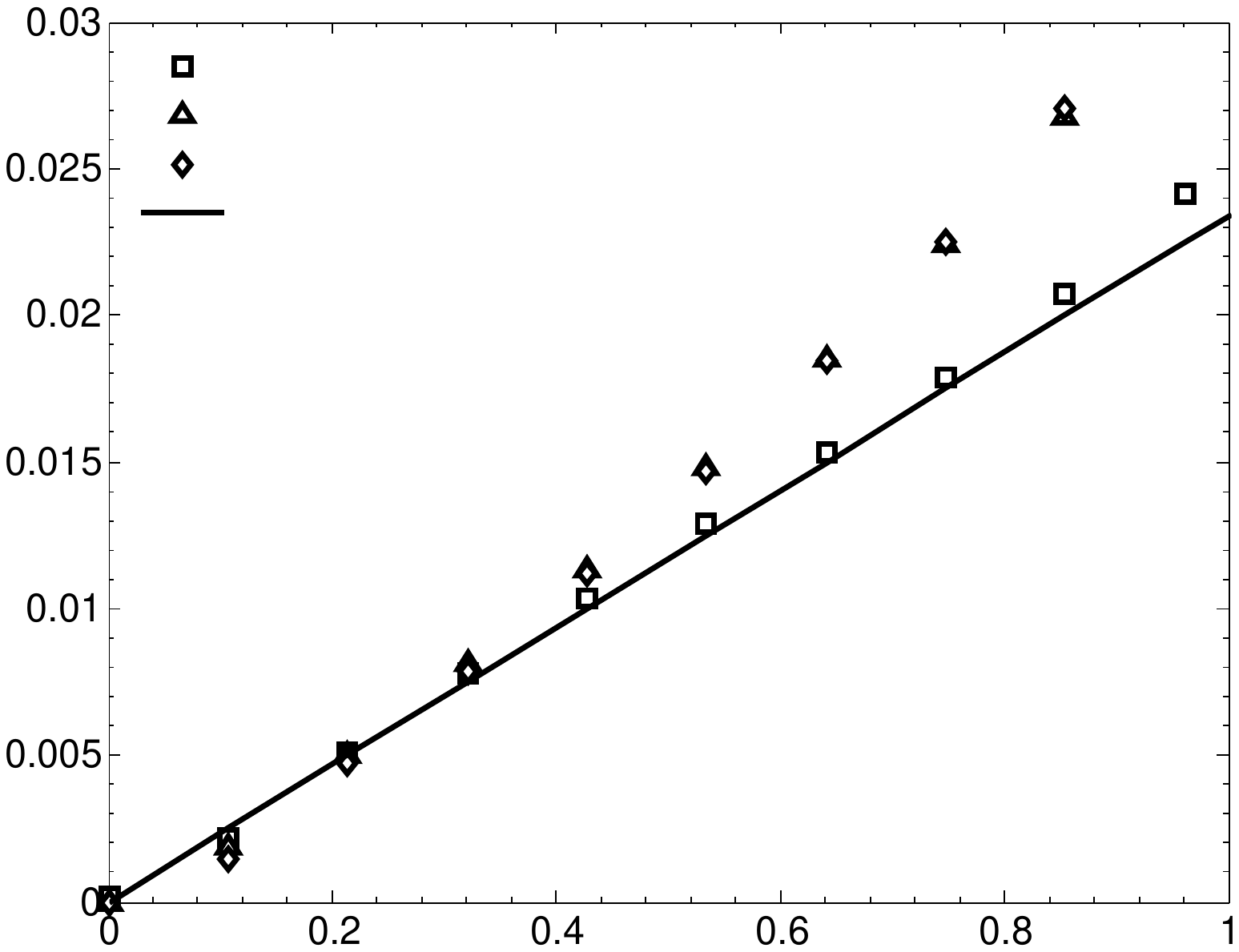}
\put(50,190){\footnotesize\text{$\dfrac{d}{d\mathcal{T}}\Big\langle\vert\bm{r}^{f}(-\mathcal{T})\vert^{2}\Big\rangle_{\bm{r}^0}\Big/\Big\langle\vert\Delta\bm{u}^{f}(0)\vert\Big\rangle_{\bm{r}^{0}}$}}
\put(108,3){$\mathcal{T}/\tau_{\eta}$}
\put(42,165){\tiny\text{$r^{0}\in[0.25\eta,0.5\eta]$}}
\put(42,156){\tiny\text{$r^{0}\in[0.5,0.75\eta]$}}
\put(42,147){\tiny\text{$r^{0}\in[0.75\eta,\eta]$}}
\put(42,138){\tiny\text{$\mathcal{T}$}}
\end{overpic}}
\caption{DNS data for (a) $(d/d\mathcal{T})\langle\vert\bm{r}^{f}(\mathcal{T})\vert^{2}\rangle_{\bm{r}^0}$ and (b) $(d/d\mathcal{T})\langle\vert\bm{r}^{f}(-\mathcal{T})\vert^{2}\rangle_{\bm{r}^0}$, scaled by $\langle\vert\Delta\bm{u}^{f}(0)\vert\rangle_{\bm{r}^{0}}$ and plotted against $\mathcal{T}/\tau_\eta$, for $r^0$ in the dissipation range.}
\label{EXP_test2}
\end{figure}

The results in figure~\ref{EXP_test} show that for $\mathcal{T}\leq\mathcal{O}(\tau_\eta)$ the dispersion in the dissipation range is described well by the ballistic law.  There are slight departures for very small $\mathcal{T}/\tau_\eta$, however, these could be caused by noise in the data which is amplified by the fact that the denominator in the plotted expression tends to zero in the limit $\mathcal{T}\to0$.  These results should be contrasted with those of Fig.~5(b) of \cite{ni13} where the same quantity is plotted: they observe that for $r=\mathcal{O}(\eta)$ and $\mathcal{T}=0.1\tau_\eta$, the quantity is $\mathcal{O}(10)$, whereas in our data it is $\mathcal{O}(1)$.  In figure~\ref{EXP_test2} we plot\[\dfrac{d}{d\mathcal{T}}\Big\langle\vert\bm{r}^{f}(\mathcal{T})\vert^{2}\Big\rangle_{\bm{r}^0}\Big/\Big\langle\vert\Delta\bm{u}^{f}(0)\vert\Big\rangle_{\bm{r}^{0}},\](and also the BIT equivalent) as a further test to examine departures from the ballistic law at small-times.  Recall that the Batchelor-type exponential growth grows with $\mathcal{T}$ in the small-$\mathcal{T}$ regime, and so the plotted quantity would be a constant at small-times if the Batchelor exponential growth were correct.  The data for this quantity shows a strong $\propto\mathcal{T}$ scaling at small-times, confirming the ballistic law and ruling out an exponential growth of the Batchelor kind at small-times, consistent with theoretical expectations.  These results then call into question the findings in \cite{ni13}.  While we are uncertain as to the full explanation for the discrepancies, we note that the discrepancy cannot arise simply as a consequence of the difference between the types of turbulent flow field  that we are considering (their experimental flow is a turbulent thermal convective flow), since the ballistic law follows in the dissipation range from the small-time approximation $\Delta\bm{u}^f(\mathcal{T})\approx\Delta\bm{u}^f(0)+\mathcal{O}(\mathcal{T}/\tau_\eta)$, for arbitrary statistical properties of the field $\Delta\bm{u}(\bm{r},t)$.  One possible explanation for the $\propto\mathcal{T}$ growth observed in \cite{ni13} concerns whether or not the tracer particles in their experiment were fully-mixed, that is, whether ${\bm{\nabla}_{\bm{r}^0}}\langle\delta(\bm{r}^f(0)-\bm{r}^0)\rangle=\bm{0}$.  In the general case where ${\bm{\nabla}_{\bm{r}^0}}\langle\delta(\bm{r}^f(0)-\bm{r}^0)\rangle\neq\bm{0}$, the FIT small-time fluid particle mean-square dispersion is (under the approximation $\Delta\bm{u}^f(\mathcal{T})\approx\Delta\bm{u}^f(0)+\mathcal{O}(\mathcal{T}/\tau_\eta)$)
\begin{align}
\Big\langle\vert\bm{r}^{f}(\mathcal{T})\vert^{2}\Big\rangle_{\bm{r}^0}\approx\vert\bm{r}^0\vert^{2}+2\mathcal{T}\bm{r}^0\bm{\cdot}\Big\langle\Delta\bm{u}^{f}(0)\Big\rangle_{\bm{r}^0}+\mathcal{T}^{2}\Big\langle\vert\Delta\bm{u}^{f}(0)\vert^{2}\Big\rangle_{\bm{r}^0}\label{BalNFM}.	
\end{align}
For fully-mixed fluid particles, $\langle\Delta\bm{u}^{f}(0)\rangle_{\bm{r}^0}=\bm{0}$, and (\ref{BalNFM}) reduces to (\ref{Bal}).  For non-fully-mixed fluid particles, $\langle\Delta\bm{u}^{f}(0)\rangle_{\bm{r}^0}\neq\bm{0}$ which gives\[\lim_{\mathcal{T}/\tau_\eta \to0}\Bigg[\Big\langle\vert\bm{r}^{f}(\mathcal{T})\vert^{2}\Big\rangle_{\bm{r}^0}-\vert\bm{r}^0\vert^{2}\Bigg]\propto\mathcal{T},\]consistent with the observation in \cite{ni13}.  Furthermore, in \cite{ni13} they also consider the quantity $\langle\vert\bm{r}^{f}(\mathcal{T})-\bm{r}^0\vert^{2}\rangle_{\bm{r}^0}$.  Applying the approximation $\Delta\bm{u}^f(\mathcal{T})\approx\Delta\bm{u}^f(0)+\mathcal{O}(\mathcal{T}/\tau_\eta)$ to the evolution equation governing $\langle\vert\bm{r}^{f}(\mathcal{T})-\bm{r}^0\vert^{2}\rangle_{\bm{r}^0}$ we obtain
\begin{align}
\Big\langle\vert\bm{r}^{f}(\mathcal{T})-\bm{r}^0\vert^{2}\Big\rangle_{\bm{r}^0}\approx\mathcal{T}^{2}\Big\langle\vert\Delta\bm{u}^{f}(0)\vert^{2}\Big\rangle_{\bm{r}^0},\label{BalNFM2}	
\end{align}
which applies \emph{irrespective} of whether the fluid particles are fully-mixed.  The quantity $\langle\vert\Delta\bm{u}^{f}(0)\vert^{2}\rangle_{\bm{r}^0}$ is in principle different for fully-mixed and non-fully-mixed fluid particles, however this does not change the $\mathcal{T}$ dependence of $\langle\vert\bm{r}^{f}(\mathcal{T})-\bm{r}^0\vert^{2}\rangle_{\bm{r}^0}$.  The important point then is that whereas in the fully-mixed case (see \S\ref{FPDT})\[\lim_{\mathcal{T}/\tau_\eta \to0}\Bigg[\Big\langle\vert\bm{r}^{f}(\mathcal{T})\vert^{2}\Big\rangle_{\bm{r}^0}-\vert\bm{r}^0\vert^{2}\Bigg]\propto\mathcal{T}^2,\quad \lim_{\mathcal{T}/\tau_\eta \to0}\Bigg[\Big\langle\vert\bm{r}^{f}(\mathcal{T})-\bm{r}^0\vert^{2}\Big\rangle_{\bm{r}^0}\Bigg]\propto\mathcal{T}^2 \]
in the non-fully mixed case\[\lim_{\mathcal{T}/\tau_\eta \to0}\Bigg[\Big\langle\vert\bm{r}^{f}(\mathcal{T})\vert^{2}\Big\rangle_{\bm{r}^0}-\vert\bm{r}^0\vert^{2}\Bigg]\propto\mathcal{T},\quad \lim_{\mathcal{T}/\tau_\eta \to0}\Bigg[\Big\langle\vert\bm{r}^{f}(\mathcal{T})-\bm{r}^0\vert^{2}\Big\rangle_{\bm{r}^0}\Bigg]\propto\mathcal{T}^2.\]These non-fully-mixed predictions seem very close to the scalings observed for the two quantities in fig.5a and fig.5b of \cite{ni13} for $r^0$ in the dissipation range. This suggests the possibility that the scaling they observed is not evidence of exponential growth in the dissipation range at small-times, but is rather a scaling arising from the dispersion of fluid particles that are not fully-mixed.  

In the experiments of \cite{ni13}, after the particles were introduced into the system, a time period of approximately $100$ large scale eddy turnover times was allowed to elapse before the dispersion statistics were recorded, which would be expected to provide sufficient time for the tracer particles to fully mix throughout the flow \cite{ni15}.  However, in their experiments, thermal plumes can rise from the bottom of the system and may bring with them an increased concentration of particles into the observation volume, thus destroying the well-mixedness of the system \cite{ni15}.  Given that the results in Fig.5 of \cite{ni13} which appear to be affected by non-well-mixedness are for $r^0=\mathcal{O}(\eta)$, and that the thermal plumes in the experiment have cross-sectional sizes $\mathcal{O}(\eta)$, this may well provide a plausible explanation for the aforementioned discrepancies.  Future experimental efforts are required to test whether or not this is in fact the case.

\begin{figure}[ht]
\centering
\subfloat[]
{\begin{overpic}
[trim = 30mm 75mm 20mm 75mm,scale=0.5,clip,tics=20]{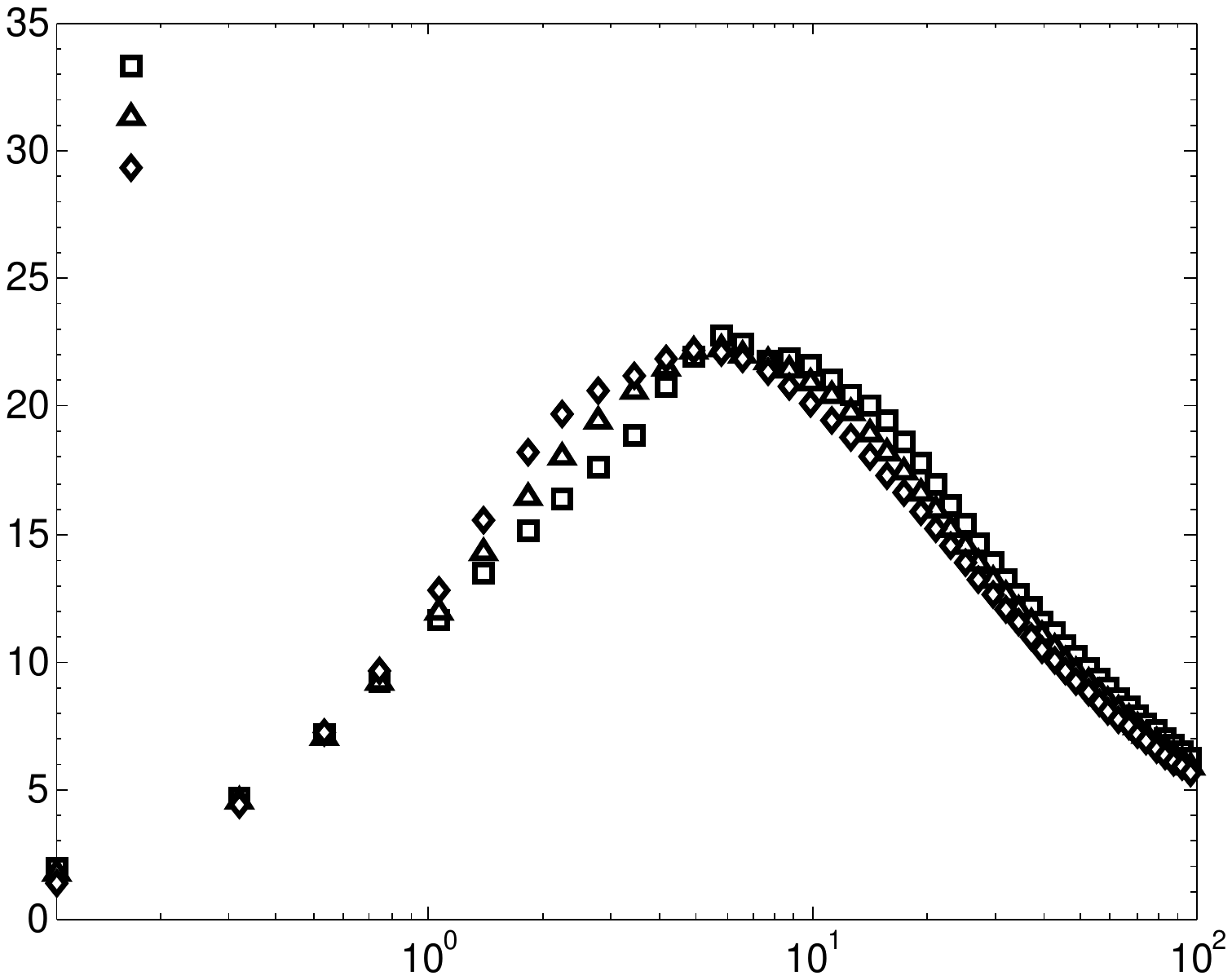}
\put(50,190){\footnotesize\text{$\mathcal{T}^{-1}\ln\Bigg(\Big\langle\vert\bm{r}^{f}(\mathcal{T})\vert^{2}\Big\rangle_{\bm{r}^0}\Big/\vert\bm{r}^{0}\vert^{2}\Bigg)$}}
\put(105,3){$\mathcal{T}/\tau_{\eta}$}
\put(32,165){\tiny\text{$r^{0}\in[0.25\eta,0.5\eta]$}}
\put(32,156){\tiny\text{$r^{0}\in[0.5,0.75\eta]$}}
\put(32,147){\tiny\text{$r^{0}\in[0.75\eta,\eta]$}}
\end{overpic}}
\subfloat[]
{\begin{overpic}
[trim = 30mm 75mm 20mm 75mm,scale=0.5,clip,tics=20]{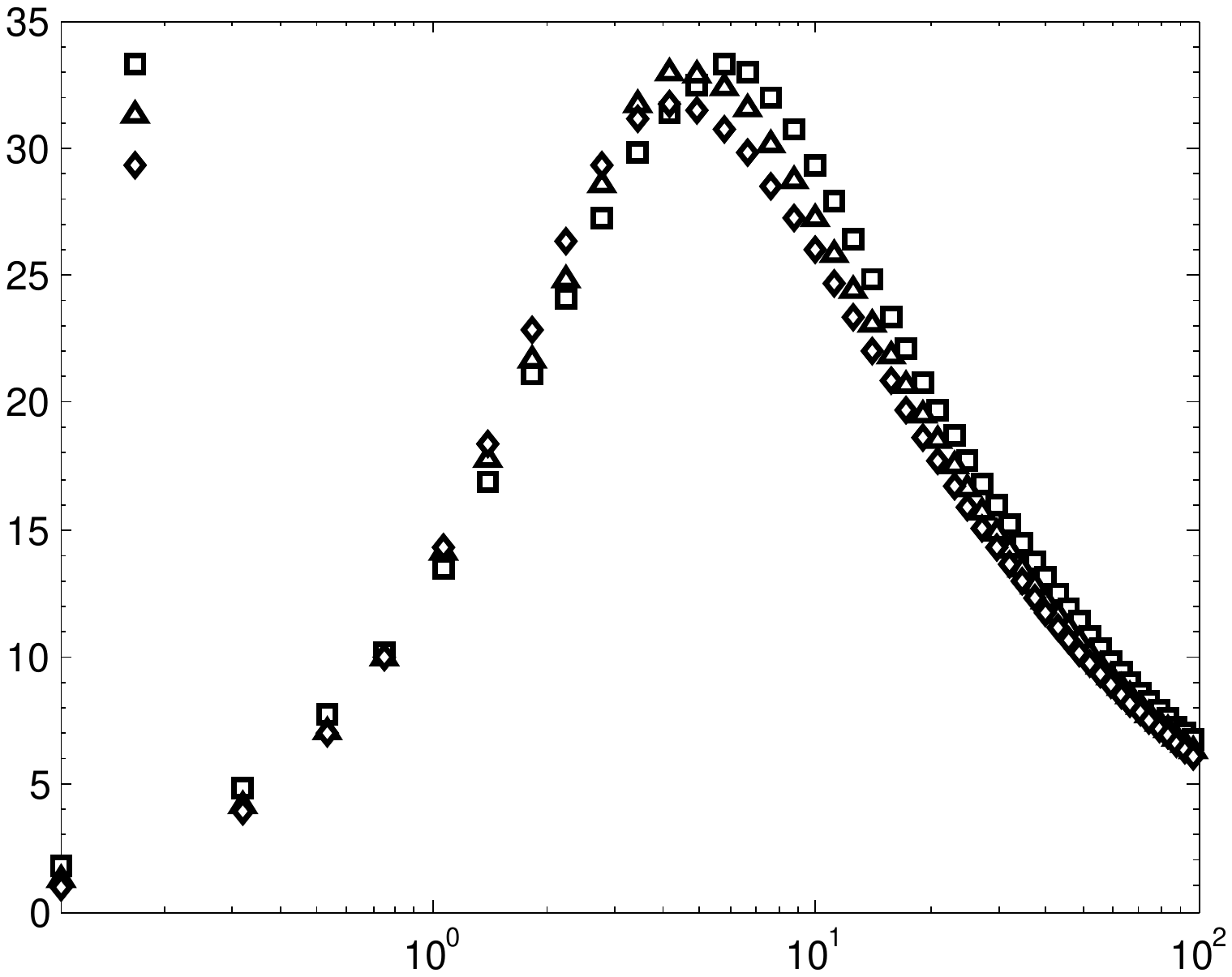}
\put(50,190){\footnotesize\text{$\mathcal{T}^{-1}\ln\Bigg(\Big\langle\vert\bm{r}^{f}(-\mathcal{T})\vert^{2}\Big\rangle_{\bm{r}^0}\Big/\vert\bm{r}^{0}\vert^{2}\Bigg)$}}
\put(105,3){$\mathcal{T}/\tau_{\eta}$}
\put(32,165){\tiny\text{$r^{0}\in[0.25\eta,0.5\eta]$}}
\put(32,156){\tiny\text{$r^{0}\in[0.5,0.75\eta]$}}
\put(32,147){\tiny\text{$r^{0}\in[0.75\eta,\eta]$}}
\end{overpic}}
\caption{DNS data for (a) $\mathcal{T}^{-1}\ln(\langle\vert\bm{r}^{f}(\mathcal{T})\vert^{2}\rangle_{\bm{r}^0}/\vert\bm{r}^{0}\vert^{2})$ and (b) $\mathcal{T}^{-1}\ln(\langle\vert\bm{r}^{f}(-\mathcal{T})\vert^{2}\rangle_{\bm{r}^0}/\vert\bm{r}^{0}\vert^{2})$, plotted against $\mathcal{T}/\tau_\eta$, for $r^0$ in the dissipation range.}
\label{EXP_test3}
\end{figure}

\begin{figure}[ht]
\centering
\subfloat[]
{\begin{overpic}
[trim = 25mm 70mm 20mm 70mm,scale=0.48,clip,tics=20]{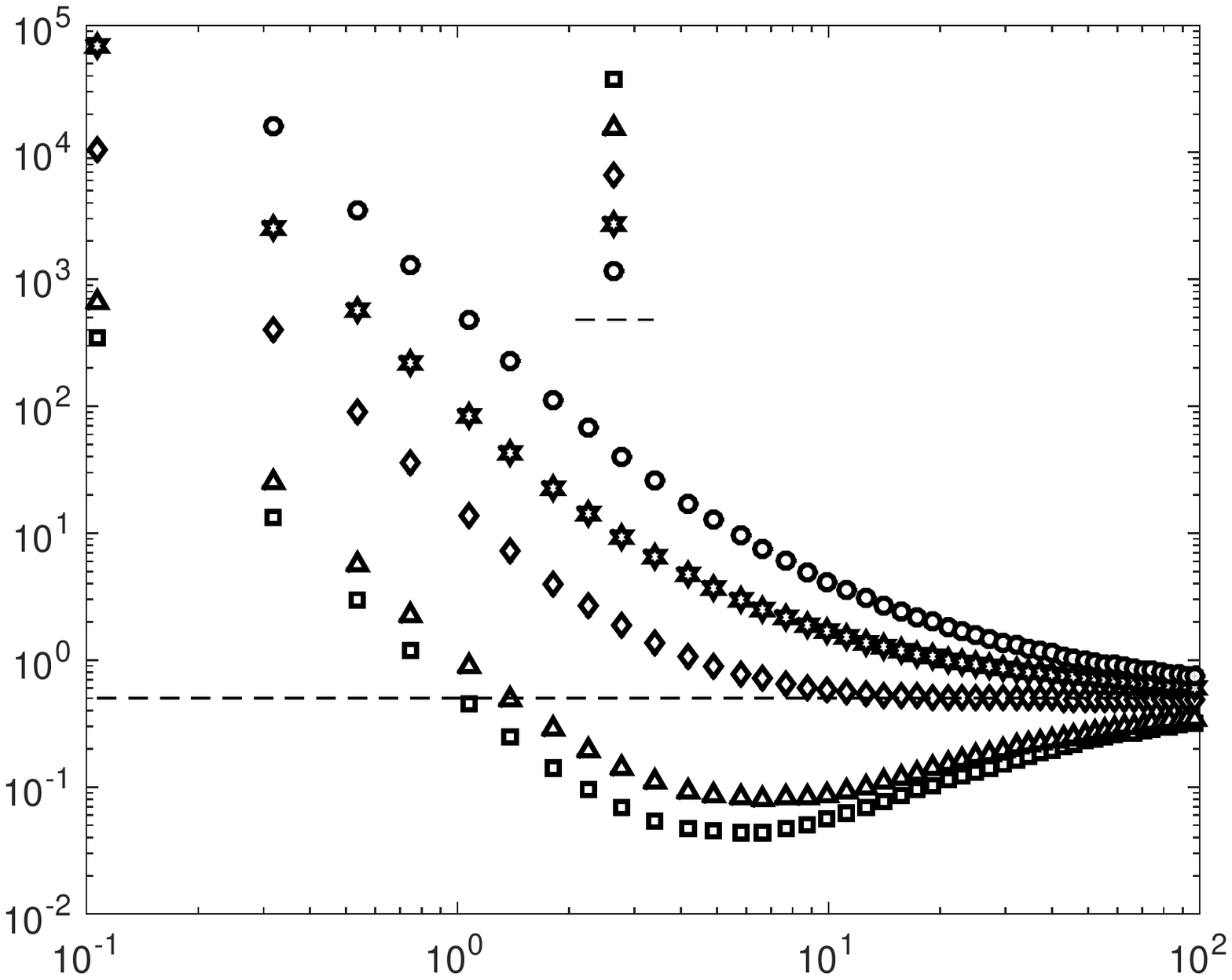}
\put(75,190){\footnotesize\text{$\Big\langle\vert\bm{r}^{f}(\mathcal{T})\vert^{2}\Big\rangle_{\bm{r}^0}\Big/\langle\epsilon\rangle\mathcal{T}^{3}$}}
\put(105,3){$\mathcal{T}/\tau_{\eta}$}
\put(121,171){\tiny\text{$r^{0}\in[0.5\eta,0.75\eta]$}}
\put(121,162){\tiny\text{$r^{0}\in[0.75,\eta]$}}
\put(121,152){\tiny\text{$r^{0}\in[3\eta,4\eta]$}}
\put(121,142){\tiny\text{$r^{0}\in[8\eta,10\eta]$}}
\put(121,134){\tiny\text{$r^{0}\in[20\eta,25\eta]$}}
\put(121,126){\tiny\text{$\mathfrak{g}^{F}=0.5$}}
\end{overpic}}
\subfloat[]
{\begin{overpic}
[trim = 25mm 70mm 20mm 70mm,scale=0.48,clip,tics=20]{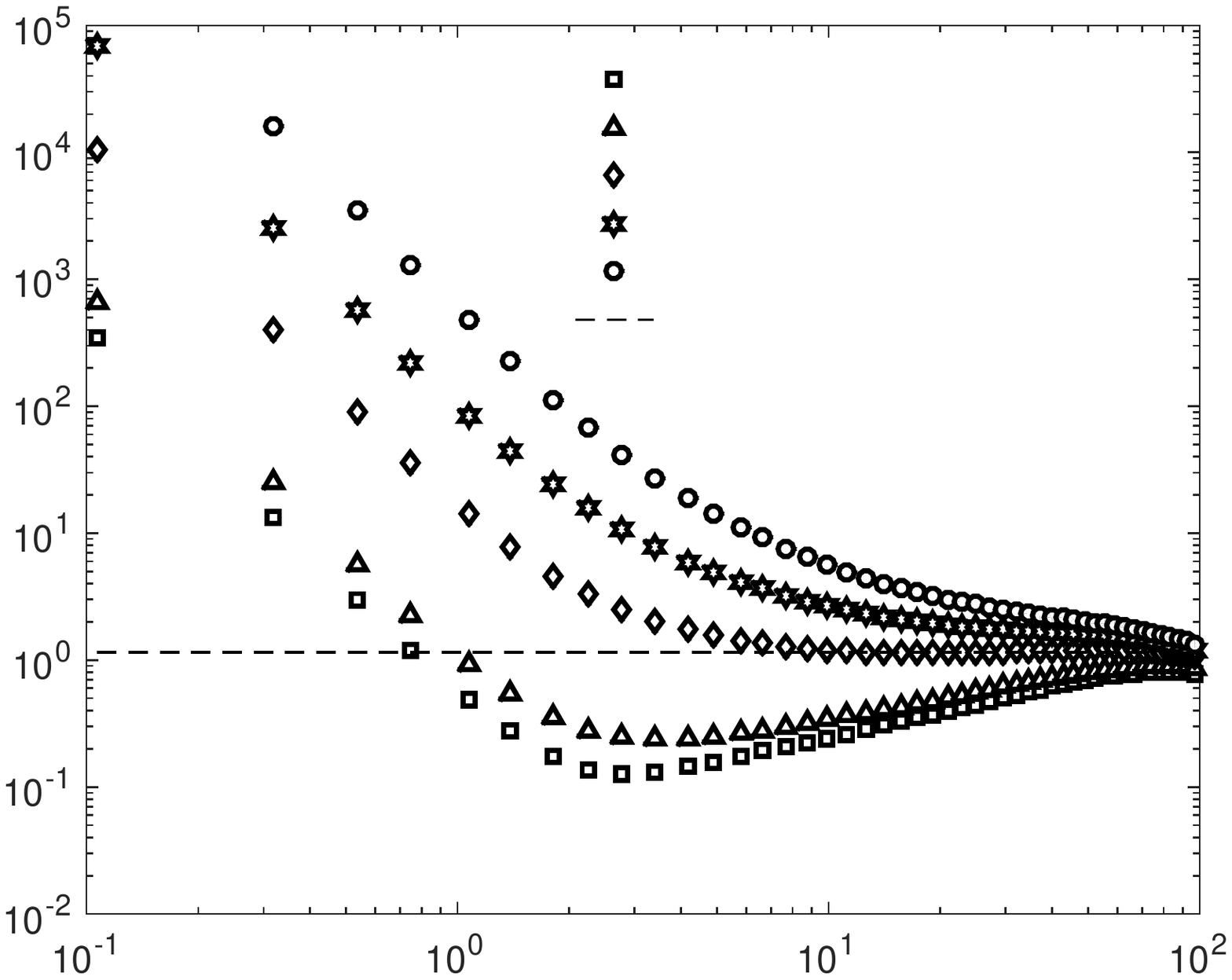}
\put(75,190){\footnotesize\text{$\Big\langle\vert\bm{r}^{f}(-\mathcal{T})\vert^{2}\Big\rangle_{\bm{r}^0}\Big/\langle\epsilon\rangle\mathcal{T}^{3}$}}
\put(105,3){$\mathcal{T}/\tau_{\eta}$}
\put(121,171){\tiny\text{$r^{0}\in[0.5\eta,0.75\eta]$}}
\put(121,162){\tiny\text{$r^{0}\in[0.75,\eta]$}}
\put(121,152){\tiny\text{$r^{0}\in[3\eta,4\eta]$}}
\put(121,142){\tiny\text{$r^{0}\in[8\eta,10\eta]$}}
\put(121,134){\tiny\text{$r^{0}\in[20\eta,25\eta]$}}
\put(121,126){\tiny\text{$\mathfrak{g}^{B}=1.15$}}
\end{overpic}}
\caption{DNS data for (a) FIT and (b) BIT fluid particle mean-square separations at various $r^{0}$ scaled by $\langle\epsilon\rangle\mathcal{T}^{3}$.}
\label{RO_scaling}
\end{figure}

In order to consider whether the fluid pairs undergo a Batchelor-type exponential growth subsequent to the initial ballistic growth, in figure~\ref{EXP_test3} we plot $\mathcal{T}^{-1}\ln(\langle\vert\bm{r}^{f}(\mathcal{T})\vert^{2}\rangle_{\bm{r}^0}/\vert\bm{r}^{0}\vert^{2})$ (and the equivalent BIT version), which would be constant in a Batchelor type exponential growth regime.  The results do not reveal any evidence of an exponential growth, for either the FIT and BIT case.  It is, however, possible that our initial separations are simply not small enough in order for the pairs to remain in the dissipation range at $\mathcal{T}/\tau_\eta\geq\mathcal{O}(1)$.
\FloatBarrier
\vspace{2mm}
We now consider the case where at sufficiently large $\mathcal{T}$ the particle separation $\bm{r}^f(\mathcal{T})$ lies in the inertial range, in which case the mean-square separation is predicted to follow the RO $\mathcal{T}^3$ law for arbitrary $r^0$ when $Re_\lambda\to\infty$.  In Fig.~\ref{RO_scaling} we plot $\langle\vert\bm{r}^{f}(\mathcal{T})\vert^{2}\rangle_{\bm{r}^0}/\langle\epsilon\rangle\mathcal{T}^{3}$ and $\langle\vert\bm{r}^{f}(-\mathcal{T})\vert^{2}\rangle_{\bm{r}^0}/\langle\epsilon\rangle\mathcal{T}^{3}$ for various $r^{0}$ in order to see whether the data shows an approach to RO scaling.  For $r^{0}\in[3\eta,4\eta]$ the data shows a clear convergence to RO scaling in both the FIT and BIT cases, yielding values of $\mathfrak{g}^{F}$ and $\mathfrak{g}^{B}$ in excellent agreement with experimental data \cite{berg06a}.  For $r^{0}\leq\eta$ the data indicates that for $\mathcal{T}\gtrsim5\tau_{\eta}$ the particles separate faster than RO scaling (indicated by the positive slope for $\mathcal{T}\gtrsim5\tau_{\eta}$).  We expect this is due to the influence of their separation in the dissipation range because of the finite temporal correlation radius of the field $\Delta\bm{u}$.  For separations larger than $r^{0}\in[3\eta,4\eta]$ the fluid mean square separation is slower than RO scaling (indicated by the negative slope) throughout the range of $\mathcal{T}$ for which we have data.  The curves do however seem to be tending to RO scaling at the largest values of $\mathcal{T}$.  For $r^{0}$ in the inertial regime we would expect an initial ballistic separation followed by RO with a transition region in between. 
\FloatBarrier
For separations larger than $r^{0}\in[3\eta,4\eta]$ it is likely that our time span of $0\leq\mathcal{T}\leq100\tau_{\eta}$ only extends to the transition region and hence we do not observe RO scaling.  To observe RO scaling over a larger range of $r^{0}$ we would need a DNS with significantly larger $Re_{\lambda}$.  The results in Fig.~\ref{RO_scaling} (a) are very similar to those in Fig.~4 (a) of \cite{sawford08}.  Note also that our data agrees with the findings in \cite{bitane12} that $r^{0}\approx 4\eta$ could be an ``optimal choice'' for the initial separation to observe RO scaling.
\vspace{5mm}
\FloatBarrier

\section{Inertial particle dispersion}
\label{IPD}

Having considered the dispersion of fluid particles, we now consider the more complex scenario of the dispersion of inertial particles.

\subsection{Irreversibility mechanisms}
\label{IPIM}

Before proceeding to derive predictions for the BIT inertial particle dispersion, we first consider how particle inertia gives rise to an additional source of irreversibility in the dispersion process.  The equations governing the time evolution of the fluid and inertial particle-pair separations may be written as (ignoring initial conditions)
\begin{align}
\dot{\bm{r}}^f(t)&=\Delta\bm{u}(\bm{r}^f(t),t),\label{rdotF}\\
\dot{\bm{r}}^p(t)&=(St\tau_\eta)^{-1}\int\limits_0^t\dot{G}(t-s)\Delta\bm{u}(\bm{r}^p(s),s)\,ds.\label{rdotP}
\end{align}
Recall that in FIT dispersion the pairs are (on average) going to greater separations as time increases (i.e. $\dot{\bm{r}}^f(t)>\bm{0}$), whereas in BIT dispersion the pairs are going to smaller separations as time increases (i.e. $\dot{\bm{r}}^f(t)<\bm{0}$).  Just as for the fluid particles, inertial particles experience a local source of irreversibility related to the asymmetry in the probability distribution of $\Delta\bm{u}$ (see \S\ref{FPIM}).  However, the inertial particle separation described by (\ref{rdotP}) contains an additional effect.  Whereas $\dot{\bm{r}}^f(t)$ is entirely governed by the local turbulent field $\Delta\bm{u}$, (\ref{rdotP}) shows that $\dot{\bm{r}}^p(t)$ is influenced by the behavior of $\Delta\bm{u}$ along the path-history of the particle pair for times up to $t-s=\mathcal{O}(\tau_p)$ in the past.  This non-local dependence of $\dot{\bm{r}}^p(t)$ on $\Delta\bm{u}$ gives rise to an additional source of irreversibility: In FIT dispersion (statistically speaking) $\dot{\bm{r}}^f(t)>\bm{0}$ and ${\bm{r}}^f(t)>{\bm{r}}^f(s)$, while in BIT dispersion $\dot{\bm{r}}^f(t)<\bm{0}$ and ${\bm{r}}^f(t)<{\bm{r}}^f(s)$.  Since $\langle\vert\Delta\bm{u}(\bm{r},t)\vert^2\rangle$ increases with $\bm{r}$, (\ref{rdotP}) implies that FIT-separating pairs are influenced by their memory of smaller $\Delta\bm{u}$ in their path-history, whereas BIT-separating pairs are influenced by their memory of larger $\Delta\bm{u}$ in their path-history.  This enhances the discrepancy between FIT and BIT dispersion.  Note that this irreversibility mechanism is intimately connected to the non-local clustering mechanism that dominates the clustering of inertial particles for ${St\gtrsim\mathcal{O}(1)}$ \cite{bragg14b,bragg14d}.

In general the inertial particles are affected by both sources of irreversibility, and we expect that there will be a value of the particle inertia for which their dispersion is optimally affected by both sources of irreversibility, likely $\tau_p/\tau_{r^0}=\mathcal{O}(1)$.  

We now make several remarks and observations about this non-local irreversibility mechanism.  First, it vanishes in the limit $\tau_p\to0$ where the particles separation is entirely governed by the local turbulence.  This is represented in (\ref{rdotP}) through the memory kernel $\dot{G}(t-s)$, which vanishes for $t\neq s$ in the limit $\tau_p\to0$.  Second, this mechanism operates and generates irreversibility in inertial particle dispersion even in fluid velocity fields which are reversible (i.e. symmetric PDF for $\Delta\bm{u}$).  The operation of the mechanism only depends upon finite $\tau_p$ and $\bm{\nabla_r}\langle\vert\Delta\bm{u}(\bm{r},t)\vert^2\rangle\neq\bm{0}$, the latter being true for any spatially correlated fluid velocity field at $\vert\bm{r}\vert<L$.  Third, for $\vert\bm{r}\vert>L$ the motion of the two particles becomes uncorrelated and the pair dispersion becomes proportional to the one-particle dispersion, and this dispersion is time-reversible in a stationary, homogeneous velocity field.  The non-local irreversibility mechanism is consistent with the reversibility of the one-particle dispersion, since at $\vert\bm{r}\vert>L$, $\Delta\bm{u}(\bm{r},t)$ is statistically independent of $\bm{r}$, and therefore the path-history interaction of separating and approaching pairs with $\Delta\bm{u}$ become statistically equivalent, restoring FIT/BIT symmetry in this regime.  This also serves to emphasize that the statistical time-irreversibility of the particle pair dispersion does not arise simply as a consequence of the dissipative dynamics of the particles, since this would suggest that one-particle dispersion should be irreversible in stationary, homogeneous turbulence, which it is not.  Fourth, since $\bm{\nabla_r}\langle\vert\Delta\bm{u}(\bm{r},t)\vert^2\rangle>\bm{0}$ for $\vert\bm{r}\vert<L$ is simply a consequence of spatial decorrelation in the fluid velocity field, the non-local irreversibility mechanism always generates faster BIT than FIT dispersion, whether the turbulence be 2D or 3D.  However, the irreversibility mechanism associated with the turbulence dynamics depends upon the direction of the energy transfer in the velocity field, as manifested in the nature of the asymmetry of the PDF of $\Delta\bm{u}$.  As discussed earlier, since $\Delta\bm{u}$ is positively skewed in 2D turbulence because of its inverse energy transfer process, then the dispersion of fluid particles is faster for FIT than for BIT, as confirmed in the numerical simulations in \cite{faber09}.  This then leads to an interesting prediction for relative dispersion in 2D turbulence, namely, that below some critical $St$, FIT is faster than BIT dispersion, but then beyond this critical $St$ value, BIT is faster than FIT dispersion.  This critical $St$ marks the point at which the non-local irreversibility mechanism begins to dominate over the irreversibilty mechanism associated with the turbulence dynamics.  We are currently testing this prediction.

\subsection{Theoretical results}

Having considered the additional irreversibility mechanism that arises because of the particle inertia, we now construct theoretical predictions to describe the inertial particle dispersion.  In deriving the results we will need to know something about the particle relative velocity statistics.  In the following, we derive results for the dispersion given only the particle velocity statistics at $\mathcal{T}=0$.  The results then describe, given the statistical state of the particles at $\mathcal{T}=0$, how the pairs disperse as a function of $\mathcal{T}$.  Since the FIT behavior has already been analyzed in \cite{bec10b}, we shall focus on developing a theory for the BIT mean-square dispersion.

We begin by considering the regime $\mathcal{T}/\tau_{r^0}\ll1$, for which we may invoke the approximation
\begin{align}
\Delta\bm{u}^{p}(\mathcal{T})\approx\Delta\bm{u}^{p}(0)+\mathcal{O}(\mathcal{T}/\tau_{r^0}),\label{uTexp2}
\end{align}
(and similarly for the terms in (\ref{FITmsr2}) and (\ref{BITmsr2}) involving $s$ and $s^\prime$ in the time arguments), which was also used to derive the fluid particle small-time ballistic separation prediction (see \S\ref{FPDT}).  Introducing this approximation into (\ref{FITmsr2}) and (\ref{BITmsr2}) and solving the integrals, we obtain 
\begin{align}
\begin{split}
\Big\langle\vert\bm{r}^{p}(\mathcal{T})\vert^{2}\Big\rangle_{\bm{r}^0}\approx\vert\bm{r}^0\vert^{2}&+G^{2}(\mathcal{T})\Big\langle\vert\bm{w}^{p}(0)\vert^{2}\Big\rangle_{\bm{r}^0}+2G(\mathcal{T})\Big[\mathcal{T}-G(\mathcal{T})\Big]\Big\langle\bm{w}^{p}(0)\bm{\cdot}\Delta\bm{u}^{p}(0)\Big\rangle_{\bm{r}^0}\\
&+\Big[\mathcal{T}^{2}-2\mathcal{T}G(\mathcal{T})+G^{2}(\mathcal{T})\Big]\Big\langle\vert\Delta\bm{u}^{p}(0)\vert^{2}\Big\rangle_{\bm{r}^0},
\label{DispFIT_r0}
\end{split}
\end{align}
and
\begin{align}
\begin{split}
\Big\langle\vert\bm{r}^{p}(-\mathcal{T})\vert^{2}\Big\rangle_{\bm{r}^0}\approx\vert\bm{r}^0\vert^{2}&+G^{2}(\mathcal{T})\Big\langle\vert\bm{w}^{p}(-\mathcal{T})\vert^{2}\Big\rangle_{\bm{r}^0}+2G(\mathcal{T})\Big[\mathcal{T}-G(\mathcal{T})\Big]\Big\langle\bm{w}^{p}(-\mathcal{T})\bm{\cdot}\Delta\bm{u}^{p}(0)\Big\rangle_{\bm{r}^0}\\
&+\Big[\mathcal{T}^{2}-2\mathcal{T}G(\mathcal{T})+G^{2}(\mathcal{T})\Big]\Big\langle\vert\Delta\bm{u}^{p}(0)\vert^{2}\Big\rangle_{\bm{r}^0}.
\label{DispBIT_r0}
\end{split}
\end{align}
Like the fluid particle ballistic result in (\ref{Bal}), (\ref{DispFIT_r0}) and (\ref{DispBIT_r0}) are valid for any initial separation $\bm{r}^0$.  We will return shortly to consider the range of $\mathcal{T}$ for which these results should be valid.

In \cite{bec10b} the FIT result $\langle\vert\bm{r}^{p}(\mathcal{T})\vert^{2}\rangle_{\bm{r}^0}\approx\vert\bm{r}^0\vert^{2}+G^{2}(\mathcal{T})\langle\vert\bm{w}^{p}(0)\vert^{2}\rangle_{\bm{r}^0}$ was derived.  Our result in (\ref{DispFIT_r0}) contains this contribution, but is more general, capturing the influence of the local fluid velocity field on the dispersion, which is important for $St\lesssim\mathcal{O}(1)$.  We must now verify that the results in (\ref{DispFIT_r0}) and (\ref{DispBIT_r0}) obey the necessary limiting cases.  First, it is simple to confirm that (\ref{DispFIT_r0}) and (\ref{DispBIT_r0}) reduce to (\ref{Bal}) in the case when $\tau_{p}=0$ (for which $G(\mathcal{T})=0$ and $\Delta\bm{u}^{p}=\Delta\bm{u}^{f}$).  Second, in the limit $Re_{\lambda}\to\infty$ and sufficiently large $\bm{r}^{0}$, the inertial particle behavior should tend to that for fluid particles, since for a given $\tau_{p}$, $\tau_{p}/\tau_{r^{0}}\to0$ as $\vert\bm{r}^{0}\vert\to\infty$.  In this limit, we would have $\bm{w}^{p}\to\Delta\bm{u}^{f}$ and $\Delta\bm{u}^{p}\to\Delta\bm{u}^{f}$; if we introduce these into (\ref{DispFIT_r0}) and (\ref{DispBIT_r0}), we once again find that the results reduce to (\ref{Bal}).  

The result in (\ref{DispBIT_r0}) contains $\bm{w}^{p}(-\mathcal{T})$; we wish to derive theoretical descriptions that require only knowledge of the particle velocity statistics at $\mathcal{T}=0$.  The solution for $\bm{w}^{p}(-\mathcal{T})$ is
\begin{align}
\bm{w}^{p}(-\mathcal{T})=\dot{G}^{-1}(\mathcal{T})\bm{w}^{p}(0)-\tau_{p}^{-1}\dot{G}^{-1}(\mathcal{T})\int\limits^{0}_{-\mathcal{T}}\dot{G}(-s)\Delta\bm{u}^{p}(s)ds,
\label{wsol}
\end{align}
where $\dot{G}(-s)=\exp[\tau_{p}^{-1}s]$.  From (\ref{wsol}) we obtain the following
\begin{align}
\begin{split}
\Big\langle\vert\bm{w}^{p}(-\mathcal{T})\vert^{2}\Big\rangle_{\bm{r}^0}&=\dot{G}^{-2}(\mathcal{T})\Big\langle\vert\bm{w}^{p}(0)\vert^{2}\Big\rangle_{\bm{r}^0}-2\tau_{p}^{-1}\dot{G}^{-2}(\mathcal{T})\int\limits^{0}_{-\mathcal{T}}\dot{G}(-s)\Big\langle\bm{w}^{p}(0)\bm{\cdot}\Delta\bm{u}^{p}(s)\Big\rangle_{\bm{r}^0}ds\\
&\quad+\tau_{p}^{-2}\dot{G}^{-2}(\mathcal{T})\int\limits^{0}_{-\mathcal{T}}\int\limits^{0}_{-\mathcal{T}}\dot{G}(-s)\dot{G}(-s^\prime)\Big\langle\Delta\bm{u}^{p}(s)\bm{\cdot}\Delta\bm{u}^{p}(s^\prime)\Big\rangle_{\bm{r}^0}ds^\prime\,ds.
\label{w2sol}
\end{split}
\end{align}
Introducing into this (\ref{uTexp2}) and evaluating the integrals, we obtain
\begin{align}
\begin{split}
\Big\langle\vert\bm{w}^{p}(-\mathcal{T})\vert^{2}\Big\rangle_{\bm{r}^0}&\approx\dot{G}^{-2}(\mathcal{T})\Big\langle\bm{w}^{p}(0)\bm{\cdot}\bm{w}^{p}(0)\Big\rangle_{\bm{r}^0}-2\tau_{p}^{-1}\dot{G}^{-2}(\mathcal{T})G(\mathcal{T})\Big\langle\bm{w}^{p}(0)\bm{\cdot}\Delta\bm{u}^{p}(0)\Big\rangle_{\bm{r}^0}\\
&\quad+\tau_{p}^{-2}\dot{G}^{-2}(\mathcal{T})G^{2}(\mathcal{T})\Big\langle\Delta\bm{u}^{p}(0)\bm{\cdot}\Delta\bm{u}^{p}(0)\Big\rangle_{\bm{r}^0}.
\label{w2sol2}
\end{split}
\end{align}
In a similar manner, we also obtain 
\begin{align}
\begin{split}
\Big\langle\bm{w}^{p}(-\mathcal{T})\bm{\cdot}\Delta\bm{u}^{p}(0)\Big\rangle_{\bm{r}^0}&\approx\dot{G}^{-1}(\mathcal{T})\Big\langle\bm{w}^{p}(0)\bm{\cdot}\Delta\bm{u}^{p}(0)\Big\rangle_{\bm{r}^0}-\tau_{p}^{-1}\dot{G}^{-1}(\mathcal{T})G(\mathcal{T})\Big\langle\Delta\bm{u}^{p}(0)\bm{\cdot}\Delta\bm{u}^{p}(0)\Big\rangle_{\bm{r}^0}.
\label{wusol}
\end{split}
\end{align}
Using these results in (\ref{DispBIT_r0}), we obtain
\begin{align}
\begin{split}
\Big\langle\vert\bm{r}^{p}(-\mathcal{T})\vert^{2}\Big\rangle_{\bm{r}^0}\approx\vert\bm{r}^0\vert^{2}&+G^{2}(-\mathcal{T})\Big\langle\vert\bm{w}^{p}(0)\vert^{2}\Big\rangle_{\bm{r}^0}-2G(-\mathcal{T})\Big(G(-\mathcal{T})+\mathcal{T}\Big)\Big\langle\bm{w}^{p}(0)\bm{\cdot}\Delta\bm{u}^{p}(0)\Big\rangle_{\bm{r}^0}\\
&+\Big[G^{2}(-\mathcal{T})+2\mathcal{T}G(-\mathcal{T})+\mathcal{T}^{2}\Big]\Big\langle\vert\Delta\bm{u}^{p}(0)\vert^{2}\Big\rangle_{\bm{r}^0}.
\label{DispBIT_r02a}
\end{split}
\end{align}
In addition, since we do not in general know the statistics of $\Delta\bm{u}^p$, we make the approximation
\begin{align}
\Big\langle\vert\Delta\bm{u}^{p}(0)\vert^{2}\Big\rangle_{\bm{r}^0}&\approx\Big\langle\vert\Delta\bm{u}^{f}(0)\vert^{2}\Big\rangle_{\bm{r}^0},\label{upufA}
\end{align}
and also
\begin{align}
\Big\langle\bm{w}^{p}(0)\bm{\cdot}\Delta\bm{u}^{p}(0)\Big\rangle_{\bm{r}^0}&\approx\sqrt{\Big\langle\vert\bm{w}^{p}(0)\vert^{2}\Big\rangle_{\bm{r}^0}}\sqrt{\Big\langle\vert\Delta\bm{u}^{f}(0)\vert^{2}\Big\rangle_{\bm{r}^0}},
\end{align}
which finally gives us 
\begin{align}
\begin{split}
\Big\langle\vert\bm{r}^{p}(-\mathcal{T})\vert^{2}\Big\rangle_{\bm{r}^0}\approx\vert\bm{r}^0\vert^{2}&+G^{2}(-\mathcal{T})\Big\langle\vert\bm{w}^{p}(0)\vert^{2}\Big\rangle_{\bm{r}^0}+\Big[G^{2}(-\mathcal{T})+2\mathcal{T}G(-\mathcal{T})+\mathcal{T}^{2}\Big]\Big\langle\vert\Delta\bm{u}^{f}(0)\vert^{2}\Big\rangle_{\bm{r}^0}\\
&-2G(-\mathcal{T})\Big[G(-\mathcal{T})+\mathcal{T}\Big]\sqrt{\Big\langle\vert\bm{w}^{p}(0)\vert^{2}\Big\rangle_{\bm{r}^0}}\sqrt{\Big\langle\vert\Delta\bm{u}^{f}(0)\vert^{2}\Big\rangle_{\bm{r}^0}}.
\label{STT}
\end{split}
\end{align}
Note that in the expression $\langle\vert\Delta\bm{u}^{f}(0)\vert^{2}\rangle_{\bm{r}^0}$, the conditionality is $\bm{r}^f(0)=\bm{r}^{0}$ (not $\bm{r}^p(0)=\bm{r}^{0}$), such that ${\langle\vert\Delta\bm{u}^{f}(0)\vert^{2}\rangle_{\bm{r}^0}=\langle\vert\Delta\bm{u}(\bm{r}^0,0)\vert^{2}\rangle}$ since the fluid particles are fully mixed.

It is possible that (\ref{STT}) may not be accurate for small $St$, since we have partially removed the effect of the preferential sampling of $\Delta\bm{u}$ by the inertial particles by invoking approximation (\ref{upufA}) (some of the effect is captured within $\langle\vert\bm{w}^{p}(0)\vert^{2}\rangle_{\bm{r}^0}$).  We consider the effects of this on (\ref{STT}) for low $St$ particles in \S\ref{IPS}.

The result in (\ref{STT}) implies that the time dependance of the mean-square dispersion BIT depends not only upon $St$ but also $\bm{r}^0$, which is distinct from the ${St=0}$ case where the mean-square dispersion in the small-time regime grows as $\mathcal{T}^2$ for any $\bm{r}^0$.  In particular, in the dissipation regime where $\langle\vert\bm{w}^{p}(0)\vert^{2}\rangle_{\bm{r}^0}\propto\vert\bm{r}^0\vert^{\xi}$ with $\xi(St)\leq 2$ \cite{gustavsson11,salazar12a}, then for a given finite $St$, $\langle\vert\bm{r}^{p}(-\mathcal{T})\vert^{2}\rangle_{\bm{r}^0}$ may grow like $\mathcal{T}^2$ at larger separations (where $\langle\vert\bm{w}^{p}(0)\vert^{2}\rangle_{\bm{r}^0}/\langle\vert\Delta\bm{u}^{f}(0)\vert^{2}\rangle_{\bm{r}^0}=\mathcal{O}(1)$) but grow like $G^2(-\mathcal{T})$ in the limit $\vert\bm{r}^0\vert/\eta\to0$.

Formally, the range of $\mathcal{T}$ over which (\ref{STT}) should remain valid is determined by the approximation in (\ref{uTexp2}).  Since this approximation leads to a good description of the fluid particle dispersion up to ${\mathcal{T}=\mathcal{O}(\tau_{r^0})}$, then we may expect that (\ref{STT}) should remain valid for ${\mathcal{T}\leq\mathcal{O}(\tau_{r^0})}$.  There is, however, an exception to this:  In regions where $\bm{w}^p\gg\Delta\bm{u}^p$, which we refer to as `caustic regions' \cite{wilkinson05,bec10a,salazar12a}, $\Delta\bm{u}^p$ is irrelevant to the particle dispersion process, and so the range of the validity of (\ref{uTexp2}) does not control the range of the validity of (\ref{STT}).  In this case, the range of the validity of (\ref{STT}) is controlled by how long $\bm{w}^p\gg\Delta\bm{u}^p$ along the pair trajectory.  We expect that this time should be $\mathcal{O}(\tau_p)$, reflecting the time it takes for the particles to dissipate their excess kinetic energy relative to that of the local fluid velocity difference field.  Therefore,  (\ref{STT}) should remain valid for ${\mathcal{T}\leq\widehat{\mathcal{T}}}$ where
\begin{align}
\widehat{\mathcal{T}}=
\left\{
	\begin{array}{ll}
		\tau_p  & \mbox{if } \langle\vert\bm{w}^{p}(0)\vert^{2}\rangle_{\bm{r}^0}\gg\langle\vert\Delta\bm{u}^{f}(0)\vert^{2}\rangle_{\bm{r}^0}\\
		\tau_{r^0} & \mbox{otherwise.}
	\end{array}
\right.
\end{align}

We now consider the regime ${\mathcal{T}>\widehat{\mathcal{T}}}$.  If we ignore the terms involving the mean of $\bm{w}^p$ and $\Delta\bm{u}^p$, under the assumption that their contribution is small relative to the other terms, we may re-write (\ref{BITmsr2}) for $\mathcal{T}>\widehat{\mathcal{T}}$ as 
\begin{align}
\begin{split}
\Big\langle\vert\bm{r}^{p}(-\mathcal{T})\vert^{2}\Big\rangle_{\bm{r}^0}\approx\, &\Big\langle\vert\bm{r}^{p}(-\widehat{\mathcal{T}})\vert^{2}\Big\rangle_{\bm{r}^0}+G^{2}(\mathcal{T})\Big\langle\vert\bm{w}^{p}(-\mathcal{T})\vert^{2}\Big\rangle_{\bm{r}^0}-G^{2}(\widehat{\mathcal{T}})\Big\langle\vert\bm{w}^{p}(-\widehat{\mathcal{T}})\vert^{2}\Big\rangle_{\bm{r}^0}\\
&+2\tau_{p}^{-1}G(\mathcal{T})\int\limits^{-\widehat{\mathcal{T}}}_{-\mathcal{T}}G(-s)\Big\langle\bm{w}^{p}(-\mathcal{T})\bm{\cdot}\Delta \bm{u}^{p}(s)\Big\rangle_{\bm{r}^0}ds\\
&+\tau_{p}^{-2}\int\limits^{-\widehat{\mathcal{T}}}_{-\mathcal{T}}\int\limits^{-\widehat{\mathcal{T}}}_{-\mathcal{T}}G(-s)G(-s^\prime)\Big\langle\Delta\bm{u}^{p}(s)\bm{\cdot}\Delta\bm{u}^{p}(s^\prime)\Big\rangle_{\bm{r}^0}ds^\prime\,ds.
\label{BITmsr2LargeT}
\end{split}
\end{align}
Determining the appropriate closure approximations to apply to (\ref{BITmsr2LargeT}) depends upon both $St$ and $r^0$.  For example, for $r^0\ll\eta$, the particles may be either still in the dissipation regime or in the inertial regime at time $\widehat{\mathcal{T}}$, depending upon the value of $St$.  We will assume that at time $\widehat{\mathcal{T}}$ the pairs are in the inertial regime and leave the other case for future work (especially since, as shown in \S\ref{FPS}, even the fluid particle separation in the dissipation range at finite $\mathcal{T}$ is not fully understood).  

Let us define a time-dependent Stokes number in the inertial range as $St_r(t)\equiv \tau_p/\tau_{r}$ where $\tau_{r}\equiv (\vert\bm{r}^p(t)\vert^2/\langle\epsilon\rangle)^{1/3}$, which satisfies ${St_r(t)\to0}$ for ${\vert\bm{r}^p(t)\vert/\eta\to\infty}$ in the limit ${Re_\lambda\to\infty}$.  This implies that the effect of the particles' inertia becomes perturbative as their separation growth increases such that ${\bm{w}^{p}(t)\approx\Delta\bm{u}^{p}(t)+\mathcal{O}(St_r(t))}$.  We will make the approximation that $\bm{w}^{p}\approx\Delta\bm{u}^{p}$ for $\mathcal{T}>\widehat{\mathcal{T}}$.  We then need to describe $\Delta\bm{u}^{p}$ in the regime $St_r(t)\ll1$, and to do this we introduce the expansion
\begin{align}
\Delta\bm{u}^p(t)=\Delta\bm{u}^{[0]}(t)+St_r(t)\Delta\bm{u}^{[1]}(t)+\mathcal{O}([St_r(t)]^2),\label{DupPer}
\end{align}
where the superscripts $[0],[1]$ denote the order of the perturbation term and $\Delta\bm{u}^{[0]}(t)\equiv\Delta\bm{u}^{f}(t)$.  Under the approximation $\bm{w}^{p}\approx\Delta\bm{u}^{p}$, the unknown terms in (\ref{BITmsr2LargeT}) all involve autocovariances of $\Delta\bm{u}^{p}$, and using (\ref{DupPer}), these are expressed as
\begin{align}
\begin{split}
\Big\langle\Delta\bm{u}^{p}(s)\bm{\cdot}\Delta\bm{u}^{p}(s^\prime)\Big\rangle_{\bm{r}^0}=&\quad\Big\langle\Delta\bm{u}^{f}(s)\bm{\cdot}\Delta\bm{u}^{f}(s^\prime)\Big\rangle_{\bm{r}^0}+\Big\langle[\Delta\bm{u}^{f}(s)\bm{\cdot}\Delta\bm{u}^{[1]}(s^\prime)]St_r(s^\prime)\Big\rangle_{\bm{r}^0}\\
&+\Big\langle[\Delta\bm{u}^{f}(s)\bm{\cdot}\Delta\bm{u}^{[1]}(s^\prime)]St_r(s)\Big\rangle_{\bm{r}^0}+\mathcal{O}\Big(St_r(s)St_r(s^\prime)\Big).\label{DupPerCov}
\end{split}	
\end{align}
The terms involving $\mathcal{T}$ instead of $s$ in the time arguments are treated similarly.  Making the crude approximation that ${\vert\bm{r}^p(s)\vert^2\approx-\mathfrak{g}^{B}\langle\epsilon\rangle s^3}$ (for ${s<0}$), we have ${St_r(s)\approx -St\tau_\eta[\mathfrak{g}^{B}]^{-1/3}s^{-1}}$.  Using K41 arguments for the Lagrangian behavior of $\Delta\bm{u}$ in the inertial range, we have
\begin{align}
\Big\langle\Delta\bm{u}^{f}(s)\bm{\cdot}\Delta\bm{u}^{f}(s^\prime)\Big\rangle_{\bm{r}^0}\approx&-(1/2)\mathcal{A}^{[0]}\langle\epsilon\rangle(s+s^\prime),\quad\text{for}\quad s<0,\, s^\prime<0,\label{LK41a}\\
\Big\langle\Delta\bm{u}^{f}(s)\bm{\cdot}\Delta\bm{u}^{[1]}(s^\prime)\Big\rangle_{\bm{r}^0}\approx&-(1/2)\mathcal{A}^{[1]}\langle\epsilon\rangle(s+s^\prime),\quad\text{for}\quad s<0,\, s^\prime<0,\label{LK41b}
\end{align}
and substituting these into (\ref{DupPerCov}), we have
\begin{align}
\begin{split}
\Big\langle\Delta\bm{u}^{p}(s)\bm{\cdot}\Delta\bm{u}^{p}(s^\prime)\Big\rangle_{\bm{r}^0}\approx-(1/2)\mathcal{A}^{[0]}\langle\epsilon\rangle(s+s^\prime)+(1/2)St\tau_\eta\langle\epsilon\rangle\mathcal{A}^{[1]}[\mathfrak{g}^{B}]^{-1/3}\Bigg(\frac{s+s^\prime}{s}+\frac{s+s^\prime}{s^\prime}\Bigg),\label{DupPerCov2}
\end{split}	
\end{align}
where $\mathcal{A}^{[0]}$ and $\mathcal{A}^{[1]}$ are positive (so that (\ref{LK41a}) and (\ref{LK41b}) increase backward-in-time in the inertial range, as we would expect), dimensionless constants to be determined later.  If we now substitute (\ref{DupPerCov2}) into (\ref{BITmsr2LargeT}), invoking $\bm{w}^{p}\approx\Delta\bm{u}^{p}$ and $G(\mathcal{T}\geq\widehat{\mathcal{T}})\approx\tau_{p}$, and solve the integrals, we obtain
\begin{align}
\begin{split}
\Big\langle\vert\bm{r}^{p}(-\mathcal{T})\vert^{2}\Big\rangle_{\bm{r}^{0}}&\approx\Big\langle\vert\bm{r}^{p}(-\widehat{\mathcal{T}})\vert^{2}\Big\rangle_{\bm{r}^{0}} +\tau_{p}^{2}\mathcal{A}^{[0]}\langle\epsilon\rangle\Big(\mathcal{T}-\widehat{\mathcal{T}}\Big)+\tau_{p}(1/2)\mathcal{A}^{[0]}\langle\epsilon\rangle\Big(3\mathcal{T}^{2}-\widehat{\mathcal{T}}^{2}-2\widehat{\mathcal{T}}\mathcal{T}\Big)\\
&\quad+(St\tau_\eta)^2\langle\epsilon\rangle\mathcal{A}^{[1]}[\mathfrak{g}^B]^{-1/3}\Big(2(\mathcal{T}-\widehat{\mathcal{T}})+(1/2\mathcal{T})(\mathcal{T}^2-\widehat{\mathcal{T}}^2)+\mathcal{T}\ln[\mathcal{T}/\widehat{\mathcal{T}}]\Big)\\
&\quad+(1/2)St\tau_\eta\langle\epsilon\rangle\mathcal{A}^{[1]}[\mathfrak{g}^B]^{-1/3}\Big(2(\mathcal{T}-\widehat{\mathcal{T}})^2+(\mathcal{T}^2-\widehat{\mathcal{T}}^2)\ln[\mathcal{T}/\widehat{\mathcal{T}}]\Big)\\
&\quad+(1/2)\mathcal{A}^{[0]}\langle\epsilon\rangle\Big(\mathcal{T}^{3}+\widehat{\mathcal{T}}^{3}-\widehat{\mathcal{T}}^{2}\mathcal{T}-\widehat{\mathcal{T}}\mathcal{T}^{2}\Big),\quad\text{for $\mathcal{T}>\widehat{\mathcal{T}}$}.
\end{split}
\label{LTT}
\end{align}
Taking the limit $\tau_{p}\to0$ and considering the regime $\mathcal{T}\gg\widehat{\mathcal{T}}$, we may identify $(1/2)\mathcal{A}^{[0]}$ in (\ref{LTT}) as the backward-in-time Richardson constant $\mathfrak{g}^{B}$.  The constant $\mathcal{A}^{[1]}$ is independent of $St$ to leading order and we will later estimate its value from DNS.   Note that by its essentially perturbative construction, (\ref{LTT}) is, like the fluid particle RO law, free from intermittency corrections, since it depends linearly upon the kinetic energy dissipation rate of the fluid \cite{novikov90,boffetta99,schmitt05}. 

In the regime $\mathcal{T}\gg\widehat{\mathcal{T}}$, the result in (\ref{LTT}) simplifies to

\begin{align}
\Big\langle\vert\bm{r}^{p}(-\mathcal{T})\vert^{2}\Big\rangle_{\bm{r}^0}\approx \mathfrak{g}^B\langle\epsilon\rangle\mathcal{T}^3\Big[1+St\mathcal{B}\mathcal{T}^{-1}\ln[\mathcal{T}/\widehat{\mathcal{T}}]\Big],\quad\text{for $\mathcal{T}\gg\widehat{\mathcal{T}}$},\label{LTT2}
\end{align}
where $\mathcal{B}\equiv(1/2)\tau_\eta\mathcal{A}^{[1]}[\mathfrak{g}^{B}]^{-4/3}\geq0$.  The result in (\ref{LTT2}) is quite similar to the equivalent FIT result derived in \cite{bec10b}, but differs in one important respect.  Whereas the FIT result derived in \cite{bec10b} predicts that in the limit $\mathcal{T}/\widehat{\mathcal{T}}\to\infty$, $\langle\vert\bm{r}^{p}(\mathcal{T})\vert^{2}\rangle_{\bm{r}^0}/\langle\vert\bm{r}^{f}(\mathcal{T})\vert^{2}\rangle_{\bm{r}^0}$ approaches unity from below, (\ref{LTT2}) suggests that in the BIT case $\langle\vert\bm{r}^{p}(-\mathcal{T})\vert^{2}\rangle_{\bm{r}^0}/\langle\vert\bm{r}^{f}(-\mathcal{T})\vert^{2}\rangle_{\bm{r}^0}$ approaches unity from above.  In other words, in the BIT case the perturbative effect of the particle inertia is to increase the separation rate relative to that of fluid particles, whereas in the FIT case it decreases the separation rate.  

\subsection{DNS results}\label{IPS}

Before considering the results for the particle-pair mean-square separation, we first present the DNS data for $\langle\vert\bm{w}^{p}(0)\vert^{2}\rangle_{\bm{r}^0}$.  This statistic is useful to consider both because it will aid in the understanding the dispersion results, and because it features in the small-time theory (\ref{STT}).  
\begin{figure}[ht]
\centering
{\begin{overpic}
[trim = 30mm 70mm 20mm 65mm,scale=0.5,clip,tics=20]{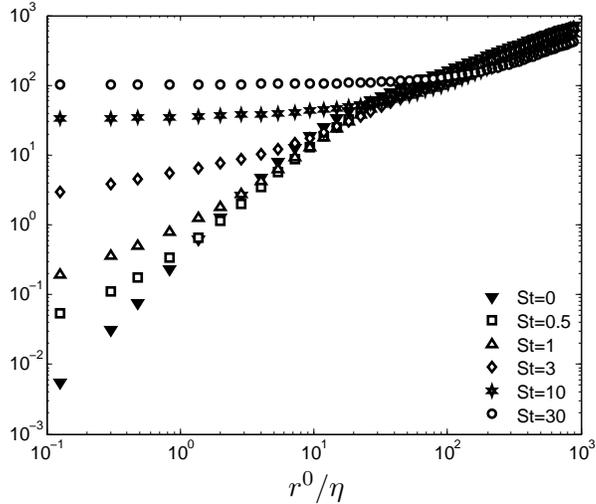}
\put(105,0){$r^0/\eta$}
\end{overpic}}
\caption{DNS data for $\langle\vert\bm{w}^{p}(0)\vert^{2}\rangle_{\bm{r}^0}/u_\eta^2$ as a function of $r^0/\eta$ for various $St$.}
\label{w2_DNS}
\end{figure}
Comparing the $St>0$ results in Figure~\ref{w2_DNS} with the $St=0$ results (i.e. fluid particles, for which $\langle\vert\bm{w}^{p}(0)\vert^{2}\rangle_{\bm{r}^0}=\langle\vert\Delta\bm{u}^{f}(0)\vert^{2}\rangle_{\bm{r}^0}$), we see that at larger separations the particle inertia gives rise to $\langle\vert\bm{w}^{p}(0)\vert^{2}\rangle_{\bm{r}^0}<\langle\vert\Delta\bm{u}^{f}(0)\vert^{2}\rangle_{\bm{r}^0}$, whereas at the smaller separations, $\langle\vert\bm{w}^{p}(0)\vert^{2}\rangle_{\bm{r}^0}>\langle\vert\Delta\bm{u}^{f}(0)\vert^{2}\rangle_{\bm{r}^0}$.  The separation at which the transition in behavior occurs is a strong function of $St$.  For a detailed explanation of the role of inertia on these statistics see \cite{bragg14c,salazar12a}; here, we summarize.  The predominant effects of inertia at these Stokes numbers are the filtering and non-local effects (preferential sampling of $\Delta\bm{u}$ has a role mainly for particles with $0<St\lesssim0.4$ and separations in the dissipation range \cite{salazar12a}).  The inertia of the particles causes them to filter out the high frequency fluctuations of $\Delta\bm{u}$, and since the inertia gives the particles a memory, their velocity dynamics at a given separation are strongly influenced by their path-history interactions with the turbulent velocity field.  For a given $\tau_{p}$, the non-local contribution becomes less important as one goes to larger and larger separations causing the filtering effect to dominate; this gives rise to $\langle\vert\bm{w}^{p}(0)\vert^{2}\rangle_{\bm{r}^0}<\langle\vert\Delta\bm{u}^{f}(0)\vert^{2}\rangle_{\bm{r}^0}$.  At the smaller scales, the non-local effect of inertia dominates and gives rise to $\langle\vert\bm{w}^{p}(0)\vert^{2}\rangle_{\bm{r}^0}>\langle\vert\Delta\bm{u}^{f}(0)\vert^{2}\rangle_{\bm{r}^0}$.
\begin{figure}[ht]
\centering
\vspace{-2mm}
\subfloat[]
{\begin{overpic}
[trim = 10mm 60mm 20mm 60mm,scale=0.45,clip,tics=20]{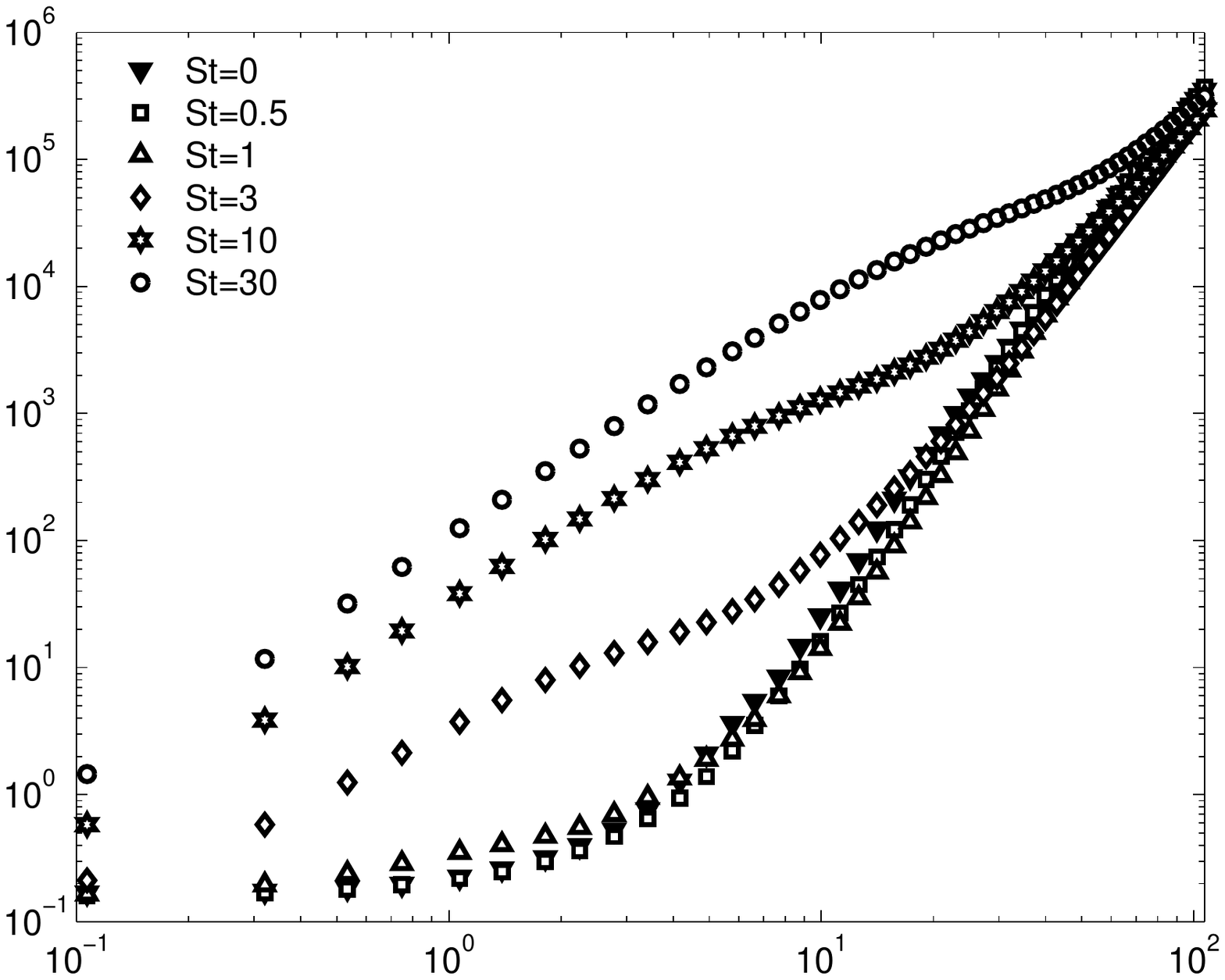}
\put(90,195){$\Big\langle\vert\bm{r}^{p}(\mathcal{T})\vert^{2}\Big\rangle_{\bm{r}^0}\Big/\eta^{2}$}
\put(115,10){$\mathcal{T}/\tau_{\eta}$}\end{overpic}}
\subfloat[]
{\begin{overpic}
[trim = 10mm 60mm 20mm 60mm,scale=0.45,clip,tics=20]{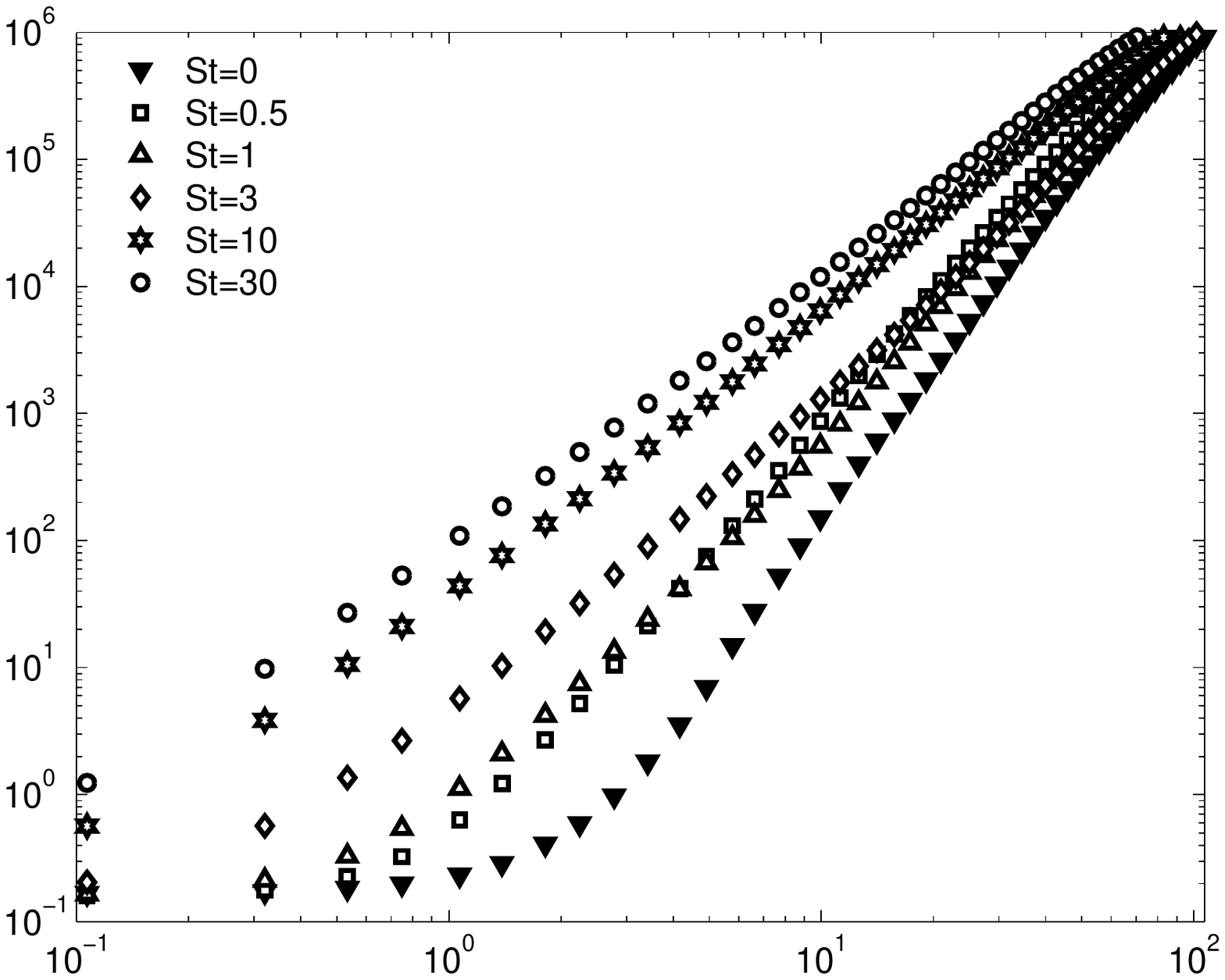}
\put(90,195){$\Big\langle\vert\bm{r}^{p}(-\mathcal{T})\vert^{2}\Big\rangle_{\bm{r}^0}\Big/\eta^{2}$}
\put(115,10){$\mathcal{T}/\tau_{\eta}$}
\end{overpic}}\\
\vspace{1mm}
\subfloat[]
{\begin{overpic}
[trim = 10mm 60mm 20mm 60mm,scale=0.45,clip,tics=20]{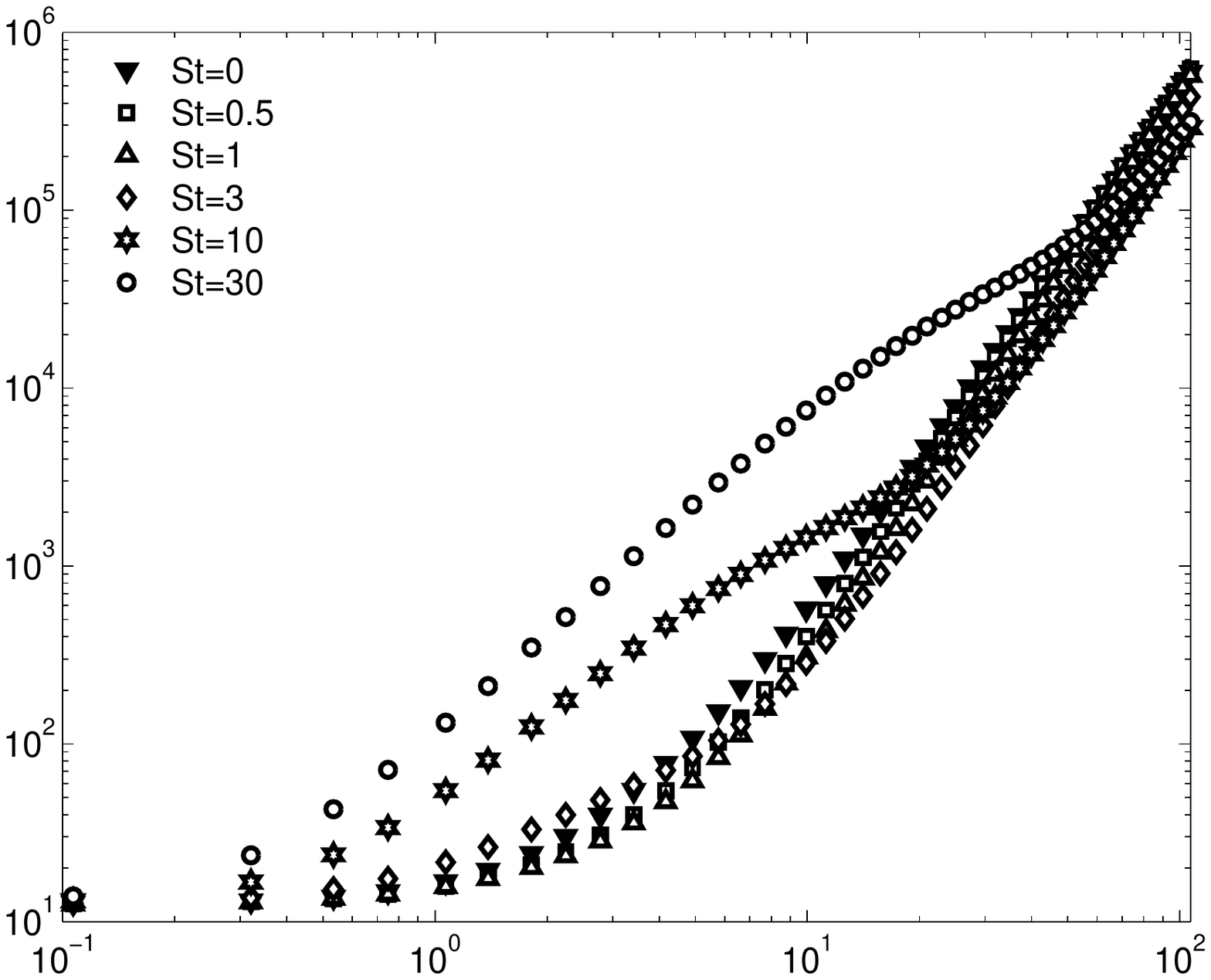}
\put(90,195){$\Big\langle\vert\bm{r}^{p}(\mathcal{T})\vert^{2}\Big\rangle_{\bm{r}^0}\Big/\eta^{2}$}
\put(115,10){$\mathcal{T}/\tau_{\eta}$}\end{overpic}}
\subfloat[]
{\begin{overpic}
[trim = 10mm 60mm 20mm 60mm,scale=0.45,clip,tics=20]{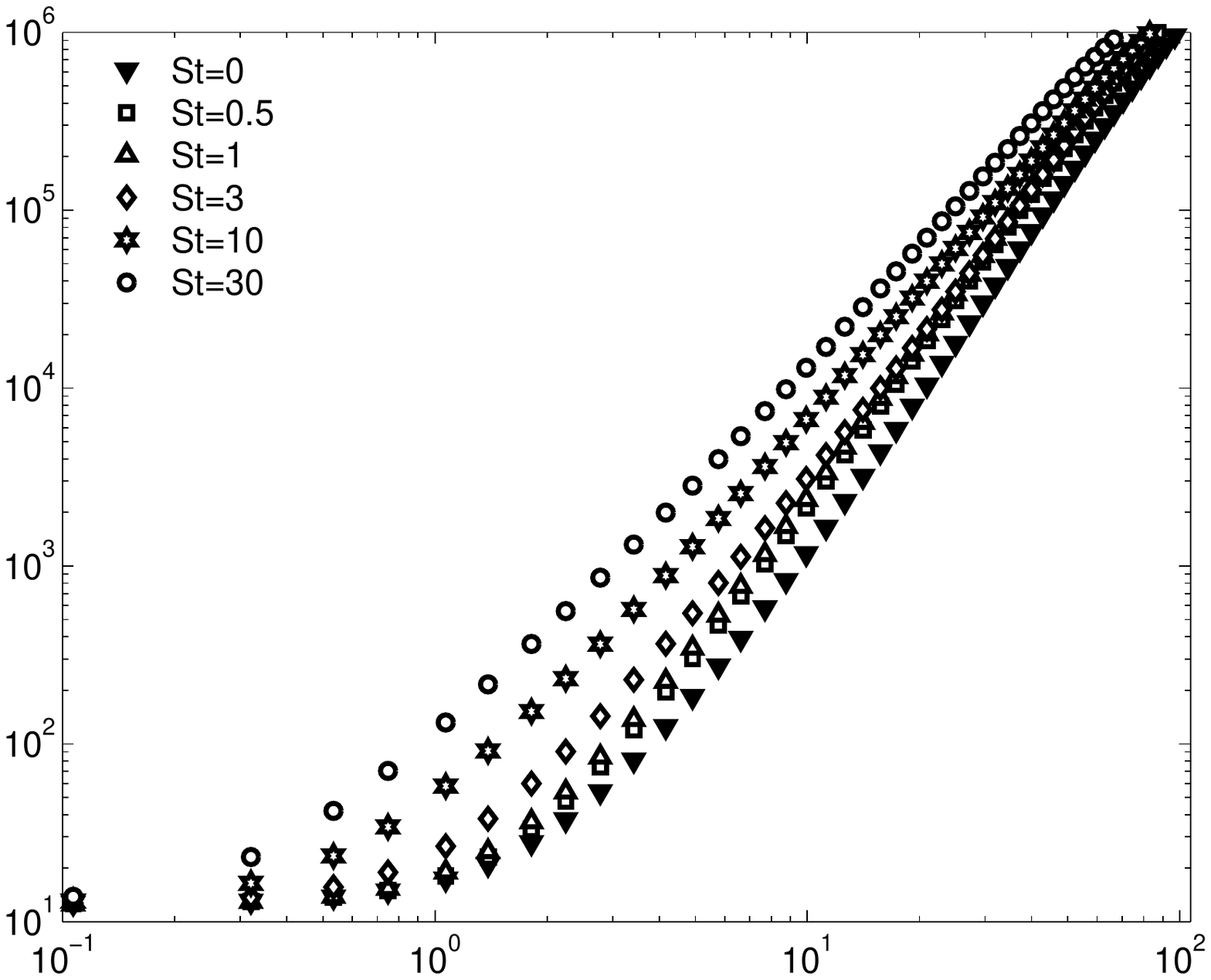}
\put(90,195){$\Big\langle\vert\bm{r}^{p}(-\mathcal{T})\vert^{2}\Big\rangle_{\bm{r}^0}\Big/\eta^{2}$}
\put(115,10){$\mathcal{T}/\tau_{\eta}$}
\end{overpic}}
\caption{FIT and BIT mean-square separation results from DNS for (a),(b) $r^{0}\in[0.25\eta,0.5\eta]$ and (c),(d) $r^{0}\in[3\eta,4\eta]$ and various $St$.}
\label{FIT_BIT_DNS}
\end{figure}
\FloatBarrier
Fig.~\ref{FIT_BIT_DNS} shows results for $\langle\vert\bm{r}^{p}(\mathcal{T})\vert^{2}\rangle_{\bm{r}^0}$ and $\langle\vert\bm{r}^{p}(-\mathcal{T})\vert^{2}\rangle_{\bm{r}^0}$ from DNS for ${r^{0}\in[0.25\eta,0.5\eta]}$, ${r^{0}\in[3\eta,4\eta]}$ and various $St$.  The results confirm the expected behavior based on the irreversibility mechanism described in \S\ref{IPIM}, namely that BIT dispersion is faster than FIT dispersion for the inertial particles.  The results also show that in addition to quantitative differences, significant qualitative differences exist between the FIT and BIT dispersion, and these may also be understood in terms of the mechanism described in \S\ref{IPIM}, as we now explain.

The FIT dispersion is explained as follows: For small $r^{0}$ and sufficiently large $St$, the inertial particles initially separate much faster than the fluid particles owing to the presence of caustics in their relative velocities at small $r^{0}$, giving rise to $\bm{w}^{p}\gg\Delta\bm{u}^{p}$ (see Fig.~\ref{w2_DNS}).  This continues up to $\mathcal{T}\sim\tau_{p}$, after which the fluid velocity field begins to significantly affect their motion.  After a few multiples of $\tau_{p}$, the fluid particles overtake the inertial particles because they are now separating faster than the inertial particles.  The inertial particle separation begins to lag behind that of the fluid particles for two reasons: First, inertia filters out high-frequency modes of $\Delta\bm{u}^{p}$ \cite{salazar12a}.  Second, because the particles are on average separating, they carry a memory of smaller fluid velocity differences in their path history.  Consequently, if we consider a set of inertial particles with a range of $\tau_{p}$, then for $\mathcal{T}\gg\max[\tau_{p}]$, $\langle\vert\bm{r}^{p}(\mathcal{T})\vert^{2}\rangle_{\bm{r}^0}$ decreases with increasing $\tau_{p}$.  This behavior is observable in Fig.~\ref{FIT_BIT_DNS} (c).

\begin{figure}[ht]
\centering
\subfloat[]
{\begin{overpic}
[trim = 10mm 60mm 20mm 60mm,scale=0.45,clip,tics=20]{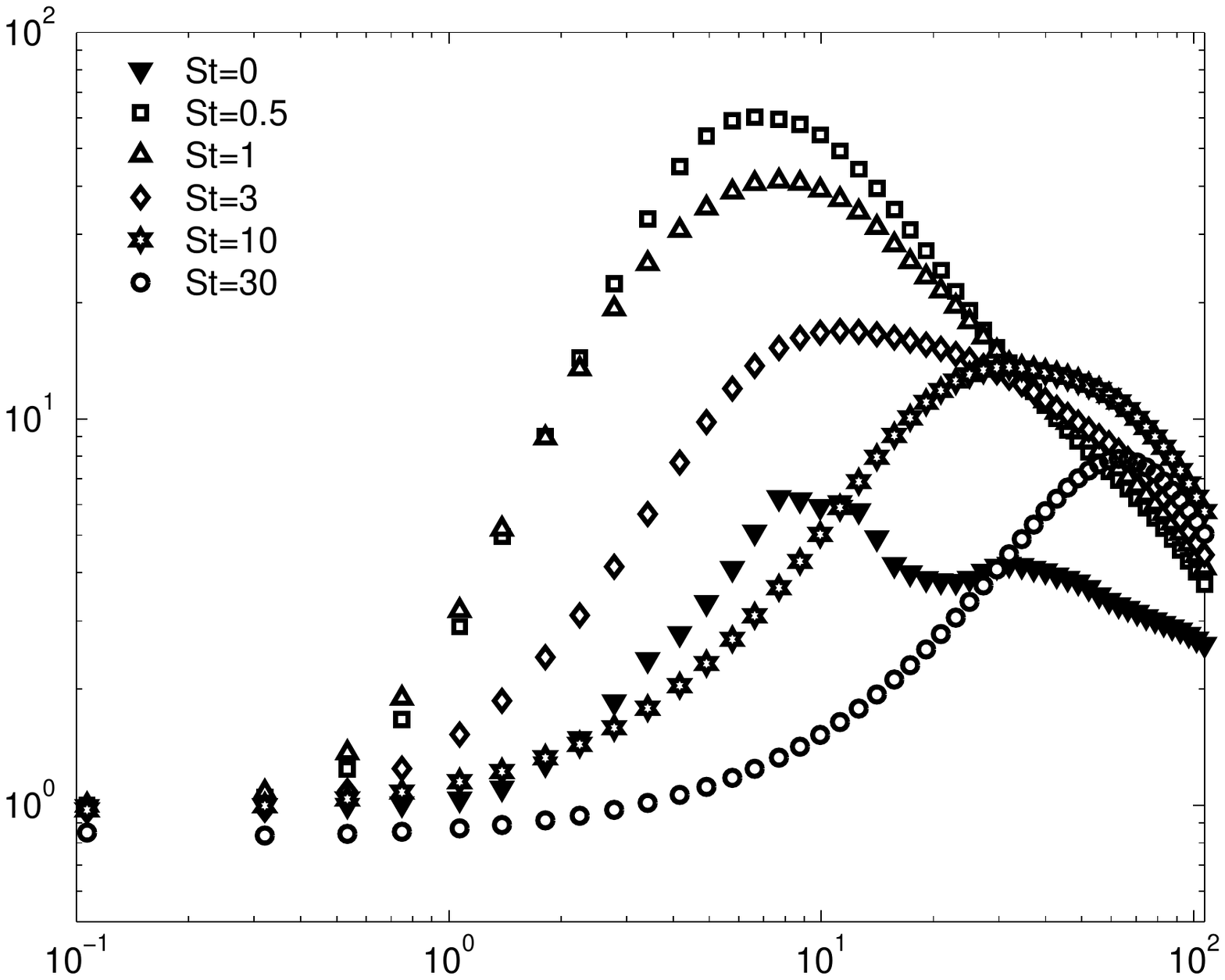}
\put(60,195){$\Big\langle\vert\bm{r}^{p}(-\mathcal{T})\vert^{2}\Big\rangle_{\bm{r}^0}\Big/\Big\langle\vert\bm{r}^{p}(\mathcal{T})\vert^{2}\Big\rangle_{\bm{r}^0}$}
\put(115,10){$\mathcal{T}/\tau_{\eta}$}
\end{overpic}}
\subfloat[]
{\begin{overpic}
[trim = 10mm 60mm 20mm 60mm,scale=0.45,clip,tics=20]{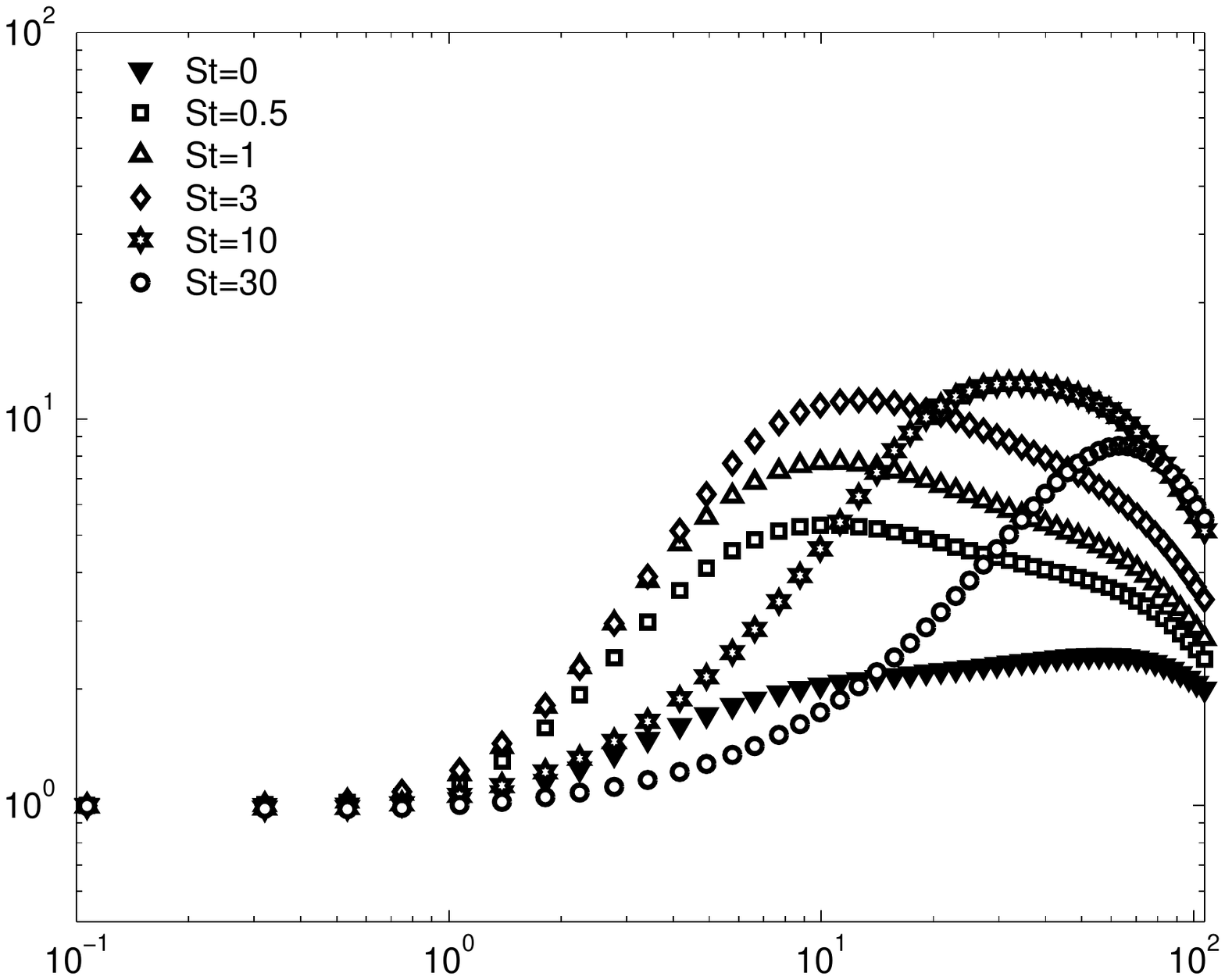}
\put(60,195){$\Big\langle\vert\bm{r}^{p}(-\mathcal{T})\vert^{2}\Big\rangle_{\bm{r}^0}\Big/\Big\langle\vert\bm{r}^{p}(\mathcal{T})\vert^{2}\Big\rangle_{\bm{r}^0}$}
\put(115,10){$\mathcal{T}/\tau_{\eta}$}
\end{overpic}}\\
\vspace{1mm}
\subfloat[]
{\begin{overpic}
[trim = 10mm 60mm 20mm 60mm,scale=0.45,clip,tics=20]{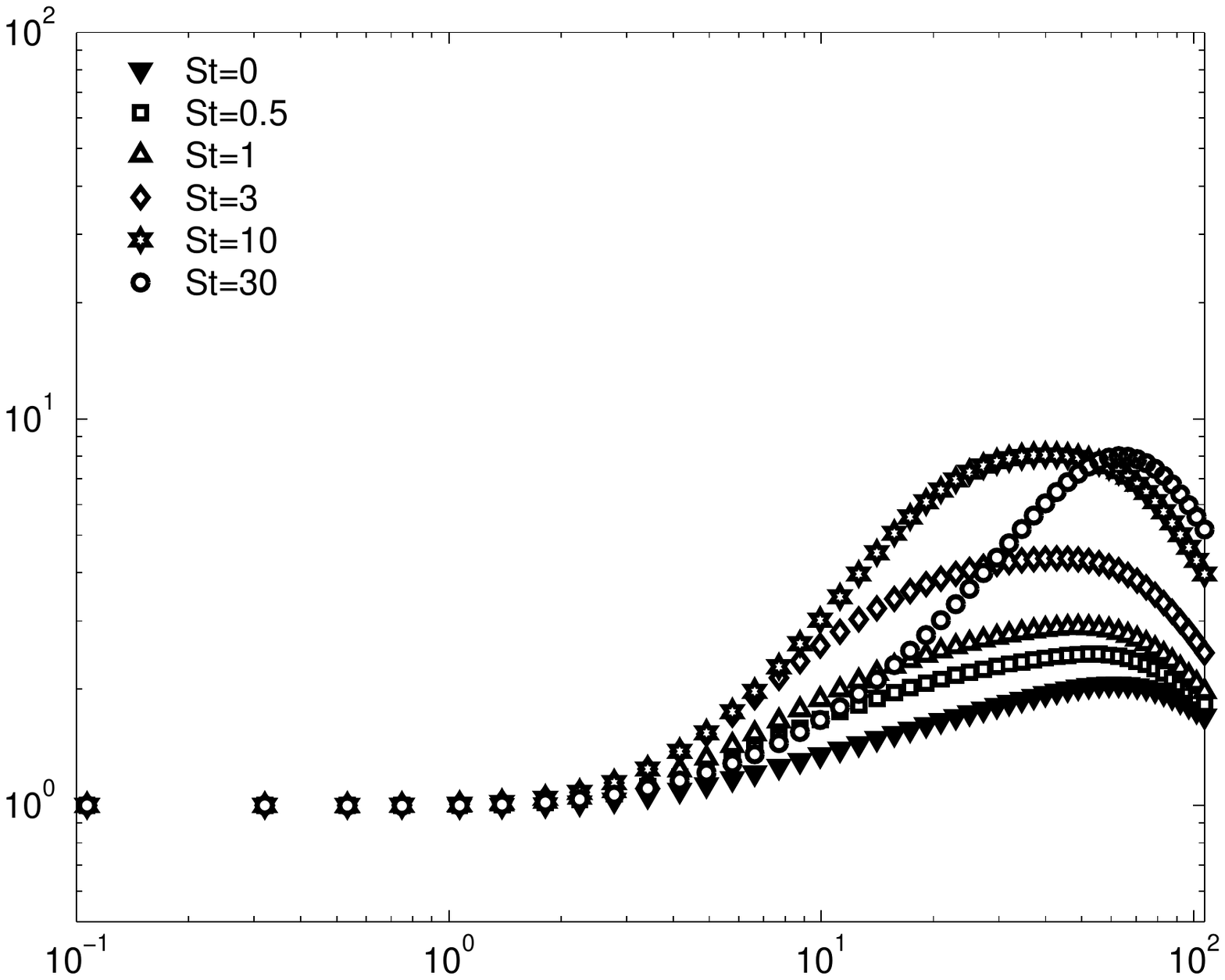}
\put(60,195){$\Big\langle\vert\bm{r}^{p}(-\mathcal{T})\vert^{2}\Big\rangle_{\bm{r}^0}\Big/\Big\langle\vert\bm{r}^{p}(\mathcal{T})\vert^{2}\Big\rangle_{\bm{r}^0}$}
\put(115,10){$\mathcal{T}/\tau_{\eta}$}
\end{overpic}}
\subfloat[]
{\begin{overpic}
[trim = 10mm 60mm 20mm 60mm,scale=0.45,clip,tics=20]{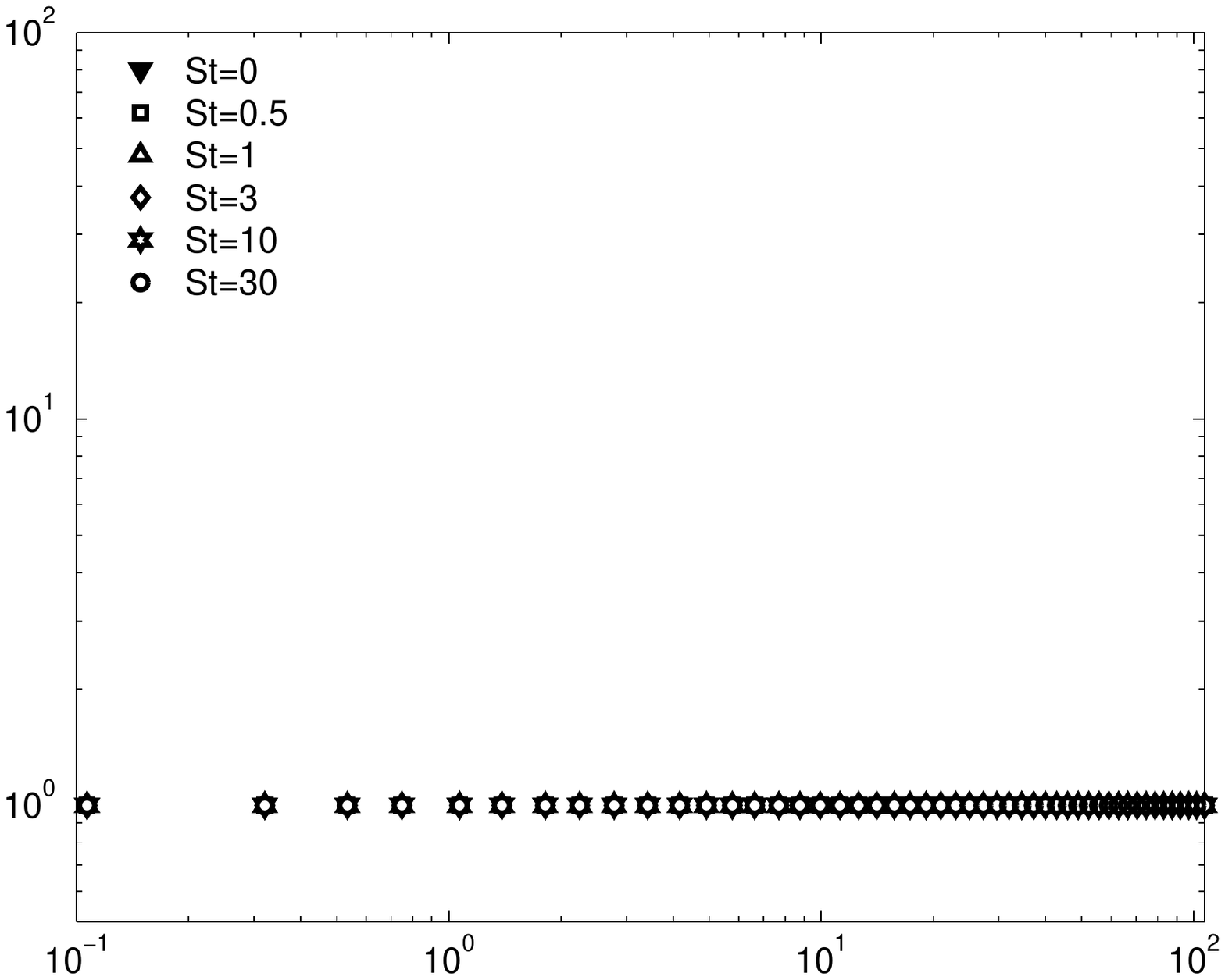}
\put(30,195){$\Big\langle\vert\bm{x}^{p}(-\mathcal{T})-\bm{x}^{p}(0)\vert^{2}\Big\rangle\Big/\Big\langle\vert\bm{x}^{p}(\mathcal{T})-\bm{x}^{p}(0)\vert^{2}\Big\rangle$}
\put(115,10){$\mathcal{T}/\tau_{\eta}$}
\end{overpic}}
\caption{Plots (a)-(c): Ratio of BIT to FIT mean-square separation from DNS for (a) $r^{0}\in[0.25\eta,0.5\eta]$, (b) $r^{0}\in[3\eta,4\eta]$, (c) $r^{0}\in[20\eta,25\eta]$ and various $St$.  Plot (d): Ratio of BIT to FIT single particle mean-square dispersion from DNS various $St$.}
\label{BIT_FIT_ratio}
\end{figure}

The BIT behavior is explained as follows:  At large separations, the particle relative velocities decrease relative to the fluid with increasing $St$, as shown in Fig.~\ref{w2_DNS}.  Consequently, for very large $\mathcal{T}$ (corresponding to pairs at large-scale separations), the rate at which particle pairs are approaching each other decreases as $St$ increases.  Note, however, that the range of $\mathcal{T}$ in our DNS data is too limited to observe this decrease.  However, as the particle pairs begin to enter the inertial range, the fluid velocity differences they experience begin to decrease (statistically), and the rate of approach for fluid particle-pairs decreases.  Since inertial particles have some memory, however, when they are in the inertial range, they will still carry a memory of their interaction with larger-scale turbulent motions in their path history.  Therefore, at some separation and some time $\mathcal{T}$, they begin to decouple from the fluid turbulence and approach each other with relative velocities greater than the fluid particle pair at the same separation.  Because BIT separating pairs retain, on average, a memory of larger-scale fluid velocity differences in their path-history, they approach each other more quickly than the corresponding fluid particle pair at both small and intermediate times, in contrast to the FIT case as discussed earlier.  Thus the particles' memory of their path-history interactions with the turbulence not only gives rise to faster BIT than FIT separation, but also gives rise to qualitative differences in their separation behavior relative to that of the corresponding fluid particle pair.
In order to show the difference between FIT and BIT dispersion more clearly,  in Figure~\ref{BIT_FIT_ratio} (a)-(c) we plot the ratio of the BIT to FIT mean-square separation for various $r^0$ and $St$.  The results clearly show that $\max[\langle\vert\bm{r}^{p}(-\mathcal{T})\vert^{2}\rangle_{\bm{r}^0}/\langle\vert\bm{r}^{p}(\mathcal{T})\vert^{2}\rangle_{\bm{r}^0}]$ is a strong function of $St$ and $r^0$, and that the irreversibility in particle dispersion is much stronger for inertial particles than for fluid particles.  
\FloatBarrier
Refer to \S\ref{IPIM} for the physical explanation.  At each $r^0$, there is an optimum value of $St$ for which the irreversibility is strongest, which corresponds to the value for which the particles experience optimally both the irreversibility associated with the turbulence dynamics and also that arising because of the history effect of their particle inertia. 

 As $r^0$ is increased, this optimum value of $St$ increases because the effects of inertia (for a given $\tau_p$) decrease with increasing separation.  As explained in \S\ref{IPIM}, for $r^0>L$ FIT and BIT dispersion become equivalent, and become related to the one-particle dispersion problem which is reversible in stationary, homogeneous turbulence.  The results in Figure~\ref{BIT_FIT_ratio} (d) for the single particle dispersion in the DNS confirm this expectation (where $\bm{x}^p(t)$ is the position of a single particle at time $t$).

\begin{figure}[ht]
\centering
\subfloat[]
{\begin{overpic}
[trim = 10mm 60mm 20mm 60mm,scale=0.45,clip,tics=20]{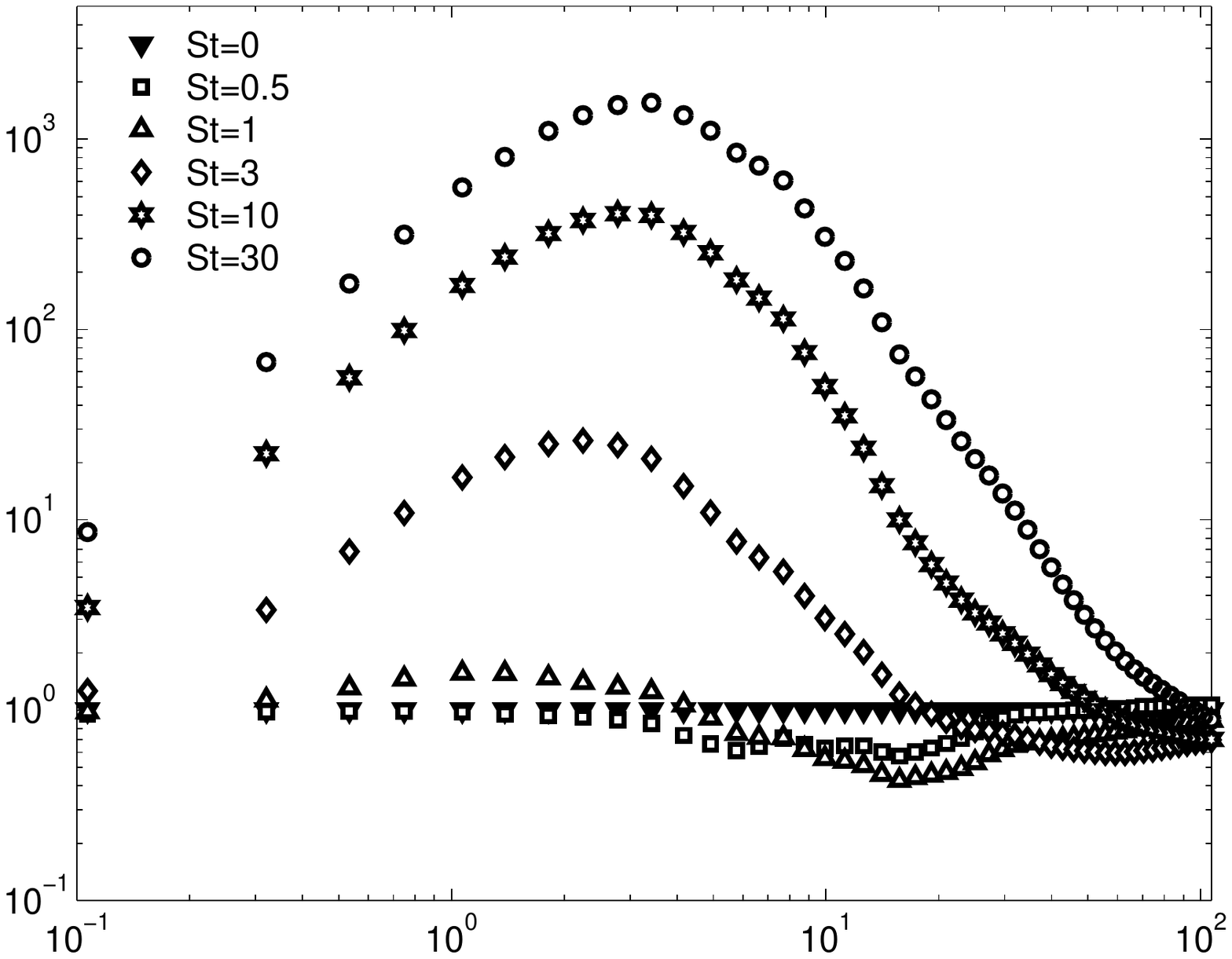}
\put(60,195){$\Big\langle\vert\bm{r}^{p}(\mathcal{T})\vert^{2}\Big\rangle_{\bm{r}^0}\Big/\Big\langle\vert\bm{r}^{f}(\mathcal{T})\vert^{2}\Big\rangle_{\bm{r}^0}$}
\put(115,10){$\mathcal{T}/\tau_{\eta}$}\end{overpic}}
\subfloat[]
{\begin{overpic}
[trim = 10mm 60mm 20mm 60mm,scale=0.45,clip,tics=20]{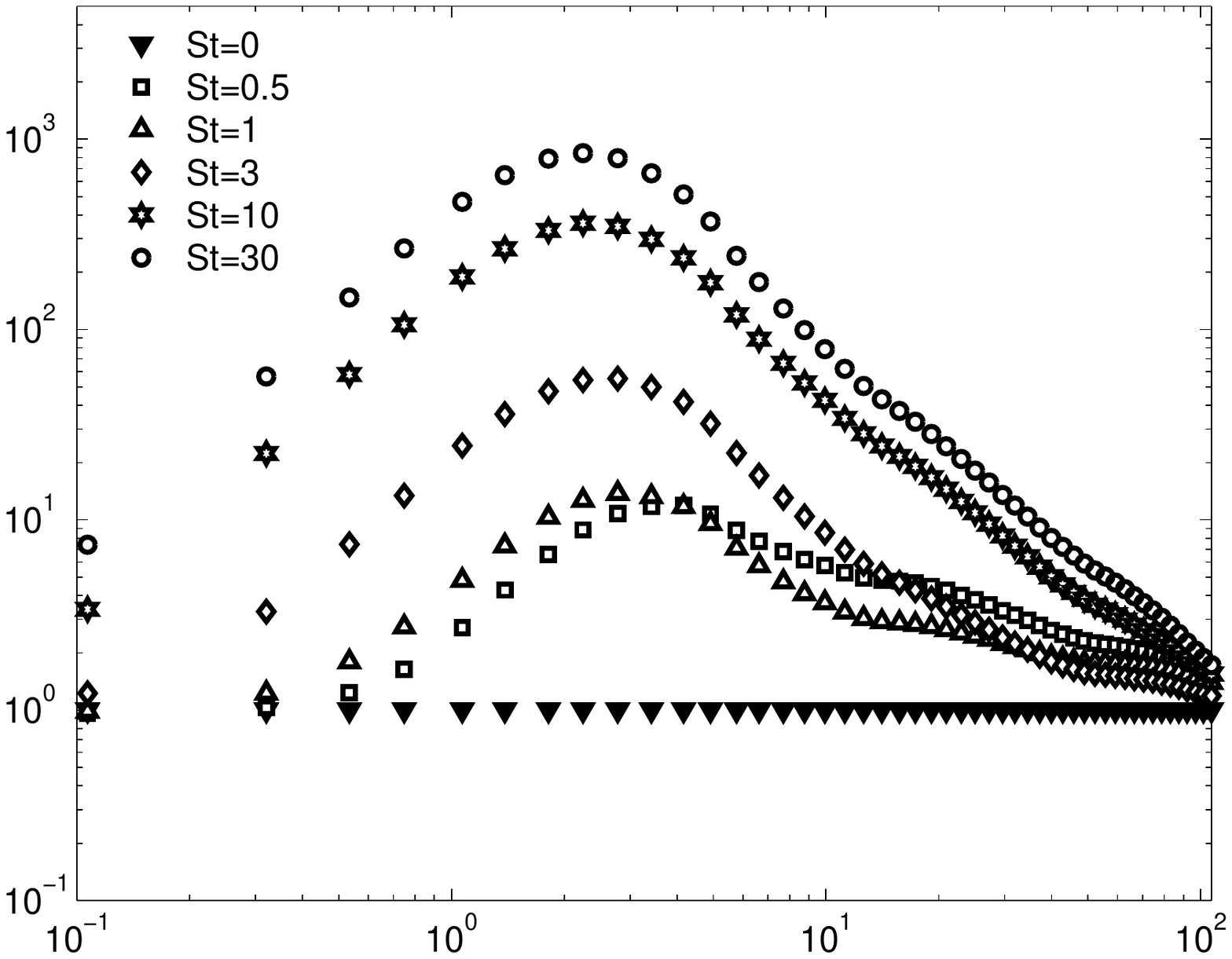}
\put(60,195){$\Big\langle\vert\bm{r}^{p}(-\mathcal{T})\vert^{2}\Big\rangle_{\bm{r}^0}\Big/\Big\langle\vert\bm{r}^{f}(-\mathcal{T})\vert^{2}\Big\rangle_{\bm{r}^0}$}
\put(115,10){$\mathcal{T}/\tau_{\eta}$}
\end{overpic}}\\
\vspace{1mm}
\subfloat[]
{\begin{overpic}
[trim = 10mm 60mm 20mm 60mm,scale=0.45,clip,tics=20]{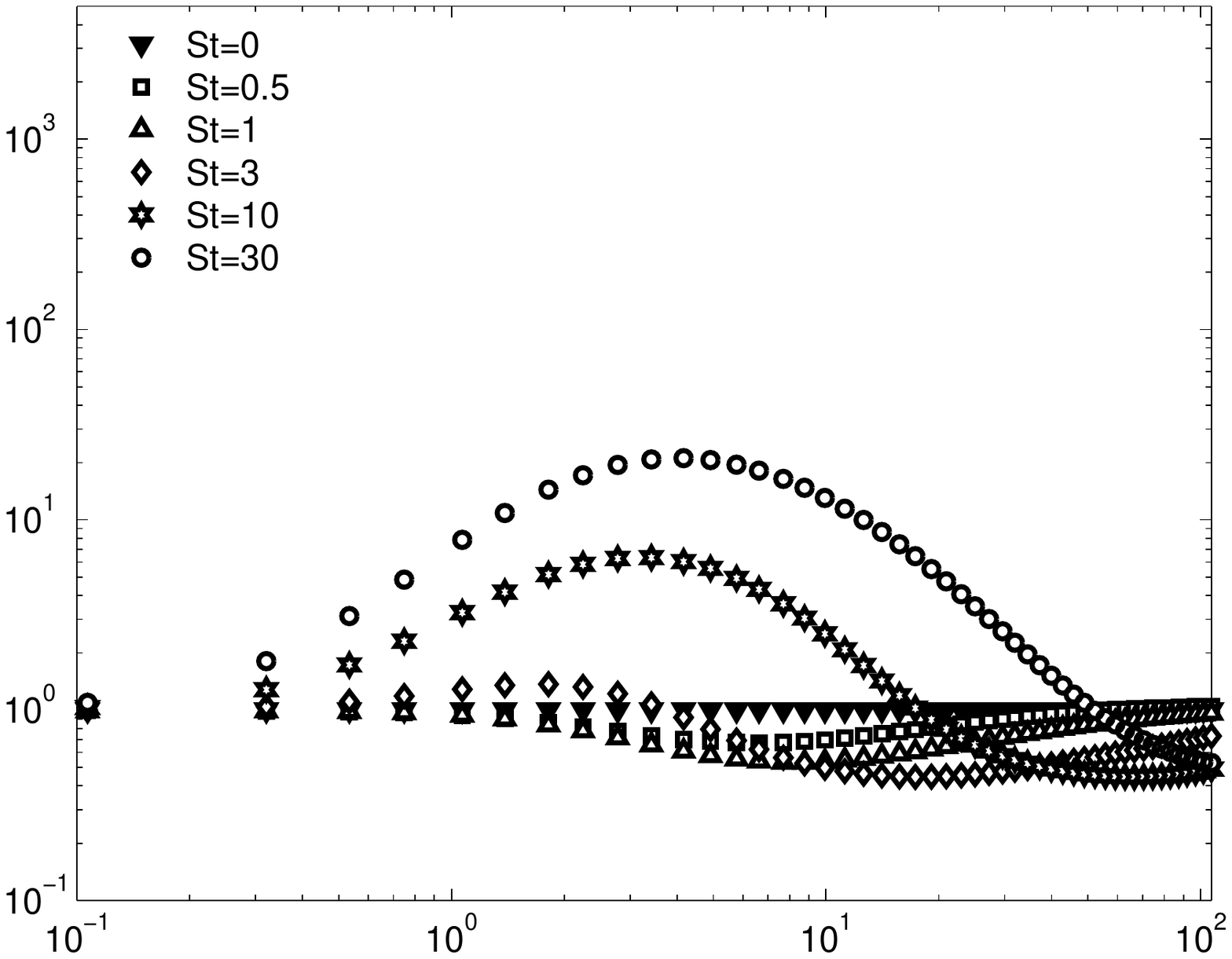}
\put(60,195){$\Big\langle\vert\bm{r}^{p}(\mathcal{T})\vert^{2}\Big\rangle_{\bm{r}^0}\Big/\Big\langle\vert\bm{r}^{f}(\mathcal{T})\vert^{2}\Big\rangle_{\bm{r}^0}$}
\put(115,10){$\mathcal{T}/\tau_{\eta}$}\end{overpic}}
\subfloat[]
{\begin{overpic}
[trim = 10mm 60mm 20mm 60mm,scale=0.45,clip,tics=20]{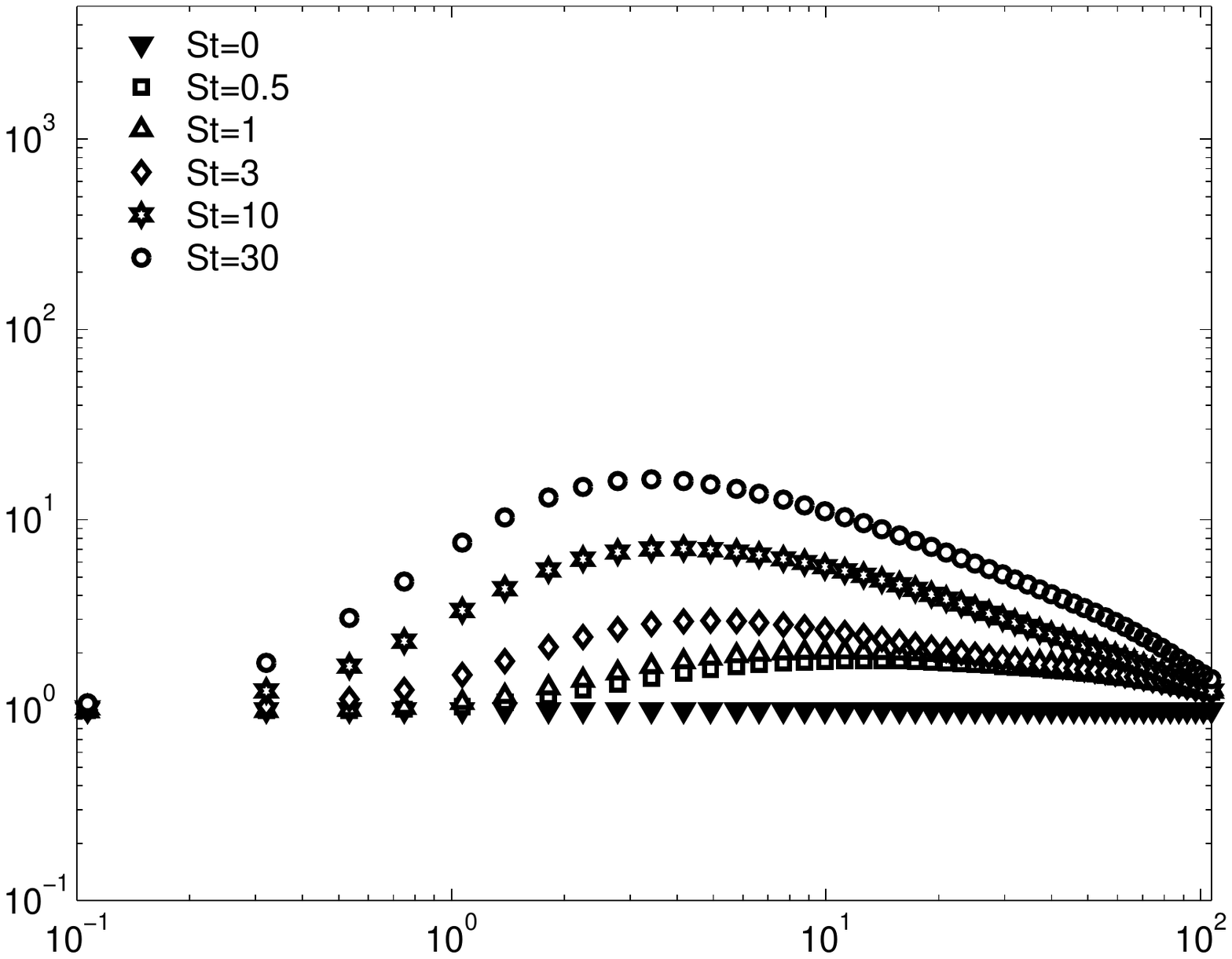}
\put(60,195){$\Big\langle\vert\bm{r}^{p}(-\mathcal{T})\vert^{2}\Big\rangle_{\bm{r}^0}\Big/\Big\langle\vert\bm{r}^{f}(-\mathcal{T})\vert^{2}\Big\rangle_{\bm{r}^0}$}
\put(115,10){$\mathcal{T}/\tau_{\eta}$}
\end{overpic}}
\caption{Ratio of inertial particle to fluid particle FIT and BIT mean-square separation from DNS for (a),(b) $r^{0}\in[0.25\eta,0.5\eta]$, (c),(d) $r^{0}\in[3\eta,4\eta]$, and various $St$.}
\label{Particle_Fluid_ratio}
\end{figure}

Having illustrated the effect of particle inertia on the dispersion irreversibility, we plot the ratio of the inertial particle to fluid particle mean-square separation in Figure~\ref{Particle_Fluid_ratio}, both FIT and BIT, in order to highlight the differences between the dispersion rates of inertial and fluid particles.  The results demonstrate the dramatic effect of inertia on the dispersion, with the inertial particle dispersion often being orders of magnitude greater than the fluid particle dispersion.  Notice also, especially in comparing Fig.~\ref{BIT_FIT_ratio} (a) with Fig.~\ref{Particle_Fluid_ratio} (a), that whereas the time of the peak value of $\max[\langle\vert\bm{r}^{p}(-\mathcal{T})\vert^{2}\rangle_{\bm{r}^0}/\langle\vert\bm{r}^{p}(\mathcal{T})\vert^{2}\rangle_{\bm{r}^0}]$ varies significantly with $St$, the time of the peak value of $\max[\langle\vert\bm{r}^{p}(-\mathcal{T})\vert^{2}\rangle_{\bm{r}^0}/\langle\vert\bm{r}^{f}(-\mathcal{T})\vert^{2}\rangle_{\bm{r}^0}]$ is approximately independent of $St$.  In the latter case, this peak time roughly corresponds to $\tau_{r^0}$.  At $\mathcal{T}<\tau_{r^0}$ the fluid particle dispersion is, relatively speaking, minimal.  The results in Fig.~\ref{Particle_Fluid_ratio} also confirm the prediction of (\ref{LTT2}) that in the large $\mathcal{T}$-regime, the inertial particles separate faster than the fluid particles BIT.  This is in contrast to the FIT case where the inertial particles lag behind the fluid particles in the large $\mathcal{T}$-regime.

\begin{figure}[ht]
\centering
\vspace{-5mm}
\subfloat[]
{\begin{overpic}
[trim = 30mm 75mm 20mm 75mm,scale=0.5,clip,tics=20]{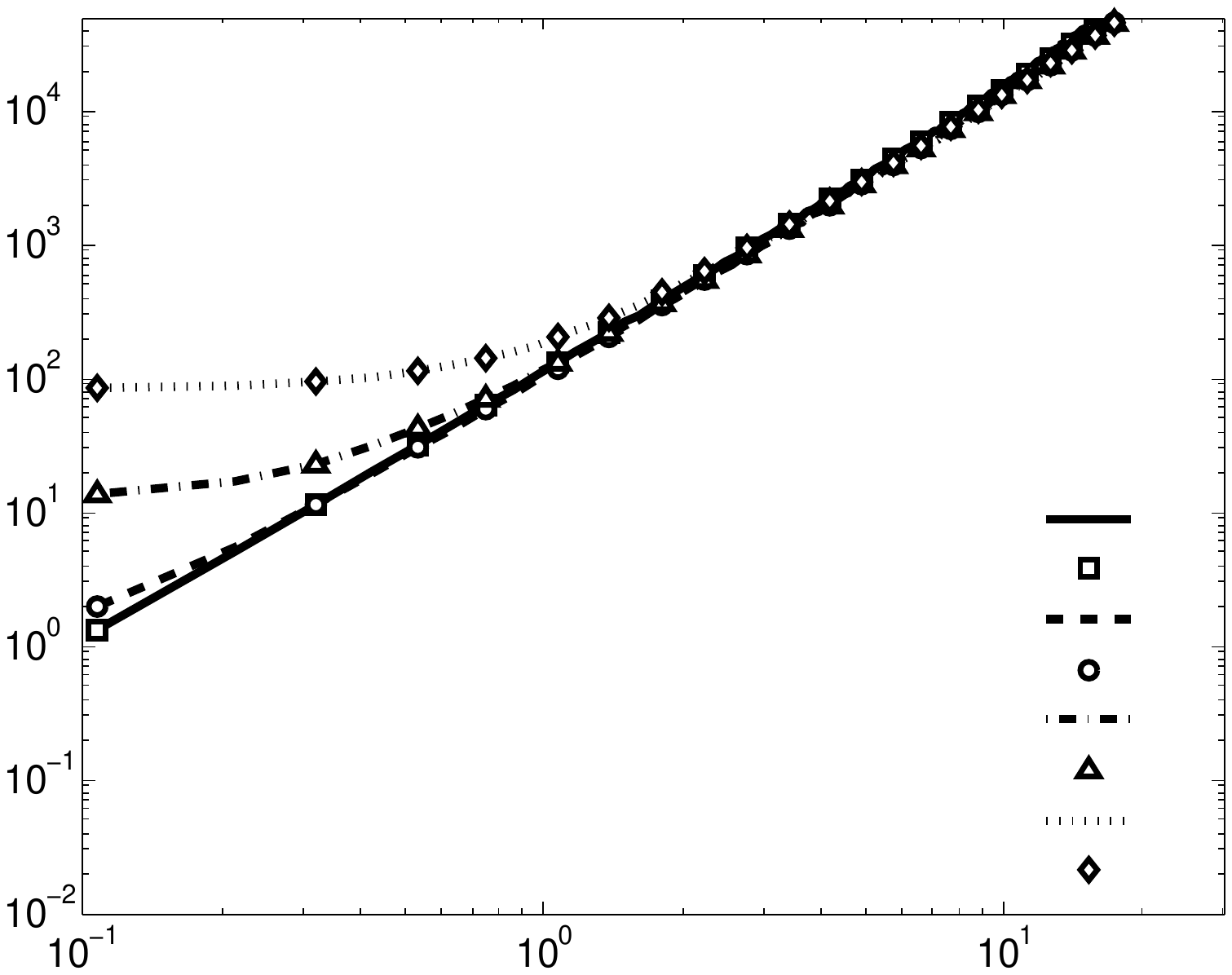}
\put(80,183){\footnotesize\text{$\Big\langle\vert\bm{r}^{p}(-\mathcal{T})\vert^{2}\Big\rangle_{\bm{r}^0}\Big/\eta^2$}}
\put(105,3){$\mathcal{T}/\tau_{\eta}$}
\put(132,83){\tiny{$r^0\in[0,0.25\eta]$}}
\put(132,75){\tiny{$r^0\in[0,0.25\eta]$}}
\put(132,66){\tiny{$r^0\in[0.75\eta,1\eta]$}}
\put(132,58){\tiny{$r^0\in[0.75\eta,1\eta]$}}
\put(132,48){\tiny{$r^0\in[3\eta,4\eta]$}}
\put(132,40){\tiny{$r^0\in[3\eta,4\eta]$}}
\put(132,30){\tiny{$r^0\in[8\eta,10\eta]$}}
\put(132,22){\tiny{$r^0\in[8\eta,10\eta]$}}
\end{overpic}}
\subfloat[]
{\begin{overpic}
[trim = 30mm 75mm 20mm 75mm,scale=0.5,clip,tics=20]{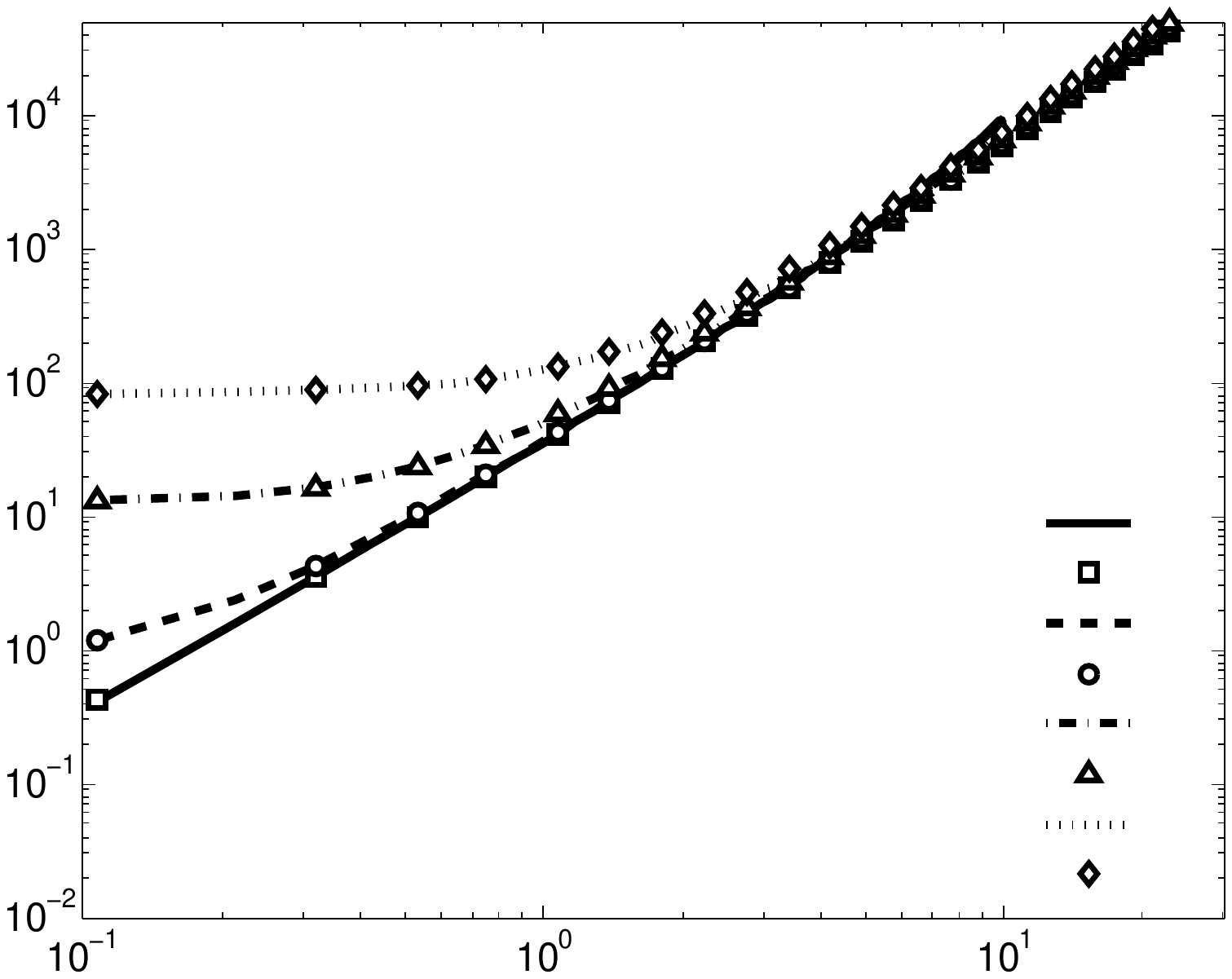}
\put(80,183){\footnotesize\text{$\Big\langle\vert\bm{r}^{p}(-\mathcal{T})\vert^{2}\Big\rangle_{\bm{r}^0}\Big/\eta^2$}}
\put(105,3){$\mathcal{T}/\tau_{\eta}$}
\put(132,83){\tiny{$r^0\in[0,0.25\eta]$}}
\put(132,75){\tiny{$r^0\in[0,0.25\eta]$}}
\put(132,66){\tiny{$r^0\in[0.75\eta,1\eta]$}}
\put(132,58){\tiny{$r^0\in[0.75\eta,1\eta]$}}
\put(132,48){\tiny{$r^0\in[3\eta,4\eta]$}}
\put(132,40){\tiny{$r^0\in[3\eta,4\eta]$}}
\put(132,30){\tiny{$r^0\in[8\eta,10\eta]$}}
\put(132,22){\tiny{$r^0\in[8\eta,10\eta]$}}
\end{overpic}}\\
\subfloat[]
{\begin{overpic}
[trim = 30mm 75mm 20mm 75mm,scale=0.5,clip,tics=20]{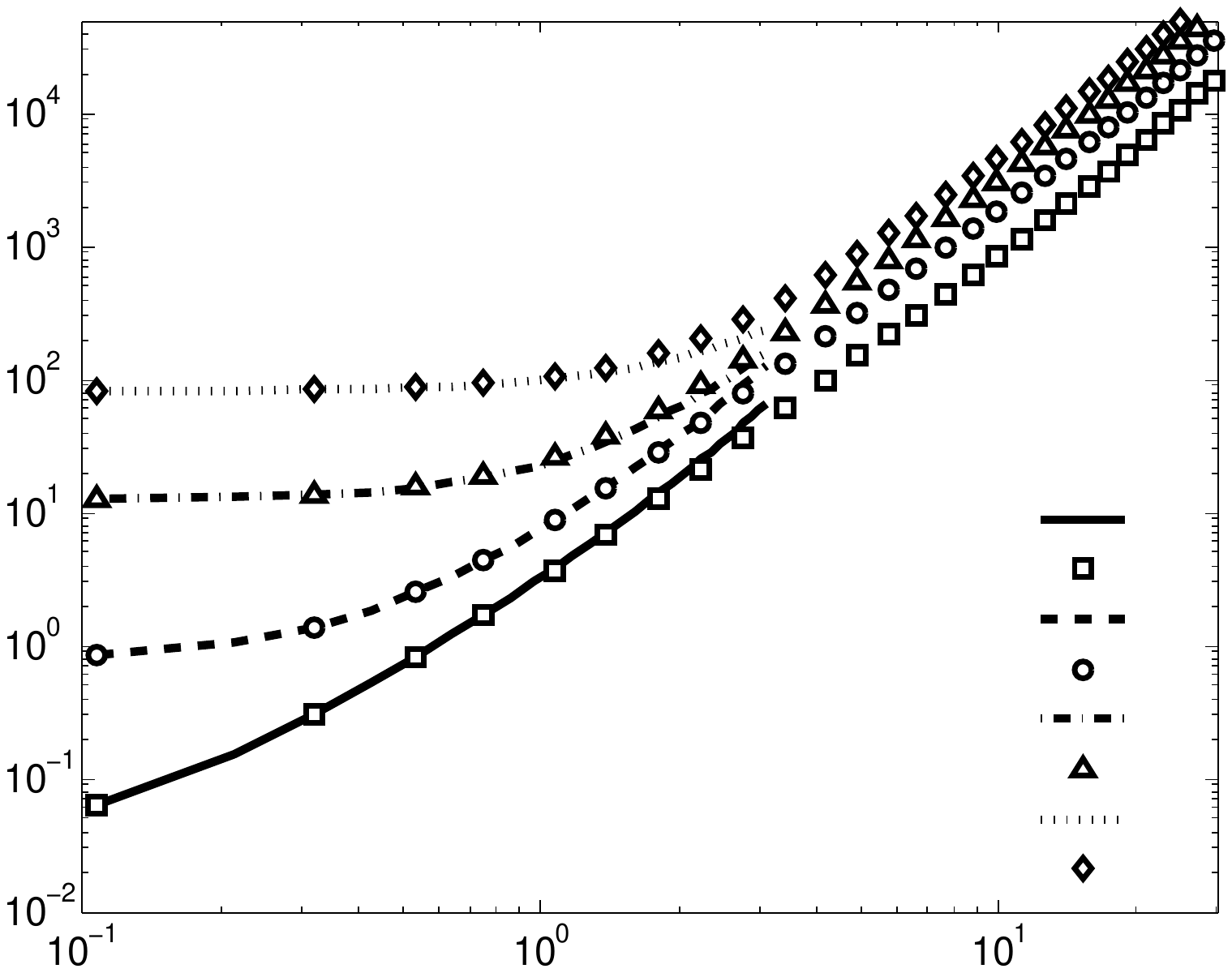}
\put(80,183){\footnotesize\text{$\Big\langle\vert\bm{r}^{p}(-\mathcal{T})\vert^{2}\Big\rangle_{\bm{r}^0}\Big/\eta^2$}}
\put(105,3){$\mathcal{T}/\tau_{\eta}$}
\put(132,83){\tiny{$r^0\in[0,0.25\eta]$}}
\put(132,75){\tiny{$r^0\in[0,0.25\eta]$}}
\put(132,66){\tiny{$r^0\in[0.75\eta,1\eta]$}}
\put(132,58){\tiny{$r^0\in[0.75\eta,1\eta]$}}
\put(132,48){\tiny{$r^0\in[3\eta,4\eta]$}}
\put(132,40){\tiny{$r^0\in[3\eta,4\eta]$}}
\put(132,30){\tiny{$r^0\in[8\eta,10\eta]$}}
\put(132,22){\tiny{$r^0\in[8\eta,10\eta]$}}
\end{overpic}}
\subfloat[]
{\begin{overpic}
[trim = 30mm 75mm 20mm 75mm,scale=0.5,clip,tics=20]{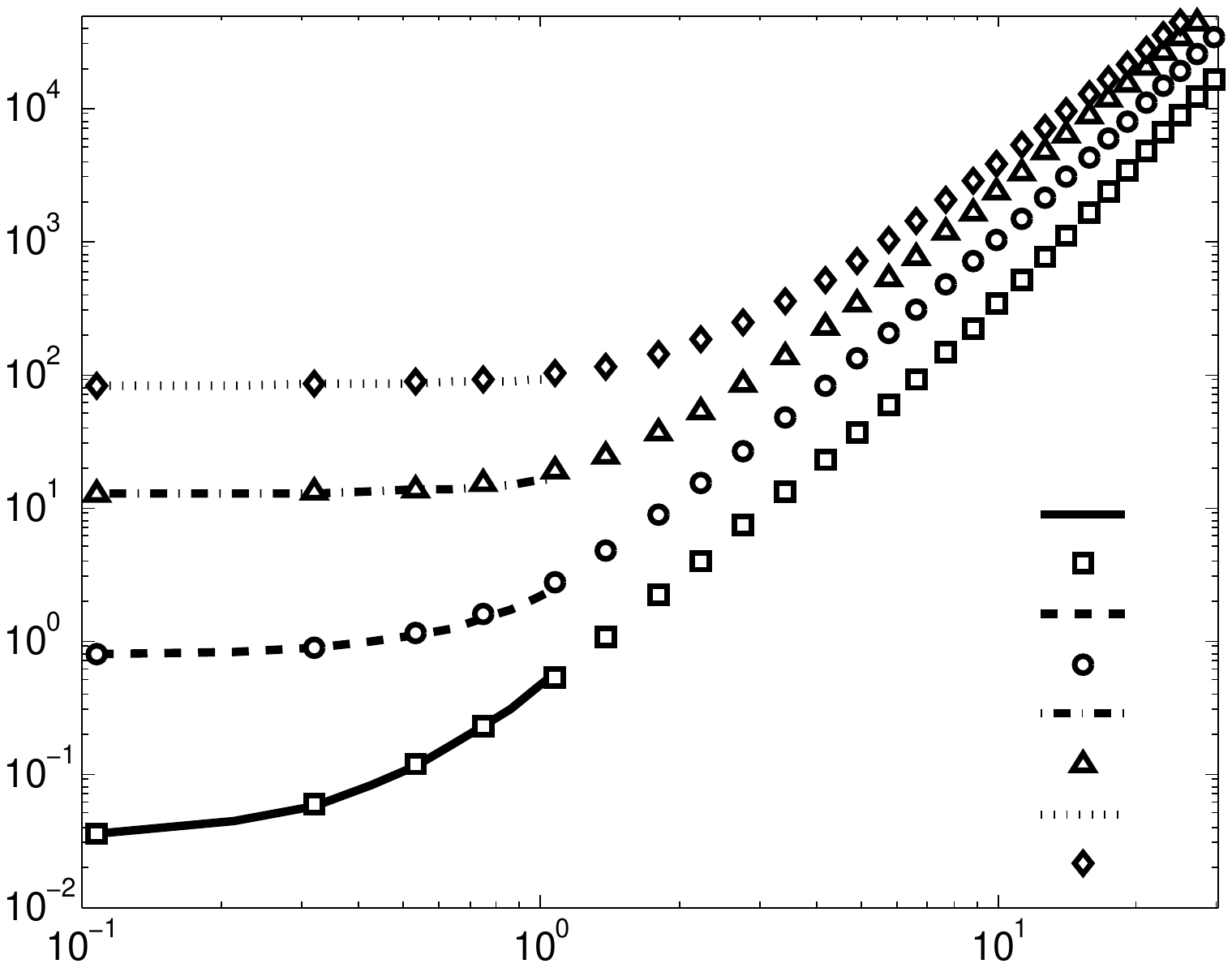}
\put(80,183){\footnotesize\text{$\Big\langle\vert\bm{r}^{p}(-\mathcal{T})\vert^{2}\Big\rangle_{\bm{r}^0}\Big/\eta^2$}}
\put(105,3){$\mathcal{T}/\tau_{\eta}$}
\put(132,83){\tiny{$r^0\in[0,0.25\eta]$}}
\put(132,75){\tiny{$r^0\in[0,0.25\eta]$}}
\put(132,66){\tiny{$r^0\in[0.75\eta,1\eta]$}}
\put(132,58){\tiny{$r^0\in[0.75\eta,1\eta]$}}
\put(132,48){\tiny{$r^0\in[3\eta,4\eta]$}}
\put(132,40){\tiny{$r^0\in[3\eta,4\eta]$}}
\put(132,30){\tiny{$r^0\in[8\eta,10\eta]$}}
\put(132,22){\tiny{$r^0\in[8\eta,10\eta]$}}
\end{overpic}}
\caption{BIT mean-square separation for (a) $St=30$, (b) $St=10$, (c) $St=3$ and (d) $St=1$ and for different initial separations $r^0$.  Lines are the predictions from Eq.(\ref{STT}) and the symbols are DNS data.}
\label{DNS_Theory}
\end{figure}

\begin{figure}[ht]
\centering
\subfloat[]
{\begin{overpic}
[trim = 25mm 75mm 20mm 75mm,scale=0.5,clip,tics=20]{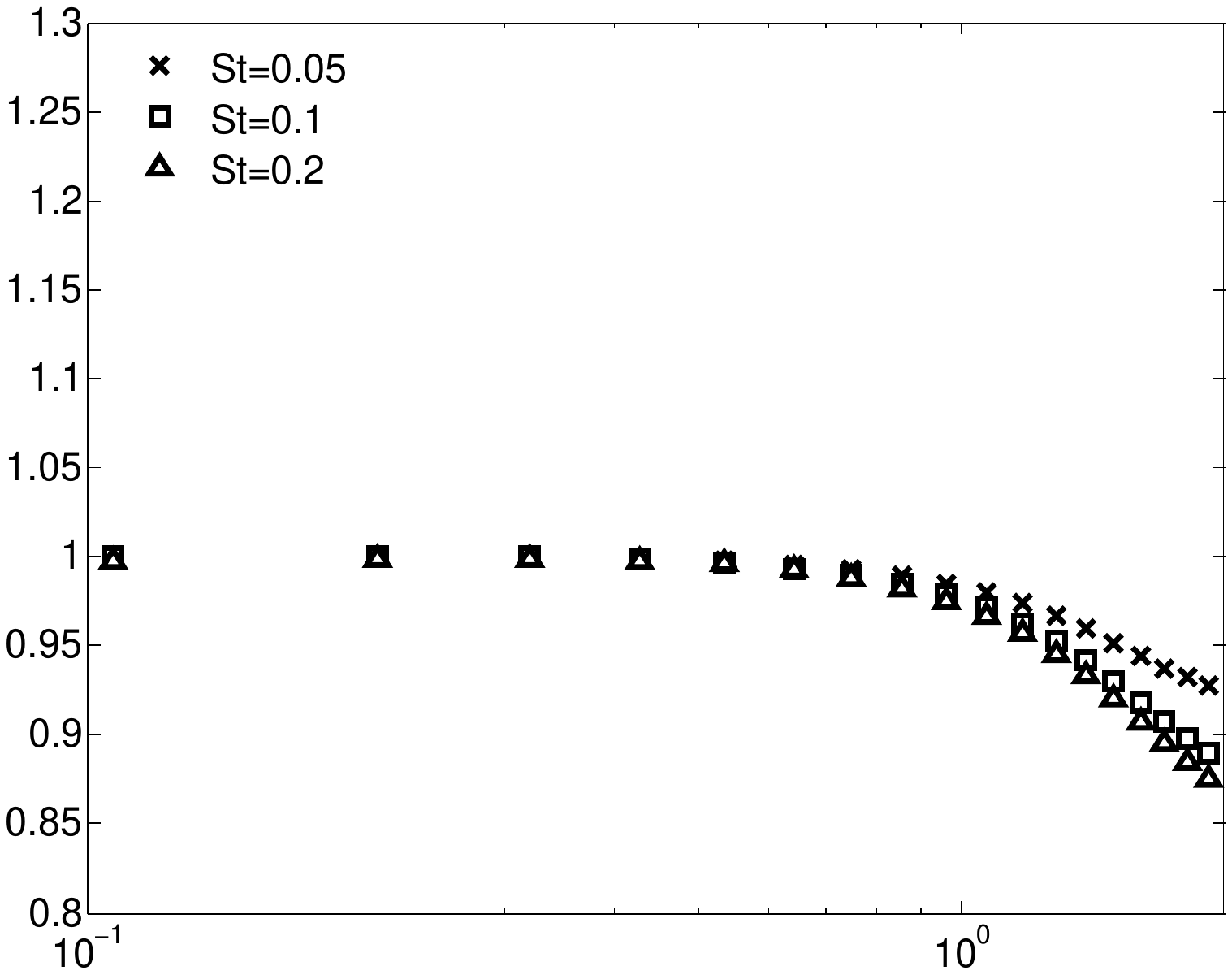}
\put(60,183){\footnotesize\text{$\Big\langle\vert\bm{r}^{p}(\mathcal{T})\vert^{2}\Big\rangle_{\bm{r}^0}\Big/\Big\langle\vert\bm{r}^{f}(\mathcal{T})\vert^{2}\Big\rangle_{\bm{r}^0}$}}
\put(105,-1){$\mathcal{T}/\tau_{\eta}$}
\end{overpic}}
\subfloat[]
{\begin{overpic}
[trim = 25mm 75mm 20mm 75mm,scale=0.5,clip,tics=20]{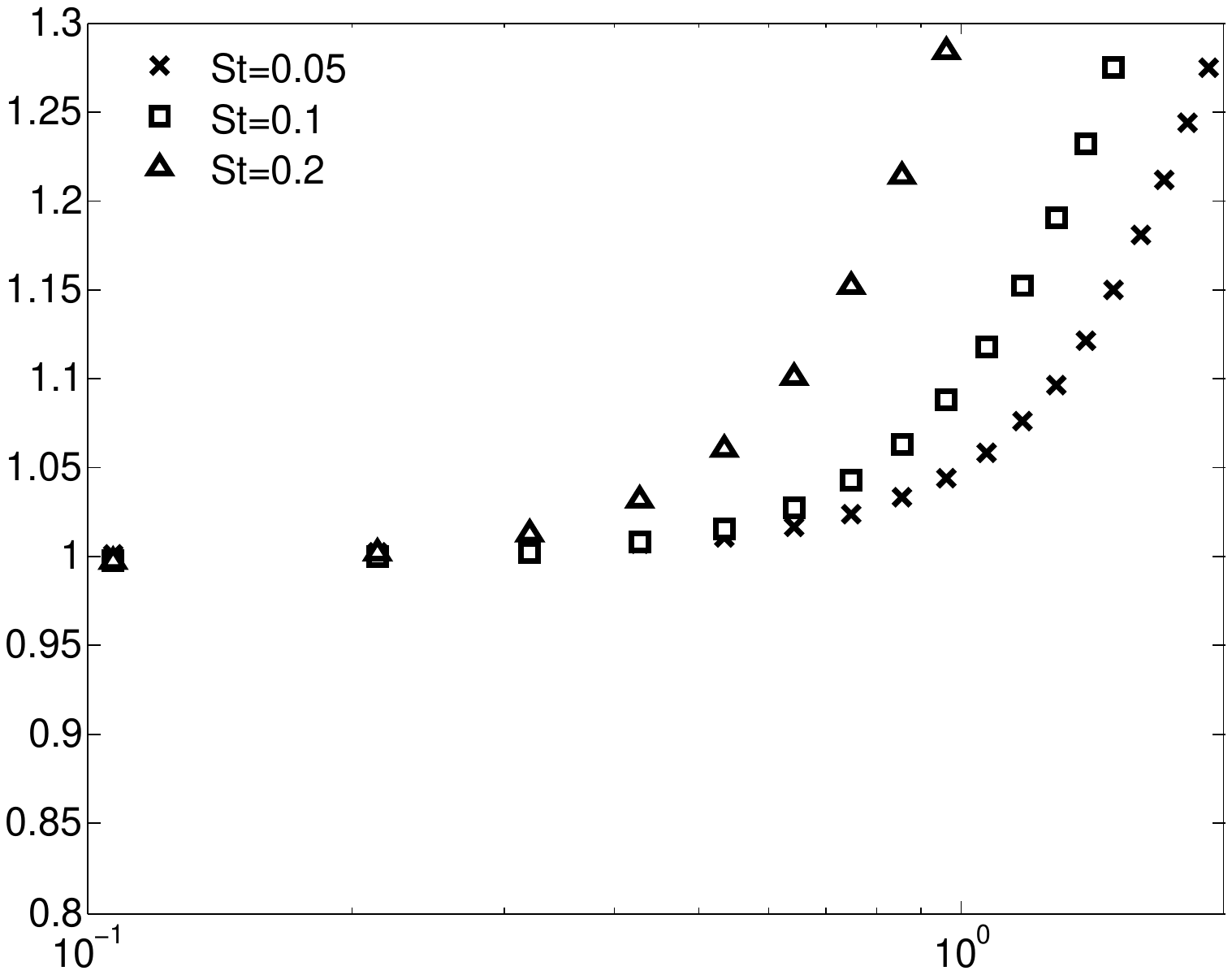}
\put(60,183){\footnotesize\text{$\Big\langle\vert\bm{r}^{p}(-\mathcal{T})\vert^{2}\Big\rangle_{\bm{r}^0}\Big/\Big\langle\vert\bm{r}^{f}(-\mathcal{T})\vert^{2}\Big\rangle_{\bm{r}^0}$}}
\end{overpic}}
\caption{Ratio of inertial to fluid particle mean-square separation (a) FIT and (b) BIT for $r^0\in[0.75\eta,\eta]$.}
\label{Ratio_Small_St}
\end{figure}

\begin{figure}[ht]
\centering
\subfloat[]
{\begin{overpic}
[trim = 25mm 75mm 20mm 75mm,scale=0.5,clip,tics=20]{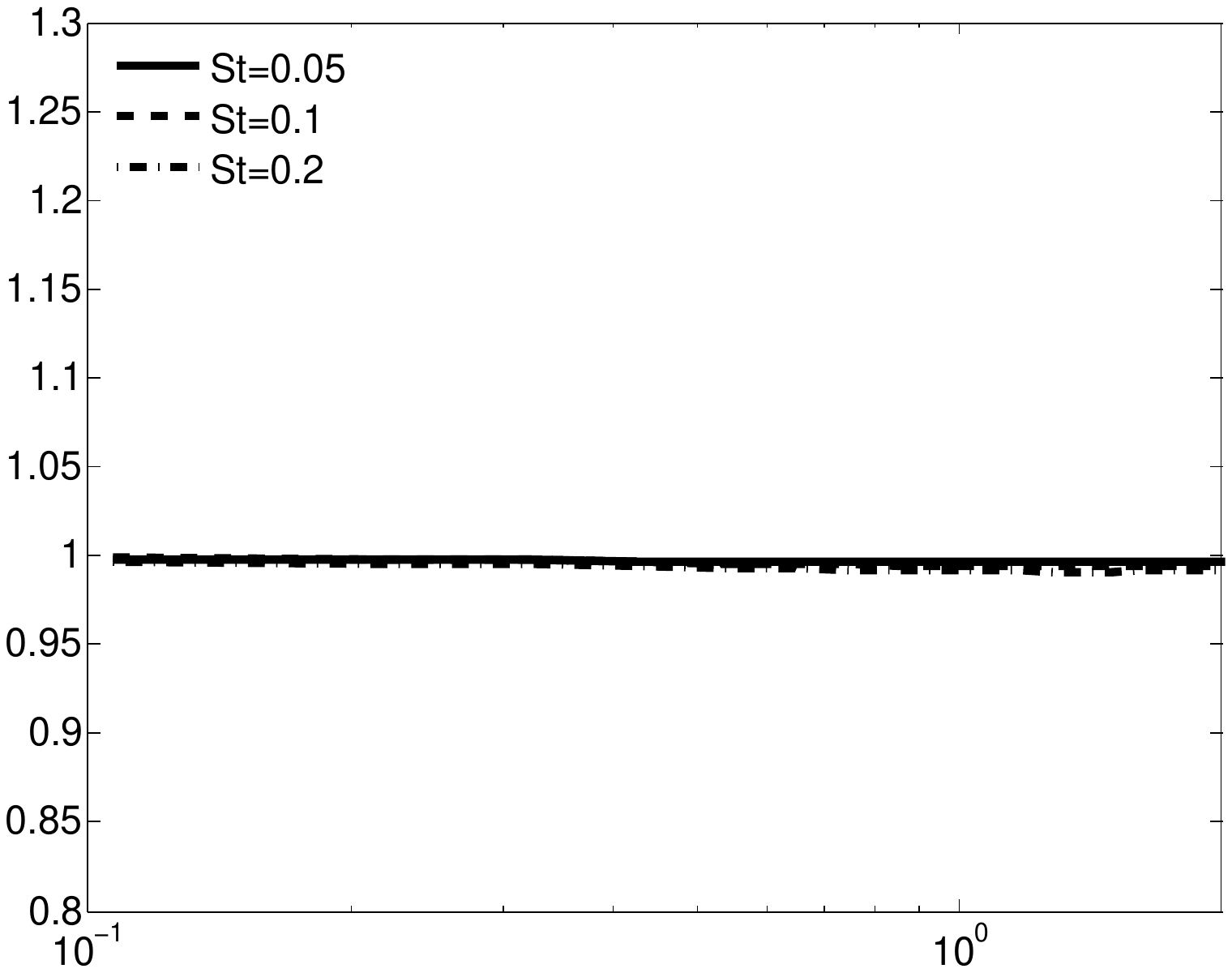}
\put(60,183){\footnotesize\text{$\Big\langle\vert\bm{r}^{p}(\mathcal{T})\vert^{2}\Big\rangle_{\bm{r}^0}\Big/\Big\langle\vert\bm{r}^{f}(\mathcal{T})\vert^{2}\Big\rangle_{\bm{r}^0}$}}
\put(105,-1){$\mathcal{T}/\tau_{\eta}$}
\end{overpic}}
\subfloat[]
{\begin{overpic}
[trim = 25mm 75mm 20mm 75mm,scale=0.5,clip,tics=20]{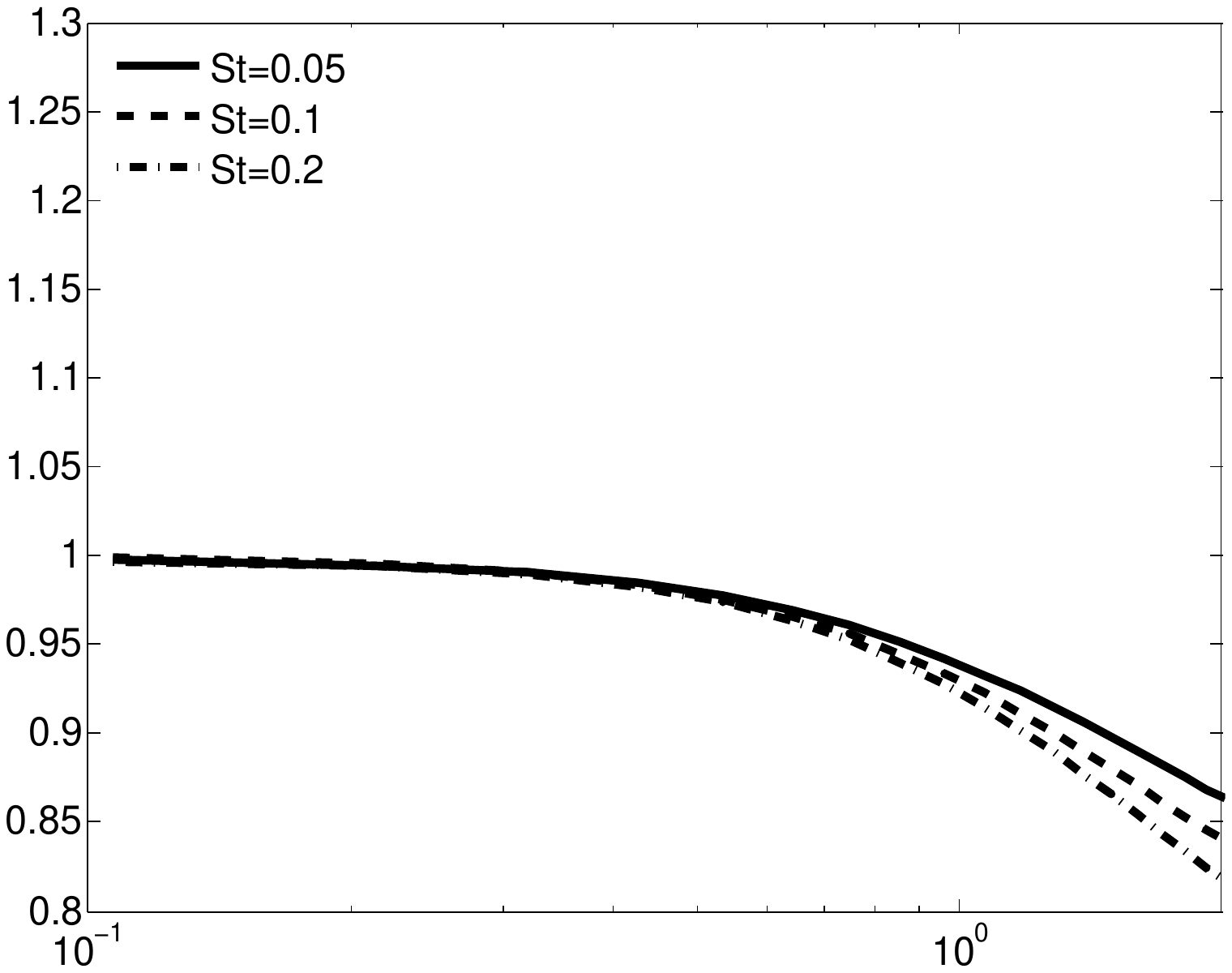}
\put(60,183){\footnotesize\text{$\Big\langle\vert\bm{r}^{p}(\mathcal{T})\vert^{2}\Big\rangle_{\bm{r}^0}\Big/\Big\langle\vert\bm{r}^{f}(\mathcal{T})\vert^{2}\Big\rangle_{\bm{r}^0}$}}
\end{overpic}}
\caption{Theoretical prediction for the ratio of inertial to fluid particle mean-square separation FIT for $r^0\in[0.75\eta,\eta]$.  The results in (a) are generated using (\ref{DispFIT_r0_2}) and those in (b) are generated using (\ref{DispFIT_r0_3}).}
\label{Theory_Ratio_Small_St}
\end{figure}

We now compare the DNS results with the theoretical prediction of (\ref{STT}) in figure~\ref{DNS_Theory}.  The DNS data for $\langle\vert\bm{w}^{p}(0)\vert^{2}\rangle_{\bm{r}^0}$ and $\langle\vert\Delta\bm{u}^{f}(0)\vert^{2}\rangle_{\bm{r}^0}$ was used in (\ref{STT}).  The results in Fig.~\ref{DNS_Theory} show that (\ref{STT}) describes the DNS very accurately for $\mathcal{T}\leq\widehat{\mathcal{T}}$, both qualitatively and quantitatively, capturing the effects of $St$ and $r^0$ on the separation behavior.  It is interesting to note that both the DNS results and the predictions of (\ref{STT}) show that the value of $\mathcal{T}$ for which $\langle\vert\bm{r}^{p}(-\mathcal{T})\vert^{2}\rangle_{\bm{r}^0}$ becomes approximately independent of $r^0$ decreases as $St$ increases.  This might seem counter-intuitive, since one might expect that the smaller $St$ is, the faster the particles forget their initial conditions.  The reason for the observed behavior, however, is due to the fact that as $St$ increases, $\langle\vert\bm{w}^{p}(0)\vert^{2}\rangle_{\bm{r}^0}$ becomes approximately independent of $\bm{r}^0$ in the dissipation range (see fig.~\ref{w2_DNS}).  For smaller $St$, $\langle\vert\bm{w}^{p}(0)\vert^{2}\rangle_{\bm{r}^0}$ remains dependent upon $\bm{r}^0$, and so particles with smaller $\bm{r}^0$ begin with both smaller separations \emph{and} smaller relative velocities, and hence the $\bm{r}^0$-dependence of $\langle\vert\bm{r}^{p}(-\mathcal{T})\vert^{2}\rangle_{\bm{r}^0}$ is apparent for longer times. 
\FloatBarrier
The results in Fig.~\ref{DNS_Theory} are for $St\geq\mathcal{O}(1)$ particles; in the regime ${St\ll1}$, where the particle motion tends to be dominated by their local interaction with $\Delta\bm{u}$, one might expect that the preferential sampling of the turbulence by the particles could influence their separation behavior.  In particular, since particles tend to avoid regions of the turbulence with high vorticity, and accumulate in the straining regions of the flow, one might expect that inertial particles in the regime ${St\ll1}$ might separate slower than fluid particle since the preferential sampling leads to $\langle\vert\Delta\bm{u}^p(0)\vert^2\rangle_{\bm{r}^0}<\langle\vert\Delta\bm{u}^f(0)\vert^2\rangle_{\bm{r}^0}$.  The result in (\ref{STT}) does not fully account for this effect since it uses the approximation $\Delta\bm{u}^p\approx\Delta\bm{u}^f$. We therefore now modify this relation to account for preferential sampling.  In the regime ${St\ll1}$, we have
\begin{align}
\Big\langle\vert\bm{w}^p(0)\vert^2\Big\rangle_{\bm{r}^0}=\Big\langle\vert\Delta\bm{u}^p(0)\vert^2\Big\rangle_{\bm{r}^0}-2St\Big\langle\Delta\bm{u}^p(0)\bm{\cdot}\Delta\bm{a}^p(0)\Big\rangle_{\bm{r}^0}+\mathcal{O}(St^2),\label{wpertsol}
\end{align}
where $\Delta\bm{a}^p$ is the fluid relative acceleration vector evaluated at the particle-pair separation.  Making the replacement ${\Delta\bm{u}^f\to\Delta\bm{u}^p}$ in (\ref{STT}), and introducing into the resulting expression (\ref{wpertsol}), we obtain the result that accounts for the effects of preferential sampling on $\langle\vert\bm{r}^p(-\mathcal{T})\vert^2\rangle_{\bm{r}^0}$ in the regime ${St\ll1}$
\begin{align}
\begin{split}
\Big\langle\vert\bm{r}^p(-\mathcal{T})\vert^2\Big\rangle_{\bm{r}^0}&=\vert\bm{r}^0\vert^2+\mathcal{C}\mathcal{T}^2+2\mathcal{T}G(-\mathcal{T})\Big(\mathcal{C}-\sqrt{\mathcal{C}\mathcal{D}}\Big)+G^2(-\mathcal{T})\Big(\mathcal{C}+\mathcal{D}-2\sqrt{\mathcal{C}\mathcal{D}}\Big)+\mathcal{O}(St^2),
\label{BIT_pert}
\end{split}
\end{align}
where
\begin{align}
\mathcal{C}&\equiv\Big\langle\vert\Delta\bm{u}^p(0)\vert^2\Big\rangle_{\bm{r}^0},\\
\mathcal{D}&\equiv\mathcal{C}-2 St\Big\langle\Delta\bm{u}^p(0)\bm{\cdot}\Delta\bm{a}^p(0)\Big\rangle_{\bm{r}^0}.
\end{align}
Assuming that ${\langle\Delta\bm{u}^p(0)\bm{\cdot}\Delta\bm{a}^p(0)\rangle_{\bm{r}^0}<0}$ in the dissipation range, as it is in the limit ${St\to0}$ and in the inertial range \cite{pumir01}, then $\mathcal{D}\geq\mathcal{C}$ and each term in (\ref{BIT_pert}) is $\geq0$.  In view of the fluid ballistic behavior, it can be seen that whether or not $\langle\vert\bm{r}^{p}(-\mathcal{T})\vert^{2}\rangle_{\bm{r}^0}$ is less than $\langle\vert\bm{r}^{f}(-\mathcal{T})\vert^{2}\rangle_{\bm{r}^0}$ will depend upon the competition between the various terms in (\ref{BIT_pert}), and this competition depends upon both $St$ and $\mathcal{T}$.  If $St\ll1$ and $\mathcal{T}\ll\tau_p$, then the leading contribution to (\ref{BIT_pert}) is ${\langle\vert\bm{r}^p(-\mathcal{T})\vert^2\rangle_{\bm{r}^0}=\vert\bm{r}^0\vert^2+\mathcal{C}\mathcal{T}^2}$.  In this case, since preferential sampling gives rise to $\langle\vert\Delta\bm{u}^p(0)\vert^2\rangle_{\bm{r}^0}<\langle\vert\Delta\bm{u}^f(0)\vert^2\rangle_{\bm{r}^0}$ then $\langle\vert\bm{r}^{p}(-\mathcal{T})\vert^{2}\rangle_{\bm{r}^0}<\langle\vert\bm{r}^{f}(-\mathcal{T})\vert^{2}\rangle_{\bm{r}^0}$.  However, it is likely that $\vert\bm{r}^0\vert^2\gg\mathcal{C}\mathcal{T}^2$ in this regime, and consequently differences between $\langle\vert\bm{r}^{p}(-\mathcal{T})\vert^{2}\rangle_{\bm{r}^0}$ and $\langle\vert\bm{r}^{f}(-\mathcal{T})\vert^{2}\rangle_{\bm{r}^0}$ may be negligible.  If ${St\ll1}$ and ${\mathcal{T}=\mathcal{O}(\tau_p)}$, then the third and fourth terms will dominate over the second term since $\vert G(-\mathcal{T})\vert$ grows exponentially fast in this regime. 
\FloatBarrier
 We then have $\langle\vert\bm{r}^{p}(-\mathcal{T})\vert^{2}\rangle_{\bm{r}^0}>\langle\vert\bm{r}^{f}(-\mathcal{T})\vert^{2}\rangle_{\bm{r}^0}$.  Consequently, the preferential sampling effect may be irrelevant for the BIT dispersion, since the explosive growth BIT associated with the time-irreversibility effect of the particle inertia may overwhelm the local preferential sampling effect.  However, the FIT dispersion for $St\ll1$ contains $G(\mathcal{T})$ which grows at a rate $\leq \mathcal{T}$, i.e. not explosive like $G(-\mathcal{T})$.  In this case the preferential sampling effect may be sufficient to cause $\langle\vert\bm{r}^{p}(\mathcal{T})\vert^{2}\rangle_{\bm{r}^0}$ to be less than $\langle\vert\bm{r}^{f}(\mathcal{T})\vert^{2}\rangle_{\bm{r}^0}$.  In Figure~\ref{Ratio_Small_St} we show the DNS results for $\langle\vert\bm{r}^{p}(\mathcal{T})\vert^{2}\rangle_{\bm{r}^0}/\langle\vert\bm{r}^{f}(\mathcal{T})\vert^{2}\rangle_{\bm{r}^0}$ and $\langle\vert\bm{r}^{p}(-\mathcal{T})\vert^{2}\rangle_{\bm{r}^0}/\langle\vert\bm{r}^{f}(-\mathcal{T})\vert^{2}\rangle_{\bm{r}^0}$.  When viewed at this scale, we can see that indeed the preferential sampling effect does cause ${St\ll1}$ particles to separate slower than fluid particles FIT.  Conversely, in the BIT case, the inertial particles always separate faster than the fluid particles, for the reasons just explained.  In order to check that the FIT result is in fact the result of preferential sampling, in figure~\ref{Theory_Ratio_Small_St}(a) we plot the small-time FIT theoretical prediction for $\langle\vert\bm{r}^{p}(\mathcal{T})\vert^{2}\rangle_{\bm{r}^0}/\langle\vert\bm{r}^{f}(\mathcal{T})\vert^{2}\rangle_{\bm{r}^0}$, 
\begin{align}
\begin{split}
\frac{\langle\vert\bm{r}^{p}(\mathcal{T})\vert^{2}\rangle_{\bm{r}^0}}{\langle\vert\bm{r}^{f}(\mathcal{T})\vert^{2}\rangle_{\bm{r}^0}}\approx\Bigg(\vert\bm{r}^0\vert^{2}&+G^{2}(\mathcal{T})\Big\langle\vert\bm{w}^{p}(0)\vert^{2}\Big\rangle_{\bm{r}^0}+2G(\mathcal{T})\Big[\mathcal{T}-G(\mathcal{T})\Big]\sqrt{\Big\langle\vert\bm{w}^{p}(0)\vert^{2}\Big\rangle_{\bm{r}^0}}\sqrt{\Big\langle\vert\Delta\bm{u}^{f}(0)\vert^{2}\Big\rangle_{\bm{r}^0}}\\
&+\Big[\mathcal{T}^{2}-2\mathcal{T}G(\mathcal{T})+G^{2}(\mathcal{T})\Big]\Big\langle\vert\Delta\bm{u}^{f}(0)\vert^{2}\Big\rangle_{\bm{r}^0}\Bigg)\Bigg/\Bigg(\vert\bm{r}^0\vert^{2}+\Big\langle\vert\Delta\bm{u}^{f}(0)\vert^{2}\Big\rangle_{\bm{r}^0}\mathcal{T}^2\Bigg),
\label{DispFIT_r0_2}
\end{split}
\end{align}
and in figure~\ref{Theory_Ratio_Small_St}(b) we plot
\begin{align}
\begin{split}
\frac{\langle\vert\bm{r}^{p}(\mathcal{T})\vert^{2}\rangle_{\bm{r}^0}}{\langle\vert\bm{r}^{f}(\mathcal{T})\vert^{2}\rangle_{\bm{r}^0}}\approx\Bigg(\vert\bm{r}^0\vert^{2}&+G^{2}(\mathcal{T})\Big\langle\vert\bm{w}^{p}(0)\vert^{2}\Big\rangle_{\bm{r}^0}+2G(\mathcal{T})\Big[\mathcal{T}-G(\mathcal{T})\Big]\sqrt{\Big\langle\vert\bm{w}^{p}(0)\vert^{2}\Big\rangle_{\bm{r}^0}}\sqrt{\Big\langle\vert\Delta\bm{u}^{p}(0)\vert^{2}\Big\rangle_{\bm{r}^0}}\\
&+\Big[\mathcal{T}^{2}-2\mathcal{T}G(\mathcal{T})+G^{2}(\mathcal{T})\Big]\Big\langle\vert\Delta\bm{u}^{p}(0)\vert^{2}\Big\rangle_{\bm{r}^0}\Bigg)\Bigg/\Bigg(\vert\bm{r}^0\vert^{2}+\Big\langle\vert\Delta\bm{u}^{f}(0)\vert^{2}\Big\rangle_{\bm{r}^0}\mathcal{T}^2\Bigg).
\label{DispFIT_r0_3}
\end{split}
\end{align}
The only difference between (\ref{DispFIT_r0_2}) and (\ref{DispFIT_r0_3}) is that (\ref{DispFIT_r0_3}), unlike (\ref{DispFIT_r0_2}), accounts for $\Delta\bm{u}^p\neq\Delta\bm{u}^f$, i.e. it accounts for preferential sampling effects (and we use DNS data to specify $\langle\vert\Delta\bm{u}^{p}(0)\vert^{2}\rangle_{\bm{r}^0}$).  Comparing figure~\ref{Theory_Ratio_Small_St} with figure~\ref{Ratio_Small_St} (a) reveals that $\langle\vert\bm{r}^{p}(\mathcal{T})\vert^{2}\rangle_{\bm{r}^0}/\langle\vert\bm{r}^{f}(\mathcal{T})\vert^{2}\rangle_{\bm{r}^0}<1$ in the small-time regime is in fact caused by preferential sampling effects.  

In an experimental study of the FIT dispersion of inertial particles \cite{gibert10}, particles with $St\leq0.5$ were considered.  However contrary to our results, they did not observe a reduction in the inertial particle dispersion due to preferential sampling, but instead observed that the inertial particles with ${0 <St \leq 0.5}$ separated faster than the fluid particles at small-times.  There are several possible explanations for this apparent disagreement.  First, their results are for $r^0$ in the inertial range, where the preferential sampling effect and clustering is weaker than in the dissipation range \cite{bragg14e}, and hence the effect of preferential sampling on the dispersion may be harder to observe than in the dissipation range.  Second, gravitational settling in their experiments is known to reduce the preferential sampling effect by reducing the interaction time between the particles and the fluid velocity field.  Third, their system consisted of heavy particles in water, with a particle-to-fluid density ratio $\leq\mathcal{O}(10)$, as compared to the $\mathcal{O}(1000)$ density ratio we consider.  At this much lower density ratio, the particle dynamics in the experiment are no longer governed only by drag forces, but involve additional forces, such as added mass and the Basset history force, which were neglected in the DNS.  The additional forces acting on the particles in the experiment may counteract the preferential sampling effect, possibly explaining why they did not observe the inertial particle dispersion for small $St$ to be slower than that of the fluid particles.

\begin{figure}[ht]
\centering
{\begin{overpic}
[trim = 30mm 75mm 20mm 75mm,scale=0.5,clip,tics=20]{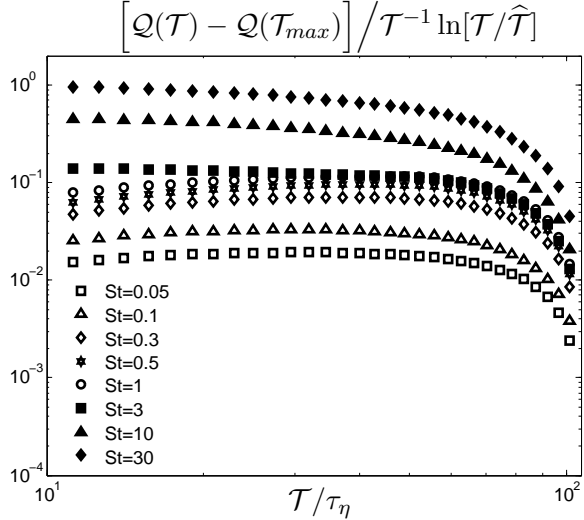}
\put(40,183){{$\Big[\mathcal{Q}(\mathcal{T})-\mathcal{Q}(\mathcal{T}_{max})\Big]\Big/\mathcal{T}^{-1}\ln[\mathcal{T}/\widehat{\mathcal{T}}]$}}
\put(105,3){$\mathcal{T}/\tau_{\eta}$}
\end{overpic}}
\caption{DNS data for $\mathcal{Q}(\mathcal{T})-\mathcal{Q}(\mathcal{T}_{max})$ at $r^0\in[3\eta,4\eta]$, where $\mathcal{T}_{max}=107\tau_\eta$.}
\label{Particle_RO_scaling}
\end{figure}

We now turn to consider the long-time behavior of the particle separation.  In figure~\ref{Particle_RO_scaling} we test the scaling prediction of (\ref{LTT2}), which predicts that the inertial particle separation approaches that of fluid particles at a rate ${\propto\mathcal{T}^{-1}\ln[\mathcal{T}/\widehat{\mathcal{T}}]}$.  Note that because of the limited $Re_\lambda$ of the DNS, our data for $\mathcal{Q}(\mathcal{T})\equiv\langle\vert\bm{r}^{p}(-\mathcal{T})\vert^{2}\rangle_{\bm{r}^0}/\langle\vert\bm{r}^{f}(-\mathcal{T})\vert^{2}\rangle_{\bm{r}^0}$ does not reach unity in the long-time limit and we therefore subtract $\mathcal{Q}(\mathcal{T}_{max})$, rather than $1$, from $\mathcal{Q}(\mathcal{T})$ when testing the prediction in Figure~\ref{Particle_RO_scaling} (something similar was also done in Fig.9 of \cite{bec10b}). 
\FloatBarrier
\vspace{2mm}
\FloatBarrier
The results confirm the scaling prediction in (\ref{LTT2}) quite well for $St\leq\mathcal{O}(1)$ over a range of $\mathcal{T}$.  The significant deviations from the predicted scaling at $\mathcal{T}=\mathcal{O}(100\tau_\eta)$ are the result of the influence of the large scales where the separation becomes diffusive in time, and the deviations for ${St=\mathcal{O}(10)}$ are because at this $Re_\lambda$, the inertial range is not large enough for $St=\mathcal{O}(10)$ to reach the regime where the particles' inertia is perturbative (i.e. where $St_r\ll1$).  From the data for $St=0.05$ in Fig.~\ref{Particle_RO_scaling} we also obtain an estimate ${\mathcal{A}^{[1]}\approx 39.13}$.  Using this estimated value for ${\mathcal{A}^{[1]}}$, we may then compare the quantitative prediction of (\ref{LTT2}) with the DNS data.
\begin{figure}[ht]
\centering
\vspace{0mm}
\subfloat[]
{\begin{overpic}
[trim = 30mm 65mm 20mm 75mm,scale=0.5,clip,tics=20]{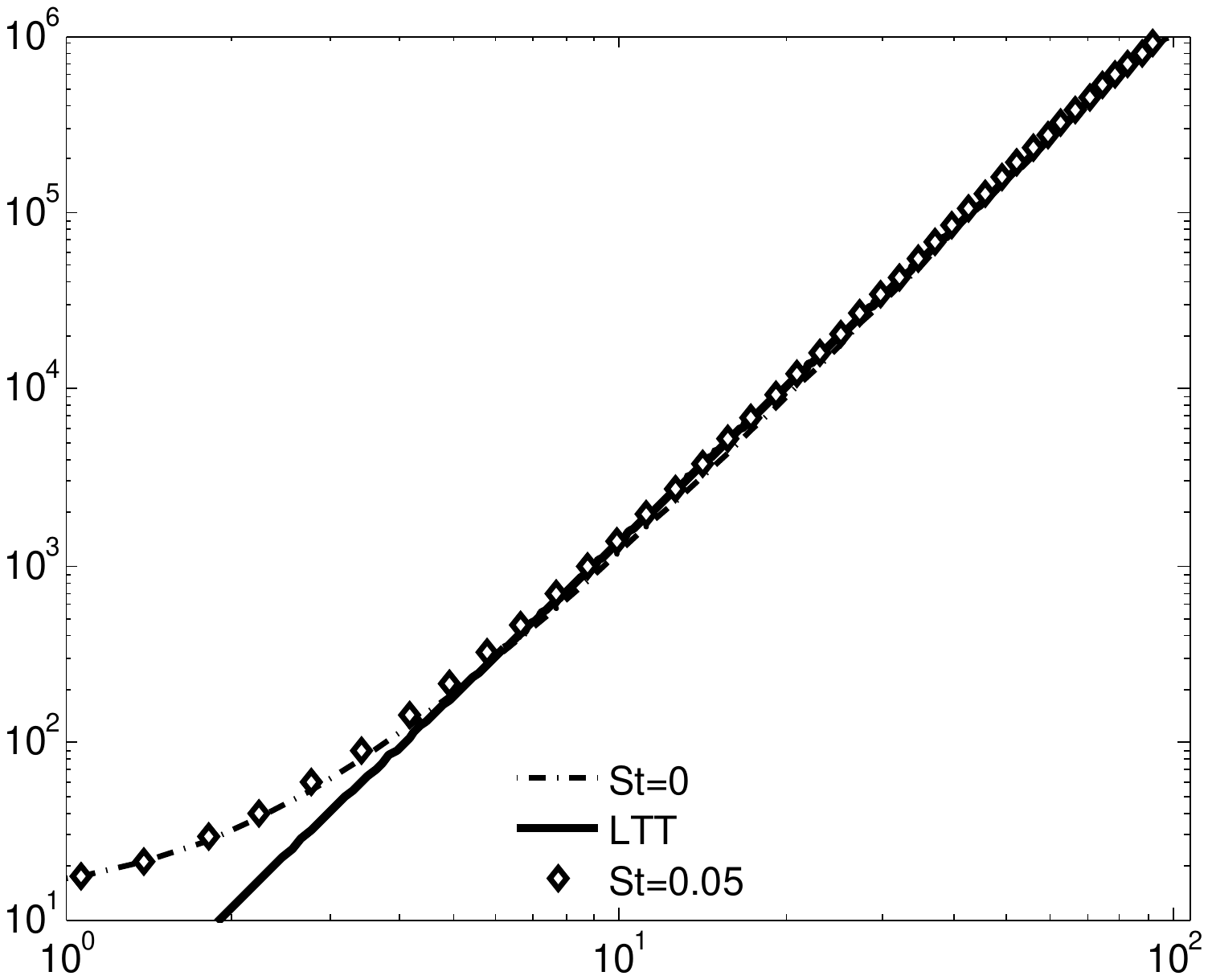}
\put(80,197){\footnotesize\text{$\Big\langle\vert\bm{r}^{p}(-\mathcal{T})\vert^{2}\Big\rangle_{\bm{r}^0}/\eta^2$}}
\put(105,6){$\mathcal{T}/\tau_{\eta}$}
\end{overpic}}
\subfloat[]
{\begin{overpic}
[trim = 30mm 65mm 20mm 75mm,scale=0.5,clip,tics=20]{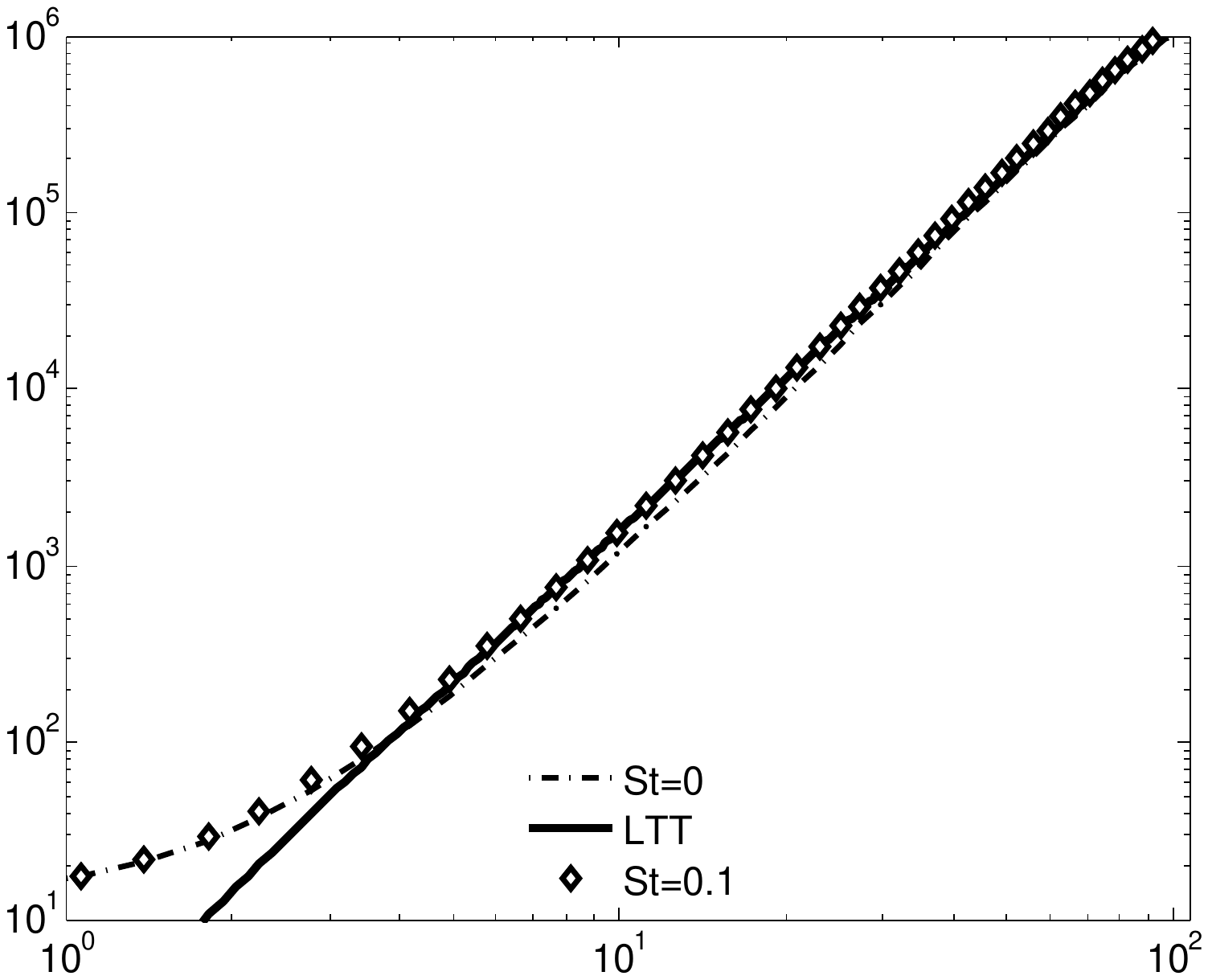}
\put(80,197){\footnotesize\text{$\Big\langle\vert\bm{r}^{p}(-\mathcal{T})\vert^{2}\Big\rangle_{\bm{r}^0}/\eta^2$}}
\put(105,6){$\mathcal{T}/\tau_{\eta}$}
\end{overpic}}\\
\subfloat[]
{\begin{overpic}
[trim = 30mm 65mm 20mm 75mm,scale=0.5,clip,tics=20]{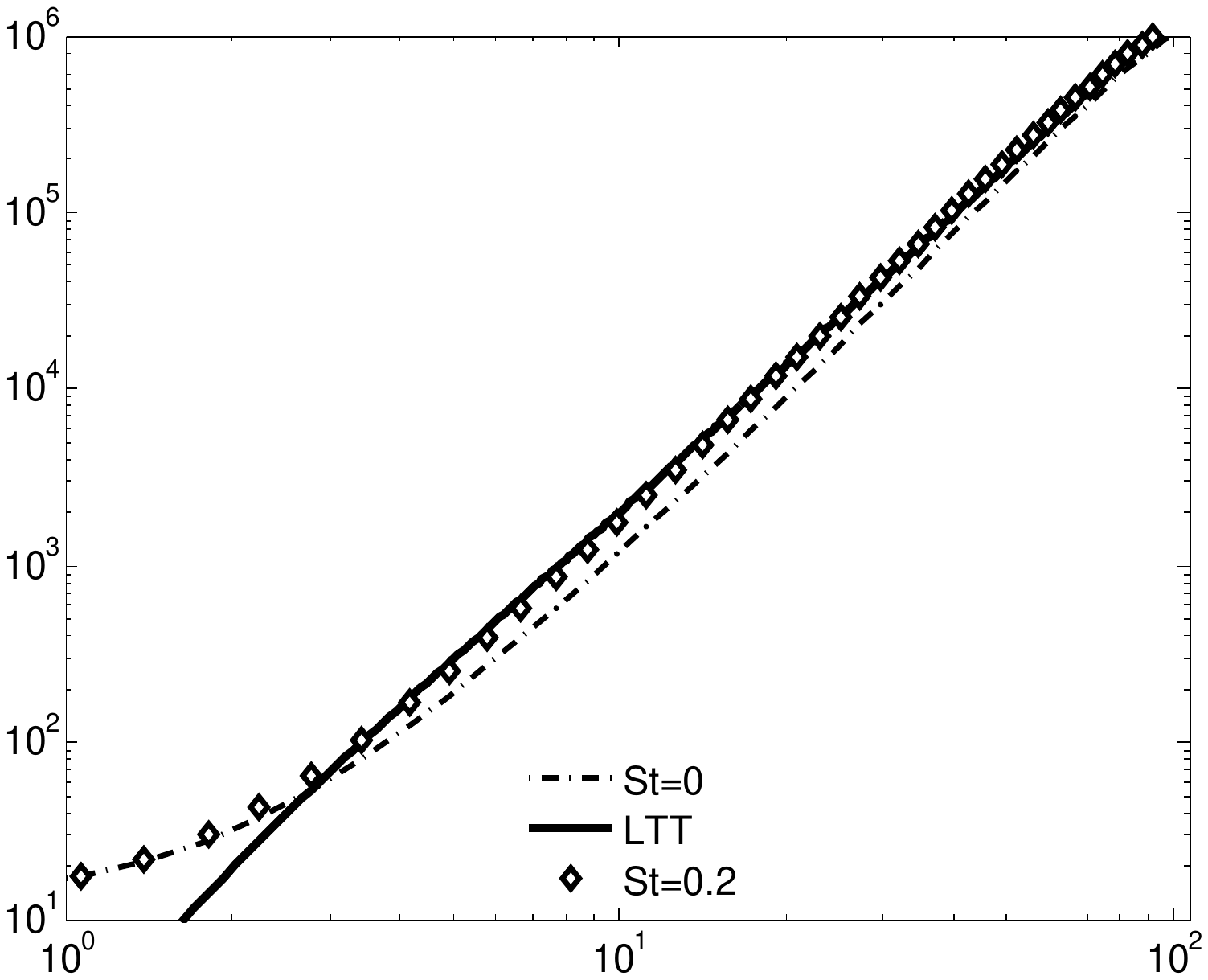}
\put(80,197){\footnotesize\text{$\Big\langle\vert\bm{r}^{p}(-\mathcal{T})\vert^{2}\Big\rangle_{\bm{r}^0}/\eta^2$}}
\put(105,6){$\mathcal{T}/\tau_{\eta}$}
\end{overpic}}
\subfloat[]
{\begin{overpic}
[trim = 30mm 65mm 20mm 75mm,scale=0.5,clip,tics=20]{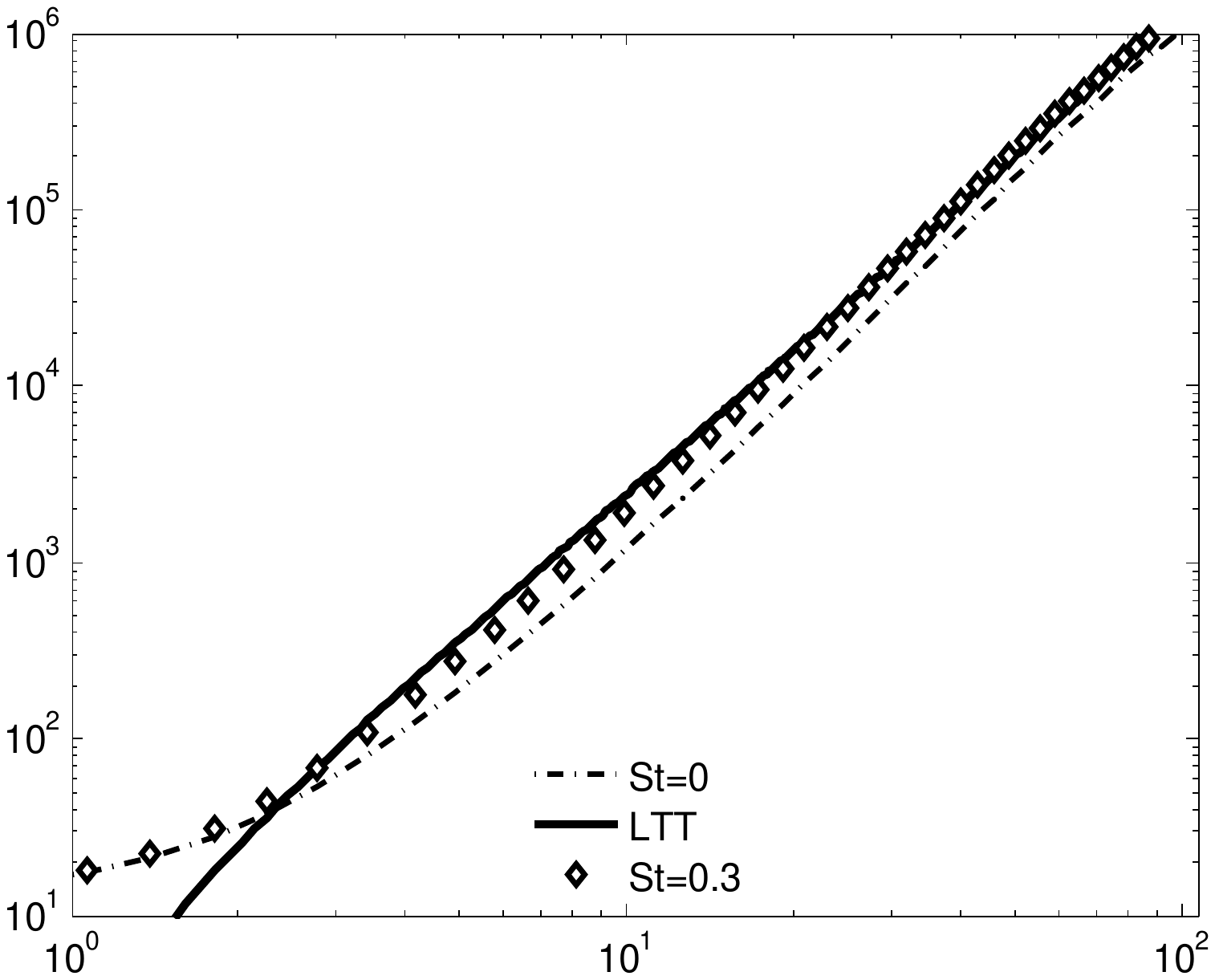}
\put(80,197){\footnotesize\text{$\Big\langle\vert\bm{r}^{p}(-\mathcal{T})\vert^{2}\Big\rangle_{\bm{r}^0}/\eta^2$}}
\put(105,6){$\mathcal{T}/\tau_{\eta}$}
\end{overpic}}
\caption{BIT mean-square separation for (a) $St=0.05$, (b) $St=0.1$, (c) $St=0.2$ and (d) $St=0.3$ and for $r^0\in[3\eta,4\eta]$.  Solid lines are the long-time theory (LTT) predictions from Eq.(\ref{LTT2}), dashed dot lines are the DNS data for $St=0$, and the symbols are DNS data for the respective $St$.}
\label{DNS_Theory2}
\end{figure}
\FloatBarrier
The results in figure~\ref{DNS_Theory2} show that the long-time theory (LTT) predictions from (\ref{LTT2}) are in good agreement with the DNS data, with the LTT accurately capturing the perturbing effect of the particle inertia BIT in the regime $St_r\ll1$, which gives rise to $\langle\vert\bm{r}^{p}(-\mathcal{T})\vert^{2}\rangle_{\bm{r}^0}/\langle\vert\bm{r}^{f}(-\mathcal{T})\vert^{2}\rangle_{\bm{r}^0}\geq 1$.  For $St>0.3$ the effect of $St$ on $\mathcal{A}^{[1]}$ becomes apparent and this affects the predictions of (\ref{LTT2}).  Figure~\ref{DNS_Theory3} shows the results for $St=0.5$ and $St=1$ using the DNS data for $\mathcal{A}^{[1]}$ for these $St$ numbers.  The results show that when the effect of $St$ on $\mathcal{A}^{[1]}$ is accounted for, the LTT describes $\langle\vert\bm{r}^{p}(-\mathcal{T})\vert^{2}\rangle_{\bm{r}^0}$ quite well.
\begin{figure}[ht]
\centering
\vspace{0mm}
\subfloat[]
{\begin{overpic}
[trim = 30mm 65mm 20mm 75mm,scale=0.5,clip,tics=20]{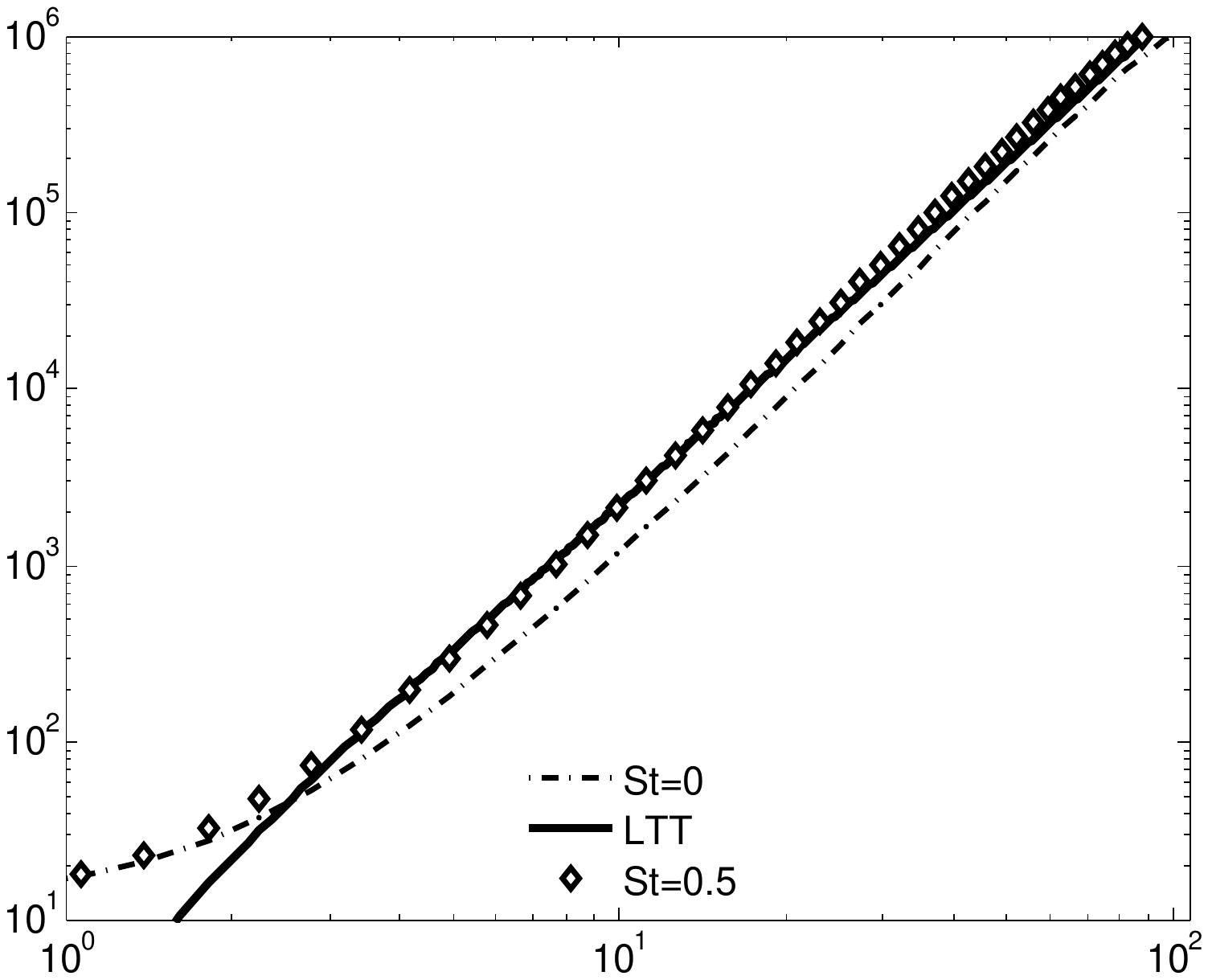}
\put(80,197){\footnotesize\text{$\Big\langle\vert\bm{r}^{p}(-\mathcal{T})\vert^{2}\Big\rangle_{\bm{r}^0}/\eta^2$}}
\put(105,3){$\mathcal{T}/\tau_{\eta}$}
\end{overpic}}
\subfloat[]
{\begin{overpic}
[trim = 30mm 65mm 20mm 75mm,scale=0.5,clip,tics=20]{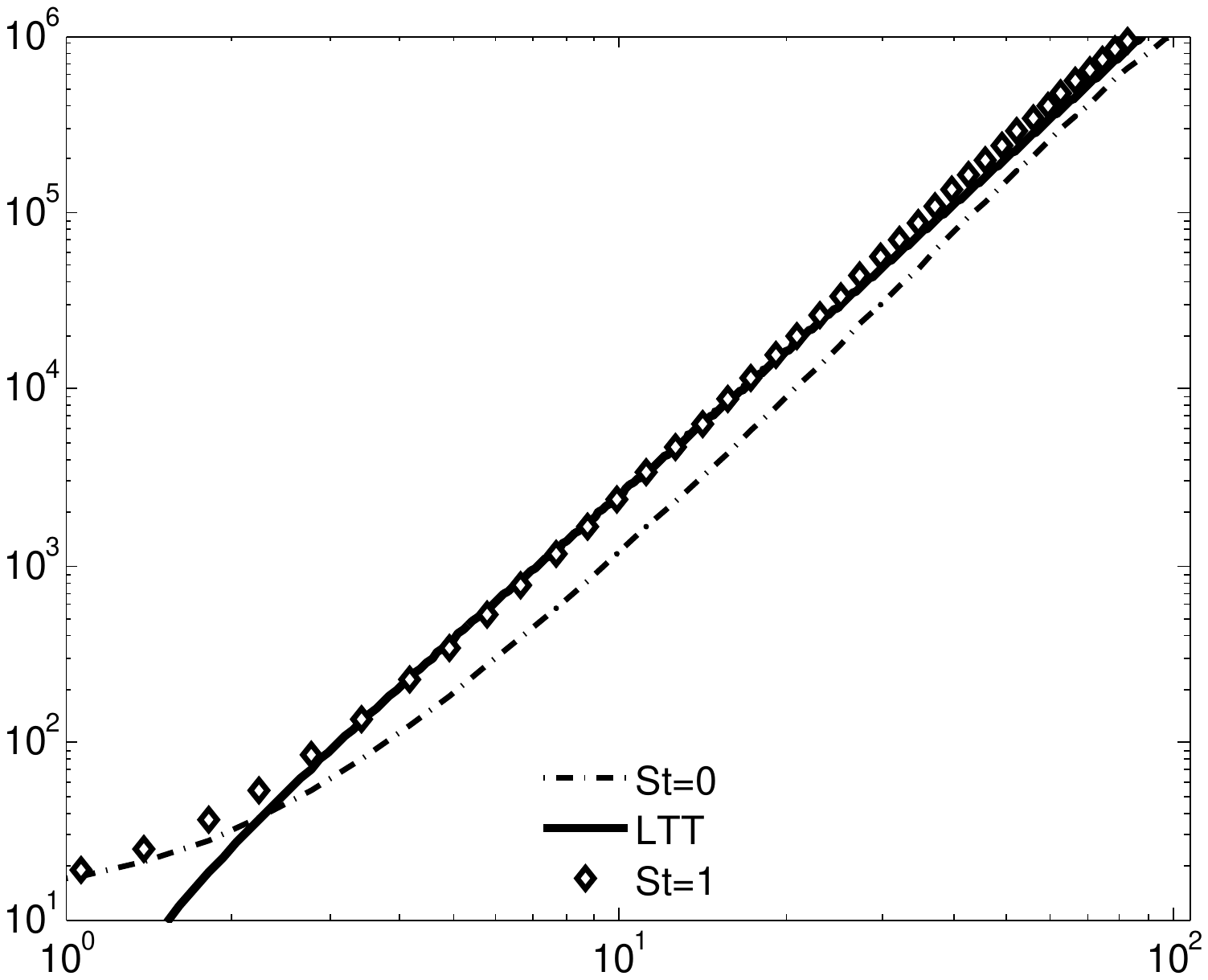}
\put(80,197){\footnotesize\text{$\Big\langle\vert\bm{r}^{p}(-\mathcal{T})\vert^{2}\Big\rangle_{\bm{r}^0}/\eta^2$}}
\put(105,3){$\mathcal{T}/\tau_{\eta}$}
\end{overpic}}
\caption{BIT mean-square separation for (a) $St=0.5$, (b) $St=1$ and for $r^0\in[3\eta,4\eta]$.  Solid lines are the long-time theory (LTT) predictions from Eq.(\ref{LTT2}), dashed dot lines are the DNS data for $St=0$ and the symbols are DNS data for the respective $St$.}
\label{DNS_Theory3}
\end{figure}
\FloatBarrier
\FloatBarrier

\section{Conclusion}\label{conc}

In this paper we have considered the FIT and BIT dispersion of fluid and inertial particles.   The FIT and BIT dispersion of inertial particles are qualitatively and quantitatively different with BIT dispersion occurring at a much greater rate.  In general the irreversibility of inertial particle relative dispersion is much greater than that for fluid particles.  This is because inertial particle pair relative dispersion is subject to both the irreversibility of the underlying turbulent velocity field and also the irreversibility mechanism arising from the non-local contribution to their velocity dynamics.

Concerning the FIT and BIT mean-square dispersion of fluid particles, our DNS data shows that the dispersion is accurately described for small-times by the ballistic law for all separations.  For finite times in the dissipation range we do not observe evidence of the simple exponential law suggested by Batchelor.  However, it is possible that the initial particle separations were not small enough to observe the exponential law before the pair left the dissipation range.  We also observe clear RO scaling for initial separations lying between 3 and 4 $\eta$, and the FIT and BIT Richardson's constants were found to be in excellent agreement with experimental data.  For smaller separations the contaminating effect of the small scale separation behavior means that the RO law is never attained for the time spans we have data, although it does appear to be approaching this asymptotically.  For larger initial separations the limited scale separation in our DNS means that the RO law is not reached by the time the fluid particle pairs are at integral scale separations.

We developed theoretical explanations and predictions for the BIT mean-square dispersion of inertial particles.  The small-time theory, which in the dissipation range describes the dispersion for times $\leq\max[St\tau_\eta,\tau_\eta]$ agrees very well with DNS data, capturing the effects of changes in both $r^0$ and $St$.  We also showed that in the small $St$ regime, the effects of preferential sampling causes the inertial particles to separate slower than fluid particles FIT, but not BIT.  The long-time theory, valid for times $\gg\max[St\tau_\eta,\tau_\eta]$, is essentially based upon an expansion around the fluid particle RO $\mathcal{T}^3$ law.  It predicts that the inertial particle mean-square separation approaches a RO law at a rate $\propto\mathcal{T}^{-1}\log[\mathcal{T}/\hat{\mathcal{T}}]$ and that the particle inertia causes the dispersion to be greater than that for fluid particles, in contrast to the FIT case in \cite{bec10b} where the inertia causes the particles to separate more slowly than the fluid particles in the long-time limit.  The DNS data confirms the predictions of the long-time theory provided that at these times the local Stokes number is small.  In our DNS this is only satisfied for $St\leq\mathcal{O}(1)$ because of the moderate $Re_\lambda$ of the flow.  However, this condition would be satisfied for arbitrary $St$ at sufficiently long-times in the limit ${Re_{\lambda}\to\infty}$. 

The research presented in this paper will be of use for understanding mixing processes of inertial particles in turbulence, which is connected to BIT and not FIT dispersion.  We have shown how dramatically the BIT and FIT dispersion can differ for inertial particles, highlighting how inaccurate it is to approximate them as being equivalent in mixing models for inertial particles.  The work is also important for the development of the theory presented in \cite{pan10} for the relative velocities of inertial particles in isotropic turbulence.  In \cite{pan10} they approximated the BIT mean square separation by its FIT counterpart since no theory or data was available to inform them of the BIT behavior.  In \cite{bragg14c} we argued that this approximation is responsible for some error in their theory predictions.  In future work we intend to develop the theory presented in \cite{pan10} by using the BIT closures developed in this paper.

\section*{Acknowledgements}

The work was supported by the National Science Foundation through CBET grant 0967349, and through a graduate research fellowship to PJI.  Additional funding was provided by Cornell University.  Computational simulations were performed on Yellowstone (ark:/85065/d7wd3xhc) at the U. S. National Center for Atmospheric Research \cite{yellowstone} under grants ACOR00001 and P35091057.

\bibliographystyle{unsrt}
\bibliography{refs_co12}

\end{document}